\documentclass[11pt,a4paper]{article}
\usepackage{amsmath,amssymb,epsfig,a4,
latexsym,axodraw}
\usepackage{slashed}
\usepackage{jheppub}
\usepackage{subfigure}

\def\BB{{\cal B}}

\def\amu{{a_\mu}}
\DeclareMathOperator{\Real}{\Re\hspace{-1pt}\mathfrak{e}}

\newcommand{\FF}[2]{{\cal T}_{#2}^{#1}}
\newcommand{\NV}{nv}
\newcommand{\CV}{cv}
\newcommand{\NS}{ns}
\newcommand{\CS}{cs}
\newcommand{\MuCV}{$\mu$cv}
\newcommand{\MuNV}{$\mu$nv}
\newcommand{\fx}{w}
\newcommand{\colorNumber}{N_C}
\newcommand{\BZ}{{f\tilde{f}_k}}
\newcommand{\BZNeut}{{f\tilde{f}_k}}
\newcommand{\BZChar}{{f'\tilde{f}_k}}

\newcommand{\MssmsNeutTWO}{m_{\tilde{\mu}_m}^2}
\newcommand{\MssmsChar}{m_{\tilde{\nu}_\mu}}
\newcommand{\MssmsCharTWO}{m_{\tilde{\nu}_\mu}^2}
\newcommand{\FTIL}{{{\cal F}}}
\newcommand{\plusswapij}{{(i\leftrightarrow j)}}
\newcommand{\LLNeut}{{L(\MssmsNeutTWO)}}
\newcommand{\LLChar}{{L(\MssmsCharTWO)}}
\newcommand{\tp}{k}
\newcommand{\abbroneloop}{g_{\fx}}

\newcommand{\xfracNeut}{N}
\newcommand{\xfracChar}{C}
\newcommand{\xNeuti}{N_i}
\newcommand{\xNeutj}{N_j}
\newcommand{\xNeutiTWO}{N^2_i}
\newcommand{\xChari}{C_i}
\newcommand{\xCharj}{C_j}

\newcommand{\Dfpf}{{\cal D}_{f' {\tilde f}_{k}}(\ell)}
\newcommand{\DfpfTWO}{{\cal D}_{f' {\tilde f}_{k}}^{\,2}(\ell)}
\newcommand{\atildepp}{\widetilde {{\cal A}} _{i j {\tilde f}_{k}} ^{n\,+}}
\newcommand{\atildepm}{\overline {{\cal A}} _{i j {\tilde f}_{k}} ^{n\,+}}
\newcommand{\atildemp}{\overline {{\cal A}} _{i j {\tilde f}_{k}} ^{n\,-}}
\newcommand{\atildemm}{\widetilde {{\cal A}} _{i j {\tilde f}_{k}} ^{n\,-}}
\newcommand{\btildepp}{\widetilde {{\cal B}} _{i j {\tilde f}_{k}} ^{n\,+}}
\newcommand{\btildepm}{\overline {{\cal B}} _{i j {\tilde f}_{k}} ^{n\,+}}
\newcommand{\btildemp}{\overline {{\cal B}} _{i j {\tilde f}_{k}} ^{n\,-}}
\newcommand{\btildemm}{\widetilde {{\cal B}} _{i j {\tilde f}_{k}} ^{n\,-}}
\newcommand{\caplus}{{\cal A}^{c\,+}_{i j {\tilde f}_{k}}}
\newcommand{\caminu}{{\cal A}^{c\,-}_{i j {\tilde f}_{k}}}
\newcommand{\cbplus}{{\cal B}^{c\,+}_{i j {\tilde f}_{k}}}
\newcommand{\cbminu}{{\cal B}^{c\,-}_{i j {\tilde f}_{k}}}
\newcommand{\acplusijsfermion}{{\cal A}^{c \, +}_{i j {\tilde f}_{k}}}
\newcommand{\acminusijsfermion}{{\cal A}^{c \, -}_{i j {\tilde f}_{k}}}
\newcommand{\bcplusijsfermion}{{\cal B}^{c \, +}_{i j {\tilde f}_{k}}}
\newcommand{\bcminusijsfermion}{{\cal B}^{c \, -}_{i j {\tilde f}_{k}}} 
\newcommand{\amuNgen}{a^{\text{(n-gen)}}_{\mu}}
\newcommand{\amuCgen}{a^{\text{(c-gen)}}_{\mu}}
\newcommand{\amuSUOL}{a_\mu^{\text{1L\,SUSY}}}
\newcommand{\amuOLCha}{a_\mu^{\text{1L\,}\tilde\chi^\pm}}
\newcommand{\amuOLNeu}{a_\mu^{\text{1L\,}\tilde\chi^0}}
\newcommand{\amuFSfclass}[2]{a_{\mu\,{#2}}^{\text{(#1)}}}
\newcommand{\amuCTclass}[2]{a_{\mu\,{#2}}^{\text{(#1-ct)}}}

\newcommand{\amuFSf}{a_{\mu}^{{\rm 2L}, f{\tilde f}}}
\newcommand{\amuSUOLapprox}{a_{\mu}^{{\text{1L\,SUSY,M.I.}}}}
\newcommand{\amuFSfL}{a_{\mu}^{{\rm 2L}, f{\tilde f}\, {\rm LL}}}

\newcommand{\muDR}{\mu_{\text{DRED}}}

\usepackage{etoolbox}

\let\theparentequation\theequation
\patchcmd{\theparentequation}{equation}{parentequation}{}{}

\renewenvironment{subequations}{%
  \refstepcounter{equation}%
  \setcounter{parentequation}{\value{equation}}%
  \setcounter{equation}{0}%
  \def\theequation{\theparentequation\alph{equation}}%
  \ignorespaces
}{%
  \setcounter{equation}{\value{parentequation}}%
  \ignorespacesafterend
}

\newcommand*{\nextParentEquation}{%
  \stepcounter{parentequation}\setcounter{equation}{0}%
}

\allowdisplaybreaks
\begin{document}

\begin{flushright}
\end{flushright}
\vspace{3em}
\begin{center}
{\Large\bf Two-Loop Corrections
 to the Muon Magnetic Moment from Fermion/Sfermion Loops in the MSSM:
 Detailed Results}
\\
\vspace{3em}
{
Helvecio Fargnoli$^{a,b}$,
Christoph Gnendiger$^a$,\\
Sebastian Pa{\ss}ehr$^c$,
Dominik St\"ockinger$^a$,
Hyejung St\"ockinger-Kim$^a$
}\\[2em]
{\sl ${}^a$Institut f\"ur Kern- und Teilchenphysik,TU Dresden, Dresden, Germany}\\
{\sl ${}^b$Universidade Federal de Lavras, Lavras, Brazil}\\
{\sl ${}^c$Max-Planck Institut f\"ur Physik,
M\"unchen, Germany}
\setcounter{footnote}{0}
\end{center}
\vspace{2ex}
\begin{abstract}
{}
Recently, first results were presented for two-loop corrections to the
muon $(g-2)$ from fermion/sfermion loops in the MSSM. These
corrections were shown to be generally large and even
logarithmically enhanced for heavy sfermions. Here, full details of
the calculation and analytical results are presented. Also, a very
compact formula is provided which can be easily implemented and
serves as a good approximation of the full result as a function of the
fourteen most important input parameters. Finally, a thorough
discussion of the numerical behaviour of the fermion/sfermion-loop
corrections to $(g-2)_{\mu}$ is given. The discussion includes the case of
very heavy SUSY masses as well as experimentally allowed scenarios
with very light SUSY masses.
\end{abstract}

\vspace{0.5cm}
\centerline
\newpage
\tableofcontents\newpage
\section{Introduction}
\label{sec:introduction}

\subsection{Current status and motivation}

The measurement of the anomalous magnetic moment of the muon, $a_\mu=(g-2)_\mu/2$,
by the Brookhaven National Laboratory
has reached an accuracy of better than one part per million, 
corresponding to an experimental uncertainty of
$6.3\times10^{-10}$~\cite{Bennett:2006}. With this accuracy,
$a_\mu$~is sensitive to quantum effects from all 
Standard Model~(SM)~interactions---electromagnetic, strong, and weak. 

The theory evaluation of the SM~prediction 
has improved very recently on all fronts. 
In Ref.~\cite{Kinoshita2012}, the
full calculation of the QED~contributions up to the 5-loop~level has been
reported, completing the effort of several decades. The hadronic
vacuum polarization contributions evaluated in
Refs.~\cite{Davier,HMNT,Benayoun:2012wc} make use of a large set of
recent, complementary experimental
data on the~$e^+e^-\to\text{hadrons}$ cross section. An earlier
discrepancy to analy\-ses based on~$\tau$-decays has been
resolved~\cite{JegerlehnerSzafron,Benayoun:2012wc}.
The latest results of hadronic light-by-light~calculations using
established methods~\cite{JegerlehnerNyffeler,dRPV} 
agree within the quoted errors.
New approaches~\cite{Goecke:2010if,Bijnens:2012an,Masjuan:2012qn,Blum:2013qu}
provide important cross-checks and promise further progress. 
The electroweak contributions benefit from the Higgs-mass
determination at the LHC~\cite{ATLAS:2013mma,CMS:yva}. Ref.~\cite{Gnendiger:2013pva}
gives an update of previous calculations of
Refs.~\cite{CKM1,CKM2,CzMV,HSW04,Gribouk}, where the exact
two-loop result for the Higgs-dependent contributions is
obtained and all known electroweak contributions up to the leading
three-loop level are consistently combined. For more details on the SM
prediction and expected further progress see
the recent 
reviews~\cite{JegerlehnerNyffeler,MdRRS}.

With this progress the theory prediction has reached an even higher
accuracy than the experiment.
The current deviation between the Brookhaven measurement and the most 
recent SM~theory evaluations, see Ref.~\cite{Gnendiger:2013pva},
is  as follows (the 
hadronic evaluation is taken either from Ref.\ \cite{Davier} or
\cite{HMNT} as indicated and does not include the evaluations of
Refs.\ \cite{JegerlehnerSzafron,Benayoun:2012wc}):
\begin{align}
\Delta a_{\mu}({\rm E821-SM}) = 
\begin{cases}
(28.7 \pm 8.0 ) \times 10^{-10} \mbox{\cite{Davier}}, \\
(26.1 \pm 8.0 ) \times 10^{-10} \mbox{\cite{HMNT}}.
\end{cases}
\label{eq:deviation}
\end{align}

The importance of this result, which corresponds to a~$3$--$4\sigma$~deviation, has motivated two new
experiments. First, the successor of the BNL experiment is
already under construction at
Fermilab~\cite{Carey:2009zzb,Roberts:2010cj}. It uses the 
same technique as at Brookhaven: high-energy muons are inserted into a storage ring at the ``magic
relativistic~$\gamma$'', for which electric focusing fields do not
perturb the muon precession.
A second experiment is planned at
J-PARC~\cite{Iinuma:2011zz}, which uses ultra-cold muons with smaller~$\gamma$, but
no electric focusing field. Both of these complementary experiments
aim to reduce the uncertainty by more than a factor four.

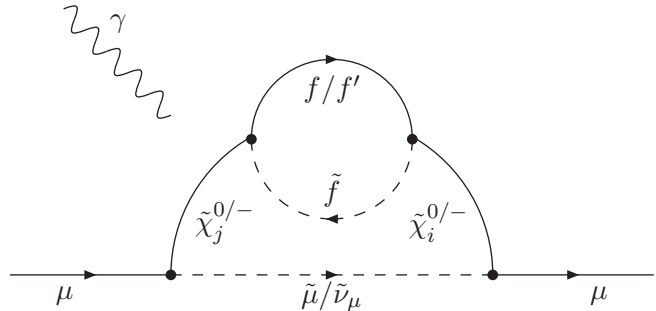
\begin{figure}
\null\hfill
\setlength{\unitlength}{1pt}
\begin{picture}(200,100)(0,0)
\CArc(0,0)(60,0,60)
\CArc(0,0)(60,120,180)
\ArrowArcn(0,51)(30,180,0)
\DashArrowArcn(0,51)(30,0,180){4}
\ArrowLine(-120,0)(-60,0)
\ArrowLine(60,0)(120,0)
\DashArrowLine(-60,0)(60,0){4}
\Photon(-100,100)(-60,60){4}{5}
\Vertex(30,51){2}
\Vertex(-30,51){2}
\Vertex(60,0){2}
\Vertex(-60,0){2}
\Text(-80,95)[]{$\gamma$}
\Text(-100,-8)[]{$\mu$}
\Text(100,-8)[]{${\mu}$}
\Text(0,-8)[c]{$\tilde{\mu}/ \tilde{\nu}_{\mu}$}
\Text(-40,20)[]{$\tilde{\chi} _{j} ^{0/-}$}
\Text(40,20)[]{$\tilde{\chi} _{i} ^{0/-}$}
\Text(0,70)[]{$f/f'$}
\Text(0,32)[]{$\tilde f$}
\end{picture}
\hfill
\caption{Prototype Feynman diagram with fermion/sfermion-loop insertion.
  The outer loop is a generalized SUSY one-loop $a_\mu$ diagram and contains either
  neutralinos~$\tilde{\chi}^{0}_{i,j}$ and a smuon~$\tilde{\mu}$ or
  charginos~$\tilde{\chi}^{-}_{i,j}$ and a sneutrino~$\tilde{\nu}_\mu$.
  The generic fermion/sfermion pair in the inner loop is denoted
  by~$(f,\tilde{f})$ and~$(f',\tilde{f})$ for neutralinos and charginos, respectively.
  The photon can couple to each charged particle.}
\label{fig:prototypes}
\end{figure}

The exciting prospect of such improved measurements motivates all
efforts to further improve the theory prediction for~$a_\mu$, both
within the SM and beyond. The present paper focusses on
supersymmetry~(SUSY) and the prediction for~$a_\mu$ in the
Minimal Supersymmetric Standard Model~(MSSM). Two-loop
diagrams with a closed fermion/sfermion loop inserted into a SUSY
one-loop diagram (see Fig.~\ref{fig:prototypes}) are computed,
together with the associated
counterterm diagrams. First results of the calculation
have been presented already in Ref.~\cite{fsf2loopA}.

It is well-known that the MSSM could easily account for the
deviation~\eqref{eq:deviation}, see e.\,g.~\cite{CzM} and~\cite{review}
for reviews. Even the recent LHC~results, including the
Higgs-mass determination and negative results from SUSY particle
searches, can be simultaneously accommodated in the
MSSM~\cite{Benbrik:2012rm,Arbey:2012dq}. In fact, combining LHC data
with $a_\mu$ motivates MSSM scenarios
which are quite distinct from the more traditionally favoured 
ones~\cite{Endo:2013bba,Ibe:2012qu,1304.2508,Cheng:2013hna,Ibe:2013oha,Mohanty:2013soa,Akula:2013ioa,Evans:2012hg,Endo:2013lva}.
Conversely, e.\,g.~in the Constrained~MSSM, the LHC~results already rule out the possibility to explain the
deviation~\eqref{eq:deviation}~\cite{Bechtle:2012zk,Balazs:2012qc,Buchmueller:2012hv}, further
highlighting the complementarity between LHC and low-energy
observables such as~$a_\mu$. Ref.~\cite{Endo:2013xka} stresses the
complementarity between $a_\mu$ and a future linear
collider. Ref.~\cite{WhitePaper,MdRRS}  
also demonstrates that the future more precise~$a_\mu$ determination
will help in measuring MSSM parameters such as~$\tan\beta$, and in
solving the LHC inverse problem~\cite{Adam:2010uz}, i.\,e.~in
discriminating between discrete choices of MSSM parameters that fit 
equally well to LHC~data. 

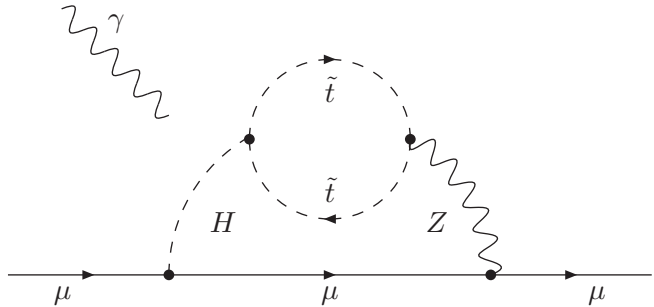
\begin{figure}
\null\hfill
\setlength{\unitlength}{1pt}
\begin{picture}(200,100)(0,0)
\PhotonArc(0,0)(60,0,60){4}{6}
\DashCArc(0,0)(60,120,180){4}
\DashArrowArcn(0,51)(30,180,0){4}
\DashArrowArcn(0,51)(30,0,180){4}
\ArrowLine(-120,0)(-60,0)
\ArrowLine(60,0)(120,0)
\ArrowLine(-60,0)(60,0)
\Photon(-100,100)(-60,60){4}{5}
\Vertex(30,51){2}
\Vertex(-30,51){2}
\Vertex(60,0){2}
\Vertex(-60,0){2}
\Text(-80,95)[]{$\gamma$}
\Text(-100,-8)[]{$\mu$}
\Text(100,-8)[]{${\mu}$}
\Text(0,-8)[c]{$\mu$}
\Text(-40,20)[]{$H$}
\Text(40,20)[]{$Z$}
\Text(0,70)[]{$\tilde t$}
\Text(0,32)[]{$\tilde t$}
\end{picture}
\hfill
\caption{Sample Feynman diagram with closed stop loop inserted into
  a SM-like one-loop diagram with Higgs- and $Z$-boson~exchange,
  computed in Refs.~\cite{HSW03,ArhribBaek,ChenGeng}. The 
  photon can couple to each charged particle.}
\label{fig:HSW03diagram}
\end{figure}

For realizing the full potential of the future~$a_\mu$~experiments,
the theory uncertainty of the~MSSM should be reduced. In Ref.~\cite{review}, it is
estimated to~\mbox{$3\times10^{-10}$} due to unknown two-loop corrections---twice as large as
the future experimental uncertainty. The current status of the MSSM
prediction for $a_\mu$ is as follows: the MSSM one-loop contributions
to $a_\mu$ have been computed and extensively documented in Refs.~\cite{moroi,MartinWells,review,Cho:2011rk}. The two-loop corrections
have been classified in Ref.~\cite{review} into two classes. 
In class~2L(a) a pure SUSY loop of either charginos, neutralinos or
sfermions is inserted into a SM-like diagram. For reference,
Fig.~\ref{fig:HSW03diagram} shows a sample diagram of this
class. Diagrams like this have been computed in Ref.~\cite{HSW03},
after approximate calculations in Refs.~\cite{ArhribBaek,ChenGeng},
and the full calculation of all class~2L(a) contributions has been
completed in Ref.~\cite{HSW04}. Diagrams of class~2L(b) correspond to
two-loop corrections to SUSY one-loop diagrams. This class of
contributions has not been computed fully yet. The
fermion/sfermion-loop corrections of Fig.~\ref{fig:prototypes} belong to this
class.
Up to now, the full
QED~corrections~\cite{amuPhotonicSUSY}, including the leading
QED~logarithms of Ref.~\cite{DG98}, and the $(\tan\beta)^2$-enhanced
corrections~\cite{Marchetti:2008hw} have been evaluated. Further 
computations of selected diagrams of classes~2L(a) and~2L(b) have
been carried out in Refs.~\cite{Feng1,Feng2,Feng3,FengLM06}. 

For a full two-loop calculation of $\amu$ in the MSSM it remains to
exactly compute all non-QED diagrams of class 2L(b), i.e.\ all
two-loop diagrams which contain at least one chargino or neutralino,
one smuon or sneutrino, and potentially further SM or SUSY particles.
These remaining diagrams can be subdivided into diagrams with and
diagrams without a closed fermion/sfermion loop.

Numerically, all the known contributions of class~2L(b) can be as
large as the future experimental uncertainty or even larger.
Hence, these are relevant corrections, and it is motivated to continue the
evaluation of the two-loop contributions to $a_\mu$ in the MSSM.

\subsection{Fermion/sfermion-loop contributions}

The fermion/sfermion-loop contributions discussed in the present work
correspond to the diagrams of
Fig.~\ref{fig:prototypes} and the associated counterterm diagrams. 
Obviously these diagrams can be regarded as
SUSY~partners to the diagrams of the type in
Fig.~\ref{fig:HSW03diagram}. Like the
latter, these 
fermion/sfermion-loop contributions form a gauge 
independent and finite class of contributions.

This class of two-loop contributions is interesting for several reasons. Of
course, its computation represents a significant step towards the full
two-loop computation of $a_\mu$ in the MSSM, and it reduces the
theory uncertainty. The present paper provides full details of the calculation
and analytical and numerical results. It also provides a very
compact approximation formula for the full result, which can be easily
implemented. 

Further, this diagram class introduces a dependence of~$a_\mu$ on
squarks and sleptons of all generations, which is
phenomenologically interesting. Most notably, if the squark masses (or
slepton masses of the first or third generation)
become large, the contributions to~$a_{\mu}$ 
do not decouple, instead they are even logarithmically enhanced.
This is a striking contrast to the Feynman diagrams
considered in the past and illustrated in
Fig.~\ref{fig:HSW03diagram}. Generally, top/stop loops or
top/sbottom loops can have a significant influence
and non-trivial parameter dependence owing to the large top-Yukawa
coupling and the potentially large stop mixing. 
The present paper thoroughly discusses the numerical behaviour of the
fermion/sfermion-loop contributions as a function of all relevant
input parameters.

Finally, the fermion/sfermion-loop contributions involve a set of interesting
counterterm contributions which are finite and do not correspond to genuine
two-loop diagrams. These are counterterm contributions to the
muon--neutralino--smuon or muon--chargino--sneutrino vertex from
counterterm insertions with a fermion/sfermion loop. These
counterterms contain in particular the large, universal~$\Delta\rho$
corrections from top (and stop) loops to the SUSY one-loop diagrams.
The influence of $\Delta\rho$ and the non-decoupling behaviour have
already been stressed and discussed in Ref.~\cite{fsf2loopA}.

Technically, the computation of the diagrams of
Fig.~\ref{fig:prototypes} is significantly more complicated than
all previously considered $a_\mu$ two-loop diagrams in the MSSM.
This is mainly because of the higher number of different mass scales.
Therefore, the diagrams have been computed in two different ways---once by
appropriately extending the standard techniques developed for Refs.\
\cite{HSW03,HSW04}, and once using an iterated one-loop calculation
similar to the simpler cases of Ref.\ \cite{BarrZee}.
A similar class of diagrams with neutralino or gluino exchange and
non-decoupling behaviour has
been considered for electric dipole moments in
Ref. \cite{Yamanaka:2012qn,Yamanaka:2012ia} in an approximation where
higgsino--gaugino mixing is neglected. In Refs.\
\cite{BarrZee,Yamanaka:2012qn,Yamanaka:2012ia} all two-loop diagrams
were ultraviolet finite, while in the present case
diagrams involve subdivergences and need to be renormalized. 

\subsection{Outline}

Our paper is organized as follows: In Sec.~\ref{sec:setup} a
systematic notation for all appearing MSSM coupling constants is
introduced, and it is shown that the one-loop contributions can be
elegantly expressed in terms of these.
Sec.~\ref{sec:cts} is devoted to the renormalization of the two-loop results.
Analytic results for the one-loop counterterm diagrams are provided,
and in particular the difference between the standard one-loop diagrams and the
counterterm diagrams with mixing between two different charginos or
neutralinos is highlighted.
In Sec.~\ref{sec:twoloop} the full analytic results for all two-loop
diagrams are given. Also
intermediate results for the one-loop subdiagrams are expressed in a form
useful for the Barr-Zee~technique. 
The numerical and phenomenological discussion is prepared in
Sec.~\ref{sec:inputparameters} with an overview of the input
parameters. In Sec.~\ref{sec:leadinglog}
 a very compact approximation
formula for the full result is provided which can be easily implemented.
Finally, in Sec.~\ref{sec:numerics} a thorough analysis of the numerical
behaviour of the 
fermion/sfermion-loop contributions
in a variety of parameter scenarios is presented.
\newpage
\section{Preparations}
\label{sec:setup}

As a preliminary step, a useful and compact notation for the MSSM coupling
constants and vertices is introduced. It generalizes the notation used in the
literature and is appropriate for all diagrams considered in the
present paper. Then, the known MSSM one-loop results for $a_\mu$ are
expressed in this simplified notation. 

\subsection{Coupling structures}
\label{sec:couplings}

All one- and two-loop $\amu$ diagrams considered in the present paper
have the structure represented by the prototypes of Fig.~\ref{fig:prototypes}.
Apart from the interaction with the external photon, only vertices of the type
fermion--sfermion--chargino/neutralino appear. The
relevant interaction Lagrangian is written as
\begin{align}
\begin{split}
{\cal L}_{\rm int} 
=\ &\overline{\tilde{\chi}^{-}_{i}}\Big(c^{L }_{i\tilde{\nu}}P_L+c^{R}_{i\tilde{\nu}}P_R\Big)l\,\tilde{\nu}^{\dagger}
 + \overline{\tilde{\chi}^{+}_{i}}\Big(c^{R*}_{i\tilde{l}_{k}^{\dagger}}P_L+c^{L*}_{i\tilde{l}_{k}^{\dagger}}P_R\Big)\nu\,\tilde{l}^{\dagger}_{k} \\
&+ \overline{\tilde{\chi}^{-}_{i}}\Big(c^{L }_{i\tilde{u}_{k}}P_L+c^{R}_{i\tilde{u}_{k}}P_R\Big)d\,\tilde{u}^{\dagger}_{k}
 + \overline{\tilde{\chi}^{+}_{i}}\Big(c^{R*}_{i\tilde{d}_{k}^{\dagger}}P_L+c^{L*}_{i\tilde{d}_{k}^{\dagger}}P_R\Big)u\,\tilde{d}^{\dagger}_{k} \\
&+ \sum_{(f,\tilde{f})} \overline{\tilde{\chi}^{0}_{i}}\Big(n^{L}_{i\tilde{f}_{k}}P_L+n^{R}_{i\tilde{f}_{k}}P_R\Big)f\tilde{f}^{\dagger}_{k} + \text{h.c.}
\end{split}
\label{unmodifiedLagrangian}
\end{align}

In this formula, family indices have been suppressed and~$\nu, l, u, d$ denote
neutrino, charged lepton, up-type and down-type quarks of any
family, respectively. The sum in the last line extends over all these fermions
and the corresponding sfermions.

The down-type sfermions in Eq.~\eqref{unmodifiedLagrangian} couple to the positively charged charginos. 
For the purposes of the muon $(g-2)$~computation it is more appropriate to rewrite these couplings in terms
of negatively charged charginos and anti-up-type fermions. This can be
achieved by using flipping rules~\cite{flippingrules} like
\begin{align}
\overline{\tilde{\chi}^{+}_{i}}\left(c^{R*}_{i\tilde{d}_{k}^{\dagger}}P_L+c^{L*}_{i\tilde{d}_{k}^{\dagger}}P_R\right)
u\,\tilde{d}^{\dagger}_{k} =
\overline{u^c}\left(c^{R*}_{i\tilde{d}_{k}^{\dagger}}P_L+c^{L*}_{i\tilde{d}_{k}^{\dagger}}P_R\right)
\tilde{\chi}^{-}_{i}\,\tilde{d}^{\dagger}_{k}.
\end{align}

Then, it is possible to rewrite the interaction Lagrangian in a
compact and unified form as 
\begin{align}
\label{modifiedLagrangian}
\begin{split}
{\cal L}_{\rm int}=&\sum_{(f',\tilde{f})}\overline{\tilde{\chi}^{-}_{i}}\left(c^{L}_{i\tilde{f}_{k}}P_L+c^{R}_{i\tilde{f}_{k}}P_R\right)
f'\tilde{f}^{\dagger}_{k} \\
&+\sum_{(f,\tilde{f})}\overline{\tilde{\chi}^{0}_{i}}\,\left(n^{L}_{i\tilde{f}_{k}}P_L+n^{R}_{i\tilde{f}_{k}}P_R\right)f\tilde{f}^{\dagger}_{k} + \text{h.c.},\\
\end{split}
\end{align}
where the sums extend over the following fermion/sfermion pairs:
\begin{subequations}
\label{sumsextent}
\begin{align}
\big(f',\tilde{f}\,\big)&=\big(l,\tilde{\nu}\big),\big(\nu^{c},\tilde{l}^{\dagger}\big),\big(d,\tilde{u}\big),\big(u^{c},\tilde{d}^{\dagger}\big),\\
\big(f,,\tilde{f}\,\big)&=\big(\nu,\tilde{\nu}\big),\big(l,\tilde{l}\,\big),\big(u,\tilde{u}\big),\big(d,\tilde{d}\big).
\end{align}
\end{subequations}

In this notation the coupling coefficients in Eq.~\eqref{modifiedLagrangian} are all systematically
indexed by the outgoing sfermion and the chirality of the incoming
fermion.\footnote{Compared to Refs.~\cite{MartinWells,amuPhotonicSUSY} the
notation has been streamlined by removing relative signs and complex
conjugations in~$n^{L,R}$ and $c^{L,R}$.} 

The relevant coupling coefficients in the MSSM are given by
\begin{subequations}
\begin{align}
c^{L}_{i\tilde{\nu}_{l}} & = -g_{2}V^{*}_{i1}, \phantom{\frac{1}{1}} \\
c^{R}_{i\tilde{\nu}_{l}} & = y_{l}U_{i2}, \phantom{\frac{1}{1}} \\
c^{R*}_{i\tilde{l}_{k}^{\dagger}} & = -g_{2}U^{*}_{i1}U^{\tilde{l}_{}}_{k1}+y_{l}U^{*}_{i2}U^{\tilde{l}_{}}_{k2}, \phantom{\frac{1}{1}} \\
c^{L*}_{i\tilde{l}_{k}^{\dagger}} & = 0, \phantom{\frac{1}{1}}\\
\nonumber \\
c^{L}_{i\tilde{u}_{k}} & = -g_{2} V^{*}_{i1}U^{\tilde{u}_{}}_{k1}+y_{u}V^{*}_{i2}U^{\tilde{u}_{}}_{k2},\phantom{\frac{1}{1}} \\
c^{R}_{i\tilde{u}_{k}} & = y_{d}U^{}_{i2}U^{\tilde{u}_{}}_{k1},\phantom{\frac{1}{1}} \\
c^{R*}_{i\tilde{d}_{k}^{\dagger}} & = -g_{2} U^{*}_{i1}U^{\tilde{d}_{}}_{k1}+y_{d}U^{*}_{i2}U^{\tilde{d}_{}}_{k2},\phantom{\frac{1}{1}} \\
c^{L*}_{i\tilde{d}_{k}^{\dagger}} & = y_{u}V^{}_{i2}U^{\tilde{d}_{}}_{k1},\\
\nonumber \\
n^{L}_{i\tilde{u}_{k}} & = \frac{1}{\sqrt{2}}\left(-\frac{1}{3}g_{1}N^{*}_{i1}-g_{2}N^{*}_{i2}\right)U^{\tilde{u}_{}}_{k1}-y_{u}N^{*}_{i4}U^{\tilde{u}_{}}_{k2}, \\
n^{R}_{i\tilde{u}_{k}} & = +\frac{2}{3}\sqrt{2}g_{1}N^{}_{i1}U^{\tilde{u}_{}}_{k2}-y_{u}N^{}_{i4}U^{\tilde{u}_{}}_{k1}, \\
n^{L}_{i\tilde{d}_{k}} & = \frac{1}{\sqrt{2}}\left(-\frac{1}{3}g_{1}N^{*}_{i1}+g_{2}N^{*}_{i2}\right)U^{\tilde{d}_{}}_{k1}-y_{d}N^{*}_{i3}U^{\tilde{d}_{}}_{k2},\\
n^{R}_{i\tilde{d}_{k}} & = -\frac{1}{3}\sqrt{2}g_{1}N^{}_{i1}U^{\tilde{d}_{}}_{k2}-y_{d}N^{}_{i3}U^{\tilde{d}_{}}_{k1},\\
\nonumber \\
n^{L}_{i\tilde{\nu}_{l}} & = \frac{1}{\sqrt{2}}\left(g_{1}N^{*}_{i1}-g_{2}N^{*}_{i2}\right), \\
n^{R}_{i\tilde{\nu}_{l}} & = 0,\phantom{\frac{1}{1}} \\
n^{L}_{i\tilde{l}_{k}} & = \frac{1}{\sqrt{2}}\left(g_{1}N^{*}_{i1}+g_{2}N^{*}_{i2}\right)U^{\tilde{l}_{}}_{k1}-y_{l}N^{*}_{i3}U^{\tilde{l}_{}}_{k2},\\
n^{R}_{i\tilde{l}_{k}} & = -\sqrt{2}g_{1}N^{}_{i1}U^{\tilde{l}_{}}_{k2}-y_{l}N^{}_{i3}U^{\tilde{l}_{}}_{k1},
\end{align}
\end{subequations}
with gauge and Yukawa couplings defined as
\begin{align}
g_{1} &= \frac{e}{c_{W}},& g_{2} &= \frac{e}{s_{W}},& 
y_{d,l} &= \frac{m_{d,l}g_{2}}{\sqrt{2}M_{W}\cos{\beta}}, & y_{u} &= \frac{m_{u}g_{2}}{\sqrt{2}M_{W}\sin{\beta}}.
\label{DefGaugeYukawaCouplings}
\end{align}
The weak mixing angle is defined via the $W$ and $Z$ pole masses:
$s_{W}^2=1-c_{W}^2=1-M_{W}^2/M_{Z}^2$. 
The unitary matrices~$U^{\tilde{f}}$ diagonalize the sfermion-mass matrices, while the unitary matrices~$U/V \text{ and } N$ 
are needed for a Singular Value Decomposition of the chargino- and a Takagi factorization~\cite{Takagi} of the neutralino-mass matrices,
respectively. For these matrices the same notation as in Refs.~\cite{fsf2loopA,review} is used; similarly for
the underlying SUSY parameters:
$\mu$ is the higgsino mass parameter,
$\tan\beta=t_{\beta}=v_u/v_d$ is the ratio of the Higgs doublet vacuum expectation values,
$M_{1,2}$ are the gaugino masses,
and the soft mass parameters for the squark and slepton doublets and singlets are
denoted by $M_{Qi}$, $M_{Ui}$, $M_{Di}$, $M_{Li}$, $M_{Ei}$ for each
generation $i\in\{1,2,3\}$. For simplicity, we choose
generation-independent masses for the first two generations, 
$M_{Q1}=M_{Q2}\equiv M_{Q}$, etc.

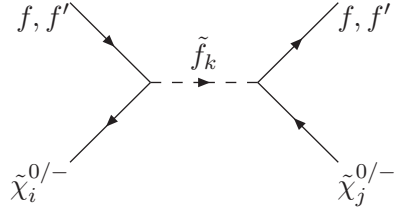
\begin{figure}
\null\hfill
\setlength{\unitlength}{1pt}
\begin{picture}(125,100)(0,0)
\ArrowLine(10,70)(40,40)
\ArrowLine(40,40)(10,10)
\DashArrowLine(40,40)(80,40){4}
\ArrowLine(110,10)(80,40)
\ArrowLine(80,40)(110,70)
\Text(60,45)[b]{$\tilde{f}_k$}
\Text(10,10)[rt]{$\tilde{\chi}^{0/-}_i$}
\Text(110,10)[lt]{$\tilde{\chi}^{0/-}_j$}
\Text(10,70)[rt]{$f,f'$}
\Text(110,70)[lt]{$f,f'$}
\end{picture}
\hfill\null
\caption{Illustration of the coupling combinations
  ${\cal A}_{ij{\tilde f}_{k}}^{z\pm}$, ${\cal B}_{ij{\tilde f}_{k}}^{z\pm}$, $z\in\{n,c\}$,
  arising from sfermion exchange between two neutralino/chargino vertices.}
\label{fig:pairs}
\end{figure}

A common feature of all $a_\mu$ diagrams considered in the present paper, 
see Fig.~\ref{fig:prototypes}, is that the above couplings always appear in pairs, 
associated with the exchange of a sfermion---either of a smuon/sneutrino
from the outer loop or of the generic sfermion from the inner loop.
The structure of these coupling pairs is illustrated in Fig.~\ref{fig:pairs}. 
For the following it is useful to abbreviate the appearing coupling combinations as
\newcommand{\acplusminus}{
 z^{L}_{i{\tilde f}_{k}} \, z^{L \, *}_{j{\tilde f}_{k}} \pm
 z^{R}_{i{\tilde f}_{k}} \, z^{R \, *}_{j{\tilde f}_{k}}
} 
\newcommand{\bcplusminus}{
 z^{L}_{i{\tilde f}_{k}} \, z^{R \, *}_{j{\tilde f}_{k}} \pm
 z^{R}_{i{\tilde f}_{k}} \, z^{L \, *}_{j{\tilde f}_{k}}
}
\begin{subequations}
\begin{align}
{\cal A}^{z \, \pm}_{i j {\tilde f}_{k}} &\equiv \acplusminus, \\ 
{\cal B}^{z \, \pm}_{i j {\tilde f}_{k}} &\equiv \bcplusminus,
\end{align}
\end{subequations}
with~$z\in\{c,n\}$.

The ${\cal A}$~combinations correspond to ``no~chirality~flip'',
and the ${\cal B}$~combinations correspond to ``one~chirality~flip''.
The ${\cal B}$ combinations are therefore always proportional
to the Yukawa coupling and mass of the fermion involved in the
vertex. The left- and right-handed couplings can be equivalently
expressed in terms of scalar and pseudoscalar coefficients as~$z^LP_L+z^RP_R=z^S-z^P\gamma_5$.
This leads to an alternative expression for the ${\cal A}$s and ${\cal B}$s,
\begin{subequations}
\begin{align}
{\cal A}_{ij{\tilde f}_{k}}^{z+}&=
  2z^S_{i{\tilde f}_{k}}z^{S*}_{j{\tilde f}_{k}} +
  2z^P_{i{\tilde f}_{k}}z^{P*}_{j{\tilde f}_{k}},\\
{\cal A}_{ij{\tilde f}_{k}}^{z-}&=
  2z^S_{i{\tilde f}_{k}}z^{P*}_{j{\tilde f}_{k}} +
  2z^P_{i{\tilde f}_{k}}z^{S*}_{j{\tilde f}_{k}},\\
{\cal B}_{ij{\tilde f}_{k}}^{z+} &=
  2z^S_{i{\tilde f}_{k}}z^{S*}_{j{\tilde f}_{k}}-
  2z^P_{i{\tilde f}_{k}}z^{P*}_{j{\tilde f}_{k}},\\
{\cal B}_{ij{\tilde f}_{k}}^{z-} &=
  2z^P_{i{\tilde f}_{k}}z^{S*}_{j{\tilde f}_{k}}-
  2z^S_{i{\tilde f}_{k}}z^{P*}_{j{\tilde f}_{k}},
\end{align}
\end{subequations}
which shows the correspondence of~${\cal A}^{z+}$ and~${\cal B}^{z+}$ to ``even~numbers~of~$\gamma_5$''
on the one hand and~${\cal A}^{z-}$ and~${\cal B}^{z-}$ to ``odd~numbers~of~$\gamma_5$'' on the other hand.
For exchanged indices these coupling combinations satisfy the relations
\begin{subequations}
\begin{align}
{\cal A}^{z \, \pm}_{i j {\tilde f}_{k}}&=
+\left({\cal A}^{z \, \pm}_{j i {\tilde f}_{k}}\right)^* ,\\
{\cal B}^{z \, \pm}_{i j {\tilde f}_{k}}&=
\pm\left({\cal B}^{z \, \pm}_{j i {\tilde f}_{k}}\right)^*.
\end{align}
\end{subequations}
%
\subsection{One-loop results up to ${\cal O}(\epsilon)$ } 
\label{sec:oneloop}
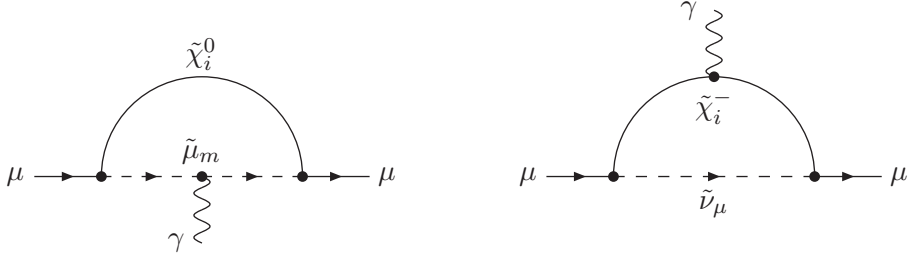
\begin{figure}
\null\hfill
\setlength{\unitlength}{1pt}
\begin{picture}(145,85)(-10,-13)
\Vertex(25,5){2}
\Vertex(100,5){2}
\Vertex(62.5,5){2}
\CArc(62.5,5)(37.5,0,180)
\ArrowLine(0,5)(25,5)
\ArrowLine(100,5)(125,5)
\DashArrowLine(25,5)(62.5,5){4}
\DashArrowLine(62.5,5)(100,5){4}
\Photon(62.5,5)(62.5,-20){3}{3}
\Text(0,5)[r]{$\mu\ $}\Text(125,5)[l]{$\ {\mu}$}
\Text(62.5,10)[b]{$\tilde{\mu}_m$}
\Text(62.5,45)[b]{$\tilde{\chi}^0_i$}
\Text(60,-20)[r]{$\gamma\ $}
\end{picture}
\hfill
\begin{picture}(145,70)(-10,-13)
\Vertex(25,5){2}
\Vertex(100,5){2}
\Vertex(62.5,42.5){2}
\CArc(62.5,5)(37.5,0,180)
\ArrowLine(0,5)(25,5)
\ArrowLine(100,5)(125,5)
\DashArrowLine(25,5)(100,5){4}
\Photon(62.5,42.5)(62.5,67.5){3}{3}
\Text(0,5)[r]{$\mu\ $}\Text(125,5)[l]{$\ {\mu}$}
\Text(62.5,0)[t]{$\tilde{\nu}_\mu$}
\Text(62.6,37.5)[t]{$\tilde{\chi}^-_i$}
\Text(60,67.5)[r]{$\gamma\ $}
\end{picture}
\hfill\null
\caption{\label{fig:oneloopdiags}  SUSY one-loop diagrams with 
neutralino--smuon and chargino--sneutrino exchange.}
\end{figure} 
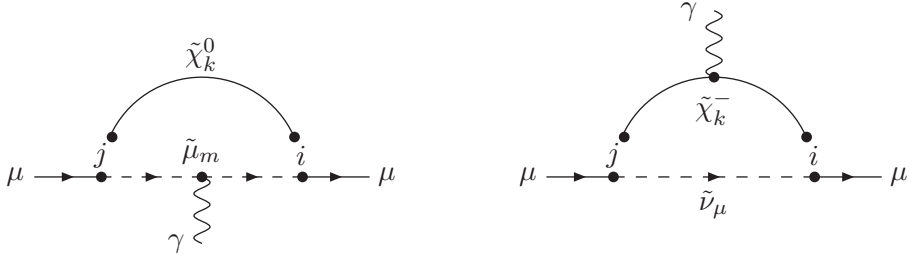
\begin{figure}
\null\hfill
\setlength{\unitlength}{1pt}
\begin{picture}(145,85)(-10,-13)
\SetColor{Black}
\Vertex(25,5){2}
\Text(25,13)[]{${j}$}
\Vertex(100,5){2}
\Text(100,13)[]{${i}$}
\SetColor{Black}
\Vertex(62.5,5){2}
\Vertex(29,20){2}
\Vertex(97,20){2}
\CArc(62.5,5)(37.5,24,156)
\ArrowLine(0,5)(25,5)
\ArrowLine(100,5)(125,5)
\DashArrowLine(25,5)(62.5,5){4}
\DashArrowLine(62.5,5)(100,5){4}
\Photon(62.5,5)(62.5,-20){3}{3}
\Text(0,5)[r]{$\mu\ $}\Text(125,5)[l]{$\ {\mu}$}
\Text(62.5,10)[b]{$\tilde{\mu}_m$}
\Text(62.5,45)[b]{$\tilde{\chi}^0_{{k}}$}
\Text(60,-20)[r]{$\gamma\ $}
\end{picture}
\hfill
\begin{picture}(145,70)(-10,-13)
\SetColor{Black} 
\Vertex(25,5){2}
\Text(25,13)[]{${j}$}
\Vertex(100,5){2}
\Text(100,13)[]{${i}$}
\SetColor{Black}
\Vertex(62.5,42.5){2}
\Vertex(29,20){2}
\Vertex(97,20){2}
\CArc(62.5,5)(37.5,24,156)
\ArrowLine(0,5)(25,5)
\ArrowLine(100,5)(125,5)
\DashArrowLine(25,5)(100,5){4}
\Photon(62.5,42.5)(62.5,67.5){3}{3}
\Text(0,5)[r]{$\mu\ $}\Text(125,5)[l]{$\ {\mu}$}
\Text(62.5,0)[t]{$\tilde{\nu}_\mu$}
\Text(62.6,37.5)[t]{$\tilde{\chi}^-_{{k}}$}
\Text(60,67.5)[r]{$\gamma\ $}
\end{picture}
\hfill\null
\caption{\label{fig:pseudodiags}Generalized diagrams, corresponding to
  Eqs.~(\ref{eq:genericaneu},~\ref{eq:genericacha}). At each vertex
  different chargino/neutralino indices are applied. These
  generalized expressions are useful building blocks for expressing
  one-loop and one-loop counterterm results.}
\end{figure}

In the following we state the SUSY one-loop results up to first order
in the dimensional regularization parameter~$\epsilon=(4-D)/2$.
For reference, the results are expressed in terms of the coupling
combinations introduced above.
The SUSY one-loop contributions are given by the Feynman diagrams of
Fig.~\ref{fig:oneloopdiags}.  
For later purposes, also slightly generalized diagrams are introduced in
Fig.~\ref{fig:pseudodiags}, where different neutralino/chargino
indices are assigned to the vertices and propagators.
It will turn out to be useful to define the following quantities,
corresponding to these generalized diagrams:
\newcommand{\snumu}{{\tilde \nu}_{\mu}}
\newcommand{\smum}{{\tilde \mu}_{m}}
\begin{subequations}
\begin{align}
\amuNgen\hspace{-.5ex}\left({\cal A}_{ji \smum}^{n \pm},
{\cal B}_{ji \smum}^{n \pm}, 
k\right)&\equiv\frac{-1}{16\pi^2}\frac{m_{\mu}^{2}}{m_{\smum}^2}
\Big\{
\frac{1}{12}{\cal A}^{n \pm}_{ji \smum}
{\cal F}_1^N(x_{k})
+\frac{m_{\tilde{\chi}^0_i}}{6m_{\mu}}
\BB^{n \pm}_{ji \smum} {\cal F}_2^N(x_{k})
\Big\}, 
\label{eq:genericaneu}
\\
\amuCgen\hspace{-.5ex}\left({\cal A}_{ji \snumu}^{c \pm},
{\cal B}_{ji \snumu}^{c \pm},
k\right) &\equiv \frac{1}{16 \pi^2}\,\frac{m_{\mu}^{2}}{m_{\snumu}^{2}}\Big\{
\frac{1}{12}{\cal A}^{c \pm}_{ji \snumu}
{\cal F}_1^C(x_k)
+\frac{m_{\tilde{\chi}^-_i}}{3m_{\mu}}
{\cal B}^{c \pm}_{ji \snumu} {\cal F}_2^C(x_k)
\Big\}. 
\label{eq:genericacha} 
\end{align}
\end{subequations}

The dimensionless mass ratios 
are defined as
$x_k = m_{\tilde{\chi}^0_k}^2 / m_{\tilde{\mu}_m}^2$ and $x_k = m_{\tilde{\chi}^-_k}^2 / m_{\tilde{\nu}_\mu}^2$ 
for neutralinos and charginos, respectively. 
The $\epsilon$-dependent loop functions ${\cal F}_{1,2}^{N,C}(x)$ have been given in Ref.~\cite{amuPhotonicSUSY} 
and are listed for reference in appendix~\ref{app:loopf}. 

The known SUSY one-loop results for neutralino--smuon and chargino--sneutrino
loops can then be expressed in terms of these generic results as
\begin{subequations}
\begin{align}
\label{eq:oneloopneu} 
\amuOLNeu&=\sum_{i,m}\amuNgen\hspace{-.5ex}\left({\cal A}_{ii \smum}^{n\,+},
{\cal B}_{ii \smum}^{n\,+}, 
i\right)
,\\
\amuOLCha&=\sum_{i}\amuCgen\hspace{-.5ex}\left({\cal A}_{ii \snumu}^{c\,+},
{\cal B}_{ii \snumu}^{c\,+},
i\right),
\end{align}
\end{subequations}
respectively. The total SUSY one-loop contribution is $\amuSUOL=\amuOLNeu+\amuOLCha$.

It shall be stressed that these one-loop contributions involve only the
``plus''-coupling combinations ${\cal A}^{z+}$, ${\cal B}^{z+}$ and only
diagonal indices $ii\smum$, $ii\snumu$. This will change later
in the case of the considered counterterm and two-loop diagrams.
\newpage
\section{Renormalization and counterterms}
\label{sec:cts}

\begin{figure}[t]
\begin{center}
\null\hfill
\fbox{
\scalebox{.85}{
\begin{picture}(125,90)(-5,-25)
\Text(0,47.5)[lb]{(\MuNV)}
\Text(25,5)[c]{\boldmath{$\times$}}
\Vertex(100,5){2}
\Vertex(62.5,5){2}
\CArc(62.5,5)(37.5,0,180)
\ArrowLine(0,5)(25,5)
\ArrowLine(100,5)(125,5)
\DashArrowLine(25,5)(62.5,5){4}
\DashArrowLine(62.5,5)(100,5){4}
\Photon(62.5,5)(62.5,-20){3}{3}
\Text(0,5)[r]{$\mu\ $}\Text(125,5)[l]{$\ {\mu}$}
\Text(62.5,10)[b]{$\tilde{\mu}_m$}
\Text(62.5,45)[b]{$\tilde{\chi}^0_i$}
\Text(60,-20)[r]{$\gamma\ $}
\end{picture}
\quad
\scalebox{1}{
\begin{picture}(5,90)(3,0)
\Text(0,45)[lb]{+}
\end{picture}
\quad
}
\begin{picture}(125,90)(10,-25)
\Text(100,5)[c]{\boldmath{$\times$}}
\Vertex(25,5){2}
\Vertex(62.5,5){2}
\CArc(62.5,5)(37.5,0,180)
\ArrowLine(0,5)(25,5)
\ArrowLine(100,5)(125,5)
\DashArrowLine(25,5)(62.5,5){4}
\DashArrowLine(62.5,5)(100,5){4}
\Photon(62.5,5)(62.5,-20){3}{3}
\Text(0,5)[r]{$\mu\ $}\Text(125,5)[l]{$\ {\mu}$}
\Text(62.5,10)[b]{$\tilde{\mu}_m$}
\Text(62.5,45)[b]{$\tilde{\chi}^0_i$}
\Text(60,-20)[r]{$\gamma\ $}
\end{picture}
}}
\hfill\null
\\
\null\hfill
\fbox{
\scalebox{.85}{
\begin{picture}(125,90)(-5,-15)
\Text(0,57.5)[lb]{(\MuCV)}
\Text(25,5)[c]{\boldmath{$\times$}}
\Vertex(100,5){2}
\Vertex(62.5,42.5){2}
\CArc(62.5,5)(37.5,0,180)
\ArrowLine(0,5)(25,5)
\ArrowLine(100,5)(125,5)
\DashArrowLine(25,5)(100,5){4}
\Photon(62.5,42.5)(62.5,67.5){3}{3}
\Text(0,5)[r]{$\mu\ $}\Text(125,5)[l]{$\ {\mu}$}
\Text(62.5,0)[t]{$\tilde{\nu}_\mu$}
\Text(62.6,37.5)[t]{$\tilde{\chi}^-_i$}
\Text(60,67.5)[r]{$\gamma\ $}
\end{picture}
\quad
\scalebox{1}{
\begin{picture}(5,90)(3,0)
\Text(0,45)[lb]{+}
\end{picture}
\quad
}
\begin{picture}(125,90)(10,-15)
\Text(100,5)[c]{\boldmath{$\times$}}
\Vertex(25,5){2}
\Vertex(62.5,42.5){2}
\CArc(62.5,5)(37.5,0,180)
\ArrowLine(0,5)(25,5)
\ArrowLine(100,5)(125,5)
\DashArrowLine(25,5)(100,5){4}
\Photon(62.5,42.5)(62.5,67.5){3}{3}
\Text(0,5)[r]{$\mu\ $}\Text(125,5)[l]{$\ {\mu}$}
\Text(62.5,0)[t]{$\tilde{\nu}_\mu$}
\Text(62.6,37.5)[t]{$\tilde{\chi}^-_i$}
\Text(60,67.5)[r]{$\gamma\ $}
\end{picture}
}}
\hfill\null
\caption{\label{fig:finctcases}
The two classes of counterterm diagrams with counterterm insertions at
the external muon vertices. To these counterterm diagrams no corresponding
two-loop diagrams exist. The crosses denote counterterm insertions. 
}
\end{center}
\end{figure}
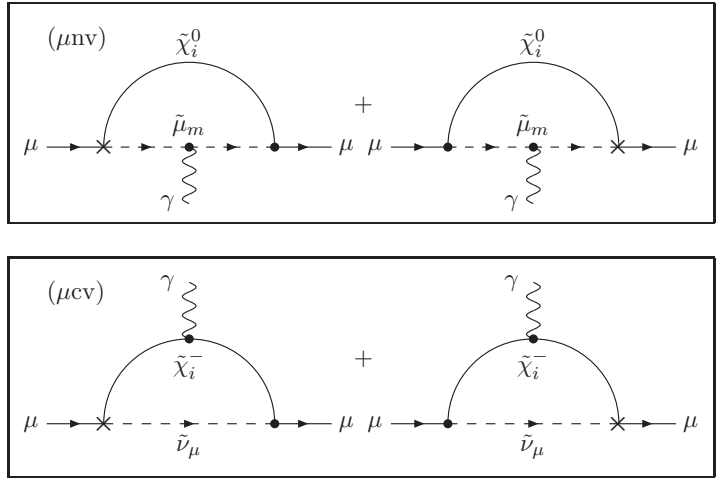

In this and the subsequent section detailed results of the calculation
of the fermion/sfermion-loop contributions to $a_\mu$ are presented.
This section is devoted to renormalization and the counterterm
contributions. The fermion/sfermion-loop contributions  
$\amuFSf$ are defined as all two-loop diagrams, where a
mixed fermion/sfermion loop is inserted into a SUSY one-loop correction
to $a_\mu$, see Fig.~\ref{fig:prototypes}, and all SUSY one-loop
counterterm diagrams with counterterm insertions from diagrams with
only fermions and/or sfermions in the loop.\footnote{%
Diagrams where a pure sfermion loop is attached to a smuon
or sneutrino propagator, and the associated counterterm
diagrams are excluded. These diagrams arise from sfermion four-point
interactions and effectively induce a shift in one of the smuon
masses; they are separately gauge independent and finite,
straightforward to compute and lead to numerically smaller results. 
They will be reported on elsewhere.}
Like pure fermion-loop contributions in the SM, this class of
contributions is gauge independent and finite by itself.

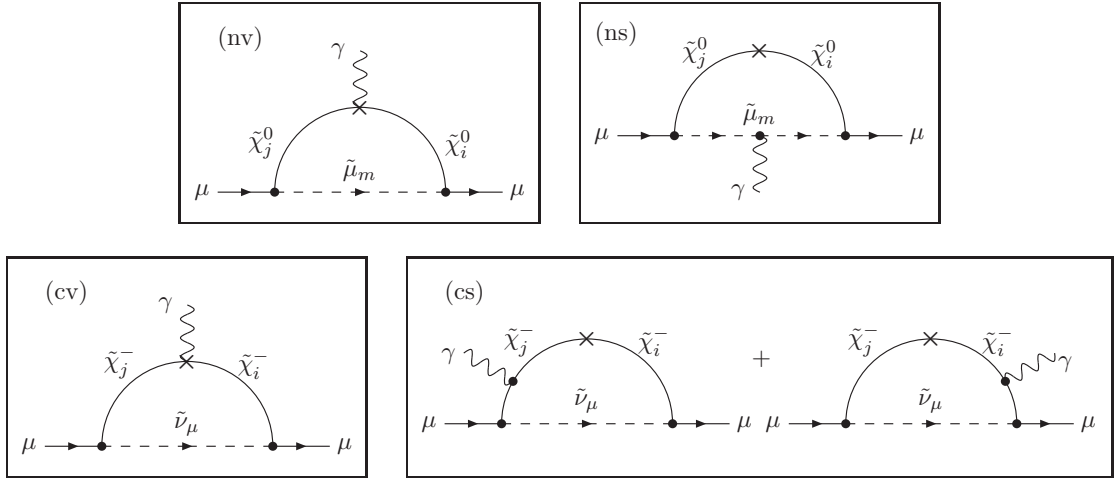
\begin{figure}[t]
\begin{center}
\null\hfill
\fbox{
\scalebox{.85}{
\begin{picture}(135,90)(-5,-5)
\Text(0,67.5)[lb]{(\NV)}
\Vertex(25,5){2}
\Vertex(100,5){2}
\CArc(62.5,5)(37.5,0,180)
\ArrowLine(0,5)(25,5)
\ArrowLine(100,5)(125,5)
\DashArrowLine(25,5)(100,5){4}
\Photon(62.5,42.5)(62.5,67.5){3}{3}
\Text(62.5,42.5)[c]{\boldmath{$\times$}}
\Text(0,5)[r]{$\mu\ $}
\Text(125,5)[l]{$\ {\mu}$}
\Text(62.5,10)[b]{$\tilde{\mu}_m$}
\Text(25,20)[br]{$\tilde{\chi}^0_j$}
\Text(100,20)[bl]{$\tilde{\chi}^0_i$}
\Text(60,67.5)[r]{$\gamma\ $}
\end{picture}
}
}
\quad
\fbox{
\scalebox{.85}{
\begin{picture}(135,90)(-5,-30)
\Text(-10,45)[lb]{(\NS)}
\Vertex(25,5){2}
\Vertex(100,5){2}
\Vertex(62.5,5){2}
\Text(62.5,42.5)[c]{\boldmath{$\times$}}
\CArc(62.5,5)(37.5,0,180)
\ArrowLine(0,5)(25,5)
\ArrowLine(100,5)(125,5)
\DashArrowLine(25,5)(62.5,5){4}
\DashArrowLine(62.5,5)(100,5){4}
\Photon(62.5,5)(62.5,-20){3}{3}
\Text(0,5)[r]{$\mu\ $}
\Text(125,5)[l]{$\ {\mu}$}
\Text(62.5,10)[b]{$\tilde{\mu}_m$}
\Text(40,35)[br]{$\tilde{\chi}^0_j$}\Text(85,35)[bl]{$\tilde{\chi}^0_i$}
\Text(60,-20)[r]{$\gamma\ $}
\end{picture}
}
}
\hfill\null
\\
\null\hfill
\fbox{
\scalebox{.85}{
\begin{picture}(135,90)(-5,-5)
\Text(0,67.5)[lb]{(\CV)}
\Vertex(25,5){2}
\Vertex(100,5){2}
\Text(62.5,42.5)[c]{\boldmath{$\times$}}
\CArc(62.5,5)(37.5,0,180)
\ArrowLine(0,5)(25,5)
\ArrowLine(100,5)(125,5)
\DashArrowLine(25,5)(100,5){4}
\Photon(62.5,42.5)(62.5,67.5){3}{3}
\Text(0,5)[r]{$\mu\ $}
\Text(125,5)[l]{$\ {\mu}$}
\Text(62.5,10)[b]{$\tilde{\nu}_\mu$}
\Text(40,35)[br]{$\tilde{\chi}^-_j$}\Text(85,35)[bl]{$\tilde{\chi}^-_i$}
\Text(60,67.5)[r]{$\gamma\ $}
\end{picture}
}
}
\quad
\fbox{
\scalebox{.85}{
\begin{picture}(125,90)(-5,-15)
\Text(0,57.5)[lb]{(\CS)}
\Vertex(25,5){2}
\Vertex(100,5){2}
\Photon(30,23.75)(8.4,36.25){3}{3}
\Text(62.5,42.5)[c]{\boldmath{$\times$}}
\CArc(62.5,5)(37.5,0,180)
\ArrowLine(0,5)(25,5)
\ArrowLine(100,5)(125,5)
\DashArrowLine(25,5)(100,5){4}
\Vertex(30,23.75){2}
\Text(0,5)[r]{$\mu\ $}
\Text(125,5)[l]{$\ {\mu}$}
\Text(62.5,10)[b]{$\tilde{\nu}_\mu$}
\Text(40,35)[br]{$\tilde{\chi}^-_j$}\Text(85,35)[bl]{$\tilde{\chi}^-_i$}
\Text(5,35)[r]{$\ \gamma$}
\end{picture}
\quad
\scalebox{1}{
\begin{picture}(5,90)(3,0)
\Text(0,45)[lb]{+}
\end{picture}
\quad
}
\begin{picture}(125,90)(10,-15)
\Vertex(25,5){2}
\Vertex(100,5){2}
\Text(62.5,42.5)[c]{\boldmath{$\times$}}
\CArc(62.5,5)(37.5,0,180)
\ArrowLine(0,5)(25,5)
\ArrowLine(100,5)(125,5)
\DashArrowLine(25,5)(100,5){4}
\Photon(95,23.75)(116.6,36.25){3}{3}
\Vertex(95,23.75){2}
\Text(0,5)[r]{$\mu\ $}
\Text(125,5)[l]{$\ {\mu}$}
\Text(62.5,10)[b]{$\tilde{\nu}_\mu$}
\Text(40,35)[br]{$\tilde{\chi}^-_j$}\Text(85,35)[bl]{$\tilde{\chi}^-_i$}
\Text(120,35)[t]{$\ \gamma$}
\end{picture}
}}
\hfill\null
\caption{\label{fig:ctcases}
The four classes of counterterm diagrams for which corresponding
non-vanishing two-loop diagrams exist.
In the (\CS)~case the sum of the two contributing diagrams is
considered. The crosses denote counterterm insertions. 
}
\end{center}
\end{figure}

The one-loop corrections to~$a_\mu$ are UV finite. However,
divergences arise in the calculation of the two-loop corrections,
shown in section~\ref{sec:twoloop}. Therefore, counterterms have to be
introduced, together with appropriate renormalization constants.
The counterterm diagrams can be classified according to
their topologies into six classes:
\begin{itemize}
  \item muon vertex counterterm diagrams with
    insertions of renormalization constants in the vertex with the
    incoming/outgoing muon of the one-loop neutralino or chargino
    diagrams, see Fig.~\ref{fig:finctcases}(\MuNV, \MuCV). There are no
    corresponding two-loop diagrams with fermion/sfermion loops, hence
    these counterterm diagrams are finite.
  \item neutralino counterterm diagrams with insertions of
    renormalization constants into a neutralino--neutralino--photon
    vertex or the neutralino self-energy, see Fig.~\ref{fig:ctcases}(\NV,\NS).
  \item chargino counterterm diagrams with insertions of
    renormalization constants into the vertex with the external photon
    or the chargino self-energy, see Fig.~\ref{fig:ctcases}(\CV,\CS).
\end{itemize}
All renormalization constants have to be computed from one-loop diagrams
involving only fermions and/or sfermions in the loop.


\subsection{Definition of~$\alpha$\label{sec:alpha}}

The anomalous magnetic moment of the muon is proportional to the
fine-structure constant~$\alpha$ at the one-loop level.
For the considered class of two-loop corrections
it is necessary to calculate the renormalization of $\alpha$ from
the photon vacuum polarization with sfermion and fermion loops, including 
light quark loops. Since the masses of the light quarks are not known exactly
and large QCD corrections arise, a perturbative evaluation of these
light quark loops is problematic. The definition of $\alpha$ in the
Thomson limit,
$\alpha(0)$, would lead to a large intrinsic uncertainty. To avoid
this issue we choose the parametrization of the electric charge in
terms of~$\alpha(M_{Z})$, defined by\footnote{This definition
  of~$\alpha$ may not be confused with the $\overline{\text{DR}}$ or
  $\overline{\text{MS}}$ definition of a running~$\alpha(\mu)$.}
\begin{align}
  \alpha(M_{Z}) = \frac{\alpha(0)}{1 - \Delta \alpha(M_{Z})}.
\end{align}
The finite shift~$\Delta \alpha(M_{Z})$ is defined as the on-shell
renormalized photon vacuum polarization from SM leptons and
quarks. The light quark contribution can be obtained from experimental
data via the optical theorem and dispersion relations. We use the
recent determination by Ref.~\cite{HMNT}: 
\begin{align}
  \begin{split}
    \Delta \alpha &= \Delta \alpha_{\text{leptonic}} + \Delta \alpha_{\text{hadronic}} + \Delta \alpha_{\text{top}}\\
                  &= 0.031498 + (0.027626\pm0.000138) - (0.0000728\pm0.0000014).
  \end{split}
\end{align}
The choice of $\alpha(M_Z)$ for the parametrization of the one-loop result leads
to the following renormalization constant for the electric charge:
\begin{align}
\frac{\delta e}{e}=  \delta Z_{e} &= -\frac{1}{2} \left(\delta Z_{AA} - \frac{s_{W}}{c_{W}} \delta Z_{ZA}\right)
\end{align}
with the photon--photon and photon--Z field renormalizations,
defined in terms of the transverse self-energies $\Sigma_{AA}$,
$\Sigma_{ZA}$,
\begin{subequations}
\begin{align}
  \label{eq:dZAA}
  \delta Z_{AA} &= -\Real{\left[\frac{\Sigma_{AA}^{\text{fermions}}(M_{Z}^{2})}{M_{Z}^{2}}\right]} -
    \Real{\left[\partial_{p^{2}}\Sigma_{AA}^{\text{others}}(p^{2})\right]}_{p^{2}=0},\\
  \delta Z_{ZA} &= -\Real{\left[\frac{2\Sigma_{ZA}(0)}{M_{Z}^{2}}\right]}.
\end{align}
\end{subequations}
The mixing self-energy $\Sigma_{ZA}(0)$ has only contributions from
the non-abelian structure of the theory; it is zero for the considered class of Feynman
diagrams. $\Sigma_{AA}^{\text{fermions}}(M_Z)$~refers to all contributions
to the photon self-energy with internal SM leptons and quarks,
evaluated at the scale $M_Z$, while~$\Sigma_{AA}^{\text{others}}$
denotes all other particle insertions, in our case sfermions.  

It should be noted that other schemes, such as replacing $\alpha(0)$
by the muon decay constant $G_{\text{F}}$, avoid the large QCD uncertainties
as well. In Ref.~\cite{fsf2loopA}, Tab.~2, we have shown that the
alternative choice of using $G_{\text{F}}$ instead of $\alpha(M_Z)$ for the
SUSY one-loop contributions would lead to significantly larger
two-loop corrections. Hence we prefer the $\alpha(M_Z)$ parametrization.
A full MSSM calculation of $\amu$ also involves the 
electroweak SM contributions to $a_\mu$. These are usually
parametrized in terms of 
$G_{\text{F}}$, whereby the renormalization scheme for the
SM~\cite{CKM1,CKM2} and SUSY~\cite{HSW03,HSW04} 
loop corrections to these SM one-loop diagrams is
defined accordingly.  It is fully consistent to
parametrize the electroweak SM one-loop contributions with $G_{\text{F}}$ and
the SUSY one-loop contributions with $\alpha(M_Z)$ at the same
time, and we assess this parametrization as optimal for a full
MSSM calculation.

\subsection{Renormalization~constants and scheme}
The necessary renormalization constants correspond to the
renormalization of the physical parameters appearing at the one-loop level, 
\begin{align}
\delta e,\ \delta M_{Z}^{2},\ \delta M_{W}^{2},\ \delta t_{\beta},\ \delta M_{1},\ \delta M_{2},\ \delta\mu,
\end{align}
and to the field renormalization of the photon, photon--$Z$ mixing,
charginos~$\tilde{\chi}^{-}_{i}$ and
neutralinos~$\tilde{\chi}^{0}_{i}$ (the latter two field
renormalization constants cancel in the sum of all counterterm diagrams),
\begin{align}
\delta Z_{AA},\ \delta Z_{ZA},\ \delta Z^{L/R}_{\tilde{\chi}^{-},ij},\ \delta Z^{L/R}_{\tilde{\chi}^{0},ij}.
\end{align}
As stated above, the renormalization constants have to be computed
from one-loop dia\-grams involving only fermions and/or sfermions in the
loop. For this reason, further renormalization constants not listed
above, such as smuon mass or muon field renormalization constants
vanish.

Charge and photon field renormalization have been defined above. The
renormalization scheme defining the remaining renormalization constants
is similar to the scheme of
Refs.~\cite{Denner93,tf,Heidi,Heinemeyer:2010mm,Fritzsche:2011nr,HeinemeyerNew}. It
implements an on-shell renormalization of the MSSM~\cite{HKRRSS} as far as possible. 
The creation and selection of the Feynman diagrams is done with
FeynArts~\cite{FeynArts}, using a preliminary model file of Ref.~\cite{HeinemeyerNew}.
The calculation of the renormalization constants is done with FormCalc~\cite{FormCalc}.
The gauge-boson masses and~$s_{W}$ are renormalized on-shell by requiring
\begin{align}
  \delta M_{Z}^{2} &= \Real\left[\Sigma_{ZZ}(M_{Z}^{2})\right],\\
  \delta M_{W}^{2} &= \Real\left[\Sigma_{WW}(M_{W}^{2})\right],\\
  \delta s_{W} &= \frac{c_{W}^{2}}{2s_{W}}\left(\frac{\delta M_{Z}^{2}}{M_{Z}^{2}} - \frac{\delta M_{W}^{2}}{M_{W}^{2}}\right)
\end{align}
in terms of the transverse self-energies $\Sigma_{ZZ}$ and $\Sigma_{WW}$. The
renormalization constant $\delta s_{W}$ contains the
leading contributions to the quantity $\Delta\rho$ from SM fermion
loops and the leading MSSM corrections to $\Delta\rho$ from sfermion
loops. This and the discussion of the previous subsection show that
the fermion/sfermion-loop corrections to $a_\mu$ are sensitive to the
two universal quantities $\Delta\alpha(M_Z)$ and $\Delta\rho$.

For the parameter $t_{\beta}$ the
$\overline{\text{DR}}$~scheme is chosen which has emerged as the best scheme in
Ref.~\cite{Freitas:2002um}; for alternative process-dependent schemes
see Ref.~\cite{Baro:2008bg}. It can be written in the form given in
Ref.~\cite{HiggsDRbar}, using self-energies of the physical Higgs
bosons $h^0,H^0$, evaluated at zero Higgs mixing angle $\alpha=0$:
\begin{align}
  \delta t_{\beta} &=
  \frac{t_{\beta}}{2}\left.\left(-\Real{\left[{\partial_{p^{2}}\Sigma_{H^{0}H^{0}}(p^{2})}{}\right]}_{\text{div.}}
      +\Real{\left[{{\partial_{p^{2}}}\Sigma_{h^{0}h^{0}}(p^{2})}\right]}_{\text{div.}}\right)\right|_{\alpha=0}.
\end{align}
Counterterms for $\cos{\beta}\equiv c_{\beta}$ and
$\sin{\beta}\equiv s_{\beta}$ can be derived from that.
The reason why $\delta t_{\beta}$ can be reduced to Higgs boson field
renormalization at the one-loop level has been clarified recently in
Refs.~\cite{Sperling:2013eva, Sperling:2013xqa}. 
 
Next, the one-loop masses of the charginos and neutralinos have to
be defined. We choose the renormalization of the chargino/neutralino
sector as detailed in Refs.~\cite{tf,Fritzsche:2011nr,Heinemeyer:2011gk,Bharucha:2012re}.
The lightest neutralino and both charginos are defined on-shell, which fixes the definitions of~$\delta M_{1}, \delta M_{2}
\text{ and } \delta \mu$ in the following way (other schemes can be found in Ref.~\cite{Chatterjee:2011wc,Baro:2009gn}): 
\begin{align}
  \begin{split}
    \delta M_{1} =& \left(N_{11}^{*}\right)^{-2}\\
    &
    \begin{aligned}\times
      \Big\{
      & \Real{\left[\Sigma^{\tilde{\chi}^{0},\text{L}}_{\text{S},11}(m_{\tilde{\chi}^{0}_{1}}^{2})\right]} + m_{\tilde{\chi}^{0}_{1}}\Real{\left[\Sigma^{\tilde{\chi}^{0},\text{L}}_{\text{V},11}(m_{\tilde{\chi}^{0}_{1}}^{2})\right]}
        - \left(N_{12}^{*}\right)^{2}\delta M_{2} + 2N_{13}^{*}N_{14}^{*}\delta\mu\\
      &
        \begin{aligned}
          - 2N_{11}^{*}\big[ & N_{13}^{*}\left(\delta{\left(M_{Z} c_{W}c_{\beta}\right)} - \delta{\left(M_{Z} s_{W}c_{\beta}\right)}\right)\\
                             & - N_{14}^{*}\left(\delta{\left(M_{Z} c_{W}s_{\beta}\right)} - \delta{\left(M_{Z} s_{W}s_{\beta}\right)}\right)\big]\Big\},
        \end{aligned}
    \end{aligned}
  \end{split}\\
  \begin{split}
  \delta M_{2} =& \frac{1}{2}\left(U_{11}^{*}U_{22}^{*}V_{11}^{*}V_{22}^{*} - U_{12}^{*}U_{21}^{*}V_{12}^{*}V_{21}^{*}\right)^{-1}\\
    &
    \begin{aligned}\times
      \Big\{
      & U_{22}^{*}V_{22}^{*}\left(m_{\tilde{\chi}^{-}_{1}}\Real{\left[\Sigma^{\tilde{\chi}^{-},\text{L}}_{\text{V},11}(m_{\tilde{\chi}^{-}_{1}}^{2}) + \Sigma^{\tilde{\chi}^{-},\text{R}}_{\text{V},11}(m_{\tilde{\chi}^{-}_{1}}^{2})\right]}
        + 2\Real{\left[\Sigma^{\tilde{\chi}^{-},\text{L}}_{\text{S},11}(m_{\tilde{\chi}^{-}_{1}}^{2})\right]}\right)\\
      & - U_{12}^{*}V_{12}^{*}\left(m_{\tilde{\chi}^{-}_{2}}\Real{\left[\Sigma^{\tilde{\chi}^{-},\text{L}}_{\text{V},22}(m_{\tilde{\chi}^{-}_{2}}^{2}) + \Sigma^{\tilde{\chi}^{-},\text{R}}_{\text{V},22}(m_{\tilde{\chi}^{-}_{2}}^{2})\right]}
        + 2\Real{\left[\Sigma^{\tilde{\chi}^{-},\text{L}}_{\text{S},22}(m_{\tilde{\chi}^{-}_{2}}^{2})\right]}\right)\\
      & - 2\left(U_{11}^{*}U_{22}^{*} - U_{12}^{*}U_{21}^{*}\right)V_{12}^{*}V_{22}^{*}\delta{\left(\sqrt{2}M_{W}s_{\beta}\right)}\\
      & - 2\left(V_{11}^{*}V_{22}^{*} - V_{12}^{*}V_{21}^{*}\right)U_{12}^{*}U_{22}^{*}\delta{\left(\sqrt{2}M_{W}c_{\beta}\right)}\Big\},
    \end{aligned}
  \end{split}\\
  \begin{split}
  \delta \mu =& \frac{1}{2}\left(U_{11}^{*}U_{22}^{*}V_{11}^{*}V_{22}^{*} - U_{12}^{*}U_{21}^{*}V_{12}^{*}V_{21}^{*}\right)^{-1}\\
    &
    \begin{aligned}\times
      \Big\{
      & U_{11}^{*}V_{11}^{*}\left(m_{\tilde{\chi}^{-}_{2}}\Real{\left[\Sigma^{\tilde{\chi}^{-},\text{L}}_{\text{V},22}(m_{\tilde{\chi}^{-}_{2}}^{2}) + \Sigma^{\tilde{\chi}^{-},\text{R}}_{\text{V},22}(m_{\tilde{\chi}^{-}_{2}}^{2})\right]}
        + 2\Real{\left[\Sigma^{\tilde{\chi}^{-},\text{L}}_{\text{S},22}(m_{\tilde{\chi}^{-}_{2}}^{2})\right]}\right)\\
      & - U_{21}^{*}V_{21}^{*}\left(m_{\tilde{\chi}^{-}_{1}}\Real{\left[\Sigma^{\tilde{\chi}^{-},\text{L}}_{\text{V},11}(m_{\tilde{\chi}^{-}_{1}}^{2}) + \Sigma^{\tilde{\chi}^{-},\text{R}}_{\text{V},11}(m_{\tilde{\chi}^{-}_{1}}^{2})\right]}
        + 2\Real{\left[\Sigma^{\tilde{\chi}^{-},\text{L}}_{\text{S},11}(m_{\tilde{\chi}^{-}_{1}}^{2})\right]}\right)\\
      & - 2\left(U_{11}^{*}U_{22}^{*} - U_{12}^{*}U_{21}^{*}\right)V_{11}^{*}V_{21}^{*} \delta{\left(\sqrt{2}M_{W}c_{\beta}\right)}\\
      & - 2\left(V_{11}^{*}V_{22}^{*} - V_{12}^{*}V_{21}^{*}\right)U_{11}^{*}U_{21}^{*} \delta{\left(\sqrt{2}M_{W}s_{\beta}\right)}\Big\}.
    \end{aligned}
  \end{split}
\end{align}

Here the chiral and covariant decomposition of the fermionic self-energies,
\begin{align}
  \Sigma^{\tilde{\chi}}_{ij}(p^{2}) &= \slashed{p}\left(P_{\text{L}}\Sigma^{\tilde{\chi}, \text{L}}_{\text{V},ij}(p^{2}) + P_{\text{R}}\Sigma^{\tilde{\chi}, \text{R}}_{\text{V},ij}(p^{2})\right)
                                     + \left(P_{\text{L}}\Sigma^{\tilde{\chi}, \text{L}}_{\text{S},ij}(p^{2}) + P_{\text{R}}\Sigma^{\tilde{\chi}, \text{R}}_{\text{S},ij}(p^{2})\right),
\end{align}
has been used. 
Knowing these quantities, the renormalization constants for all
chargino and neutralino masses~$\delta m_{\tilde{\chi}^{-}_{ij}}
\text{ and } \delta m_{\tilde{\chi}^{0}_{ij}}$ can be derived by
applying the usual renormalization procedure for the tree-level mass
matrices $\mathcal{X}$ and $\mathcal{Y}$
\begin{subequations}
\begin{align}
  U^{*}\mathcal{X}V^{\dagger} &= \mathrm{diag}{\left(m_{\tilde{\chi}^{-}_{1}},\,m_{\tilde{\chi}^{-}_{2}}\right)}, &
\left[  U^{*}\delta\mathcal{X}V^{\dagger} \right]_{ij}&= \delta
m_{\tilde{\chi}^{-}_{ij}} ,\\
  N^{*}\mathcal{Y}N^{\dagger} &= \mathrm{diag}{\left(m_{\tilde{\chi}^{0}_{1}},\,m_{\tilde{\chi}^{0}_{2}},m_{\tilde{\chi}^{0}_{3}},\,m_{\tilde{\chi}^{0}_{4}}\right)}, &
\left[  N^{*}\delta\mathcal{Y}N^{\dagger} \right]_{ij}&= \delta
m_{\tilde{\chi}^{0}_{ij}},
\end{align}
\end{subequations}
with
\begin{align}
  \mathcal{X} &= \begin{pmatrix} M_{2} & \sqrt{2}M_{W}s_{\beta}\\
                                 \sqrt{2}M_{W}c_{\beta} & \mu \end{pmatrix},
  & \mathcal{Y} &= \begin{pmatrix} M_{1} & 0 & -M_{Z}s_{W}c_{\beta} & M_{Z}s_{W}s_{\beta}\\
                                  0 & M_{2} & M_{Z}c_{W}c_{\beta} & -M_{Z}c_{W}s_{\beta}\\
                                  -M_{Z}s_{W}c_{\beta} & M_{Z}c_{W}c_{\beta} & 0 & -\mu\\
                                  M_{Z}s_{W}s_{\beta} & -M_{Z}c_{W}s_{\beta} & -\mu & 0\end{pmatrix},
\end{align}
and the matrices $U$, $V$ and $N$ of the tree-level Singular Value Decomposition
and Takagi factorization, respectively. 

It should be pointed out that this renormalization scheme is not a good choice for
all parameter scenarios. It leads to artificially large corrections if
the lightest neutralino is not bino-like which has been discussed in Refs.~\cite{Chatterjee:2011wc,Baro:2009gn} before;
it also introduces an artificial singularity for $M_{2} = \mu$, because of the explicitly appearing combination of
mixing-matrix elements in the denominators of $\delta M_{2}$ and
$\delta \mu$; a solution by using a different renormalization scheme
would be provided in Ref.~\cite{Chatterjee:2011wc}, however in our
numerical examples the exact equality $M_{2}=\mu$ does not appear.

Finally, the chargino and neutralino fields are renormalized according to the
$\overline{\text{DR}}$ definitions
\begin{subequations}
\begin{align}
  \delta Z^{\text{L/R}}_{\tilde{\chi}^{0/-},ii} =& -\Real{\left[\Sigma^{\tilde{\chi}^{0/-},\text{L/R}}_{\text{V},ii}(m_{\tilde{\chi}^{0/-}_{i}})\right]}_{\text{div.}}\\
  \begin{split}
  \delta Z^{\text{L/R}}_{\tilde{\chi}^{0/-},ij} =&\frac{2}{m_{\tilde{\chi}^{0/-}_{i}}^{2} - m_{\tilde{\chi}^{0/-}_{j}}^{2}}\\
    &
    \begin{aligned}\times
      \biggr\{
        & m_{\tilde{\chi}^{0/-}_{j}}\left(m_{\tilde{\chi}^{0/-}_{j}} \Real{\left[\Sigma^{\tilde{\chi}^{0/-}, \text{L/R}}_{\text{V},ij}(m_{\tilde{\chi}^{0/-}_{j}})\right]}
          + m_{\tilde{\chi}^{0/-}_{i}}\Real{\left[\Sigma^{\tilde{\chi}^{0/-}, \text{R/L}}_{\text{V},ij}(m_{\tilde{\chi}^{0/-}_{j}})\right]}\right)\\
        & + m_{\tilde{\chi}^{0/-}_{i}}\Real{\left[\Sigma^{\tilde{\chi}^{0/-}, \text{L/R}}_{\text{S},ij}(m_{\tilde{\chi}^{0/-}_{j}})\right]}
          + m_{\tilde{\chi}^{0/-}_{j}}\Real{\left[\Sigma^{\tilde{\chi}^{0/-}, \text{R/L}}_{\text{S},ij}(m_{\tilde{\chi}^{0/-}_{j}})\right]}\\
        & - m_{\tilde{\chi}^{0/-}_{i}}\delta m^{\text{L/R}}_{\tilde{\chi}^{0/-}_{ji}} - m_{\tilde{\chi}^{0/-}_{j}}\left(\delta m^{\text{L/R}}_{\tilde{\chi}^{0/-}_{ij}}\right)^{*}\biggr\}_{\text{div.}},
    \end{aligned}
  \end{split}
\end{align}
with                       
\begin{align}
\delta m^{\text{L}}_{\tilde{\chi}^{-}_{ij}} &= \delta m_{\tilde{\chi}^{-}_{ij}},&
  \delta m^{\text{R}}_{\tilde{\chi}^{-}_{ij}} &= \delta m_{\tilde{\chi}^{-}_{ji}}^{*},&
  \delta m^{\text{L}}_{\tilde{\chi}^{0}_{ij}} &= \delta m_{\tilde{\chi}^{0}_{ji}}.
\end{align}
\end{subequations}

The $\overline{\text{DR}}$~definition of~$\delta
Z^{\text{L/R}}_{\tilde{\chi}^{-},ij} \text{ and } \delta
Z^{\text{L}}_{\tilde{\chi}^{0},ij}$ leads to results which are numerically more stable
than a corresponding on-shell definition. In any case, the field renormalization
constants of charginos and neutralinos cancel in the sum of
all diagrams, but their inclusion renders the self-energy and vertex
diagrams individually finite.

\subsection{Counterterms}

Now, the explicit results for the six classes of counterterm
diagrams are given in terms of the renormalization contants defined above.
We use the compact notation for the coupling combinations ${\cal A}$ and ${\cal B}$
and the abbreviations 
$\amuNgen({\cal A},{\cal B},k)$ and $\amuCgen({\cal A},{\cal B},k)$
for generalized one-loop results, introduced in
Secs.~\ref{sec:couplings}~and~\ref{sec:oneloop}.
The automated calculation and implementation of the counterterm
diagrams has been done using FeynArts~\cite{FeynArts}, OneCalc (part
of the TuCalc package)~\cite{OneCalc} and the $a_\mu$-specific
routines developed for~\cite{HSW03,HSW04}.


The contribution to~$a_\mu$ from the finite muon-vertex counterterm
diagrams for neutralinos and charginos, see Fig.~\ref{fig:finctcases}(\MuNV,\MuCV),
can be expressed in terms of the abbreviations (\ref{eq:genericaneu}) and (\ref{eq:genericacha})
\begin{subequations}
\begin{align}
\begin{split}
\label{amuMuonVertexCT}
\amuCTclass{\MuNV}{i\smum}=\amuNgen\hspace{-.5ex}\Bigg(
\delta\mathcal{A}^{n+}_{ii\smum} &+ 2\Real{\sum_{k=1}^{4}\Big[v^{n}_{ik} \mathcal{A}^{n+}_{ki\smum}
                                 + a^{n}_{ik}
                                 \mathcal{A}^{n-}_{ki\smum}\Big]},
\\
\delta\mathcal{B}^{n+}_{ii\smum} &+ 2\Real{\sum_{k=1}^{4}\Big[v^{n}_{ik} \mathcal{B}^{n+}_{ki\smum}
                                 + a^{n}_{ik} \mathcal{B}^{n-}_{ki\smum}\Big]},i\Bigg),
\end{split}
\\
\begin{split}
\amuCTclass{\MuCV}{i}=\amuCgen\hspace{-.5ex}\ \Bigg(
\delta\mathcal{A}^{c+}_{ii\tilde{\nu}_\mu} &+ 2\Real{\sum_{j=1}^{2}\Big[v^{c}_{ij} \mathcal{A}^{c+}_{ji\tilde{\nu}_\mu}
                                 + a^{c}_{ij} \mathcal{A}^{c-}_{ji\tilde{\nu}_\mu}\Big]},
\\
\delta\mathcal{B}^{c+}_{ii\tilde{\nu}_\mu} &+ 2\Real{\sum_{j=1}^{2}\Big[v^{c}_{ij} \mathcal{B}^{c+}_{ji\tilde{\nu}_\mu}
                                 + a^{c}_{ij} \mathcal{B}^{c-}_{ji\tilde{\nu}_\mu}\Big]},
i\Bigg) 
\end{split}
\end{align}
\end{subequations}
with
\begin{subequations}
\begin{align}
  \delta\mathcal{A}^{z+}_{ij\tilde{f}} &= \left(z^{L}_{i\tilde{f}} \delta z^{L*}_{j\tilde{f}} + z^{L*}_{j\tilde{f}} \delta z^{L}_{i\tilde{f}}\right) +
                                           \left(z^{R}_{i\tilde{f}} \delta z^{R*}_{j\tilde{f}} + z^{R*}_{j\tilde{f}} \delta z^{R}_{i\tilde{f}}\right),\\*
  \delta\mathcal{A}^{z-}_{ij\tilde{f}} &= \left(z^{L}_{i\tilde{f}} \delta z^{L*}_{j\tilde{f}} + z^{L*}_{j\tilde{f}} \delta z^{L}_{i\tilde{f}}\right) -
                                           \left(z^{R}_{i\tilde{f}} \delta z^{R*}_{j\tilde{f}} + z^{R*}_{j\tilde{f}} \delta z^{R}_{i\tilde{f}}\right),\\*
  \delta\mathcal{B}^{z+}_{ij\tilde{f}} &= \left(z^{L}_{i\tilde{f}} \delta z^{R*}_{j\tilde{f}} + z^{L*}_{j\tilde{f}} \delta z^{R}_{i\tilde{f}}\right) +
                                           \left(z^{R}_{i\tilde{f}} \delta z^{L*}_{j\tilde{f}} + z^{R*}_{j\tilde{f}} \delta z^{L}_{i\tilde{f}}\right),\\*
  \delta\mathcal{B}^{z-}_{ij\tilde{f}} &= \left(z^{L}_{i\tilde{f}} \delta z^{R*}_{j\tilde{f}} + z^{L*}_{j\tilde{f}} \delta z^{R}_{i\tilde{f}}\right) -
                                           \left(z^{R}_{i\tilde{f}} \delta z^{L*}_{j\tilde{f}} + z^{R*}_{j\tilde{f}} \delta z^{L}_{i\tilde{f}}\right),\\
  v^{n}_{ij} &= \frac{1}{4}\left(\delta Z^{L}_{\tilde{\chi}^{0},ji} + \delta Z^{L*}_{\tilde{\chi}^{0},ji}\right),\\*
  a^{n}_{ij} &= \frac{1}{4}\left(\delta Z^{L}_{\tilde{\chi}^{0},ji} - \delta Z^{L*}_{\tilde{\chi}^{0},ji}\right),\\*
  v^{c}_{ij} &= \frac{1}{4}\left(\delta Z^{R*}_{\tilde{\chi}^{-},ji} + \delta Z^{L*}_{\tilde{\chi}^{-},ji}\right),\\*
  a^{c}_{ij} &= \frac{1}{4}\left(\delta Z^{R*}_{\tilde{\chi}^{-},ji} - \delta Z^{L*}_{\tilde{\chi}^{-},ji}\right),\\
\intertext{for $  z \in \{c, n\}$ and $\tilde{f}\in\{\tilde{\nu}_\mu,\tilde{\mu}_m\}$, and}
  \nextParentEquation
  \delta c^{L}_{i\tilde{\nu}_{\mu}} &= -\delta g_{2} V^{*}_{i1},\\*
  \delta c^{R}_{i\tilde{\nu}_{\mu}} &= \delta y_{\mu} U_{i2},\\
  \delta n^{L}_{i\tilde{\mu}_m} &= \frac{1}{\sqrt{2}}\left(\delta g_{1}N^{*}_{i1} + \delta g_{2} N^{*}_{i2}\right) U^{\tilde{\mu}}_{m1} - \delta y_{\mu} N^{*}_{i3}U^{\tilde{\mu}}_{m2},\\*
  \delta n^{R}_{i\tilde{\mu}_m} &= -\sqrt{2}\delta g_{1}N_{i1}U^{\tilde{\mu}}_{m2} - \delta y_{\mu} N_{i3} U^{\tilde{\mu}}_{m1},\\*
  \delta y_{\mu} &= \frac{m_{\mu} g_{2}}{\sqrt{2}M_{W}c_{\beta}}\left(\frac{\delta m_{\mu}}{m_{\mu}} + \frac{\delta g_{2}}{g_{2}} - \frac{\delta M_{W}}{M_{W}} - \frac{\delta c_{\beta}}{c_{\beta}}\right),\\*
  \delta g_{1} &= \frac{e}{c_{W}}\left(\delta Z_{e} - \frac{\delta c_{W}}{c_{W}}\right),\\*
  \delta g_{2} &= \frac{e}{s_{W}}\left(\delta Z_{e} - \frac{\delta s_{W}}{s_{W}}\right),\\*
  \delta c_{W} &= -\frac{s_{W}}{c_{W}}\delta s_{W}.
\end{align}
\end{subequations}
The quantities~$\delta\mathcal{A}^{z\pm}$
and~$\delta\mathcal{B}^{z\pm}$ correspond to the
renormalization of the coupling combinations and contain the entire
effect of the parameter renormalization constants. As stressed in the beginning,
these counterterm diagrams are finite by themselves, since there are no
corresponding two-loop diagrams (as long as field renormalization is included).

Next, the four counterterm classes of Fig.~\ref{fig:ctcases} are considered,
starting with the neutralino vertex and neutralino self-energy
counterterm diagrams. In each case, only the result for
off-diagonal neutralino/chargino indices $ij$ is given; the result for $j=i$
can be obtained by a limiting procedure. The neutralino vertex
counterterm diagram of Fig.~\ref{fig:ctcases}(\NV) is zero, 
\begin{align}
\amuCTclass{\NV}{ij\smum}&=0.
\end{align}
In general it would be proportional to $\delta Z_{ZA}$ which,
however, vanishes for the considered class of fermion/sfermion-loop
diagrams. 

The neutralino self-energy counterterm diagram of
Fig.~\ref{fig:ctcases}(\NS) can be expressed easily
with the help of Eq.~\eqref{eq:genericaneu}:  
\newcommand{\aNPjii}{
\amuNgen\hspace{-.5ex}\left({\cal A}_{ji\tilde{\mu}_m}^{n+},
{\cal B}_{ji\tilde{\mu}_m}^{n+},
i\right)
}
\newcommand{\aNPjij}{
\amuNgen\hspace{-.5ex}\left(
{\cal A}_{ji\tilde{\mu}_m}^{n+},
{\cal B}_{ji\tilde{\mu}_m}^{n+},
j\right)
}
\newcommand{\aNMjii}{
\amuNgen\hspace{-.5ex}\left(
{\cal A}_{ji\tilde{\mu}_m}^{n-},
{\cal B}_{ji\tilde{\mu}_m}^{n-},
i\right)
}
\newcommand{\aNMccijj}{
\amuNgen\hspace{-.5ex}\left(
{\cal A}_{ji\tilde{\mu}_m}^{n-},
-
{\cal B}_{ji\tilde{\mu}_m}^{n-},
j\right)
}
\newcommand{\aCmPjii}{
\amuCgen\hspace{-.5ex}\left({\cal A}_{ji\snumu}^{c+},
{\cal B}_{ji\snumu}^{c+},
i\right)
}
\newcommand{\aCmPjij}{
\amuCgen\hspace{-.5ex}\left(
{\cal A}_{ji\snumu}^{c+},
{\cal B}_{ji\snumu}^{c+},
j\right)
}
\newcommand{\aCmMjii}{
\amuCgen\hspace{-.5ex}\left(
{\cal A}_{ji\snumu}^{c-},
{\cal B}_{ji\snumu}^{c-},
i\right)
}
\newcommand{\aCmMccijj}{
\amuCgen\hspace{-.5ex}\left(
{\cal A}_{ji\snumu}^{c-},
-
{\cal B}_{ji\snumu}^{c-},
j\right)
}
\newcommand{\aYPjii}{
a_{f}^Y\hspace{-.5ex}\left(
{\cal A}_{ji X_k}^{Y+},
{\cal B}_{ji X_k}^{Y+},
i\right)
}
\newcommand{\aYPjij}{
a_{f}^Y\hspace{-.5ex}\left(
{\cal A}_{ji X_k}^{Y+},
{\cal B}_{ji X_k}^{Y+},
j\right)
}
\newcommand{\aYMjii}{
a_{f}^Y\hspace{-.5ex}\left(
{\cal A}_{ji X_k}^{Y-},
{\cal B}_{ji X_k}^{Y-},
i\right)
}
\newcommand{\aYMccijj}{
a_{f}^Y\hspace{-.5ex}\left(
{\cal A}_{ji X_k}^{Y-},
-
{\cal B}_{ji X_k}^{Y-},
j\right)
}
\begin{align} 
\begin{split}
\label{eq:countertermNS}
\amuCTclass{\NS}{ij\smum}=&
\frac{v}{m_{\tilde{\chi}^0_i}-m_{\tilde{\chi}^0_j}}
\Bigg[m_{\tilde{\chi}^0_j} \aNPjij-m_{\tilde{\chi}^0_i}
   \aNPjii\Bigg]
\\
&+
\frac{a}{m_{\tilde{\chi}^0_i}+m_{\tilde{\chi}^0_j}}
\Bigg[m_{\tilde{\chi}^0_j} \aNMccijj+m_{\tilde{\chi}^0_i}
   \aNMjii\Bigg]
\\
&+
\frac{s}{m_{\tilde{\chi}^0_i}-m_{\tilde{\chi}^0_j}}
\Bigg[\aNPjij-\aNPjii\Bigg]
\\
&+
\frac{p}{m_{\tilde{\chi}^0_i}+m_{\tilde{\chi}^0_j}}
\Bigg[\aNMjii-\aNMccijj\Bigg]
. 
\end{split}
\end{align}
Here the constants $v,a,s,p$ correspond to a generic counterterm
Feynman rule given by $i\slashed \ell(v-a\gamma_5)+i(s-p\gamma_5)$; they have
to be replaced by the following renormalization constants:
\begin{subequations}
\begin{align}
  v &= \frac{1}{4}\left(\delta Z^{L}_{\tilde{\chi}^{0},ij} + \delta Z^{L*}_{\tilde{\chi}^{0},ji} + \delta Z^{L}_{\tilde{\chi}^{0},ji} + \delta Z^{L*}_{\tilde{\chi}^{0},ij}\right),\\
  a &= \frac{1}{4}\left(\delta Z^{L}_{\tilde{\chi}^{0},ij} + \delta Z^{L*}_{\tilde{\chi}^{0},ji} - \delta Z^{L}_{\tilde{\chi}^{0},ji} - \delta Z^{L*}_{\tilde{\chi}^{0},ij}\right),\\
  \begin{split}
    s &= - \frac{1}{4} m_{\tilde{\chi}^{0}_{i}} \left(\delta Z^{L*}_{\tilde{\chi}^{0},ij} + \delta Z^{L}_{\tilde{\chi}^{0},ij}\right)
         - \frac{1}{4} m_{\tilde{\chi}^{0}_{j}} \left(\delta Z^{L*}_{\tilde{\chi}^{0},ji} + \delta Z^{L}_{\tilde{\chi}^{0},ji}\right)\\
      &\quad - \frac{1}{2}\left(\delta m_{\tilde{\chi}^{0},ji}^{*} + \delta m_{\tilde{\chi}^{0},ij}\right),
  \end{split}\\
  \begin{split}
    p &= - \frac{1}{4} m_{\tilde{\chi}^{0}_{i}} \left(\delta Z^{L*}_{\tilde{\chi}^{0},ij} - \delta Z^{L}_{\tilde{\chi}^{0},ij}\right)
         - \frac{1}{4} m_{\tilde{\chi}^{0}_{j}} \left(\delta Z^{L*}_{\tilde{\chi}^{0},ji} - \delta Z^{L}_{\tilde{\chi}^{0},ji}\right)\\
      &\quad - \frac{1}{2}\left(\delta m_{\tilde{\chi}^{0},ji}^{*} - \delta m_{\tilde{\chi}^{0},ij}\right).
  \end{split}
\end{align}
\end{subequations}
Several differences between the counterterm result~\eqref{eq:countertermNS}
and the standard one-loop result~\eqref{eq:oneloopneu} are noteworthy.
Off-diagonal couplings with indices~$ji\smum$ and the ``minus''-coupling combinations
~${\cal A}^-$ and~${\cal B}^-$ appear.    
Characteristically, only the~$a$- and~$p$-terms involve the ``minus''-coupling combinations. 
These terms also involve denominators with sums of the neutralino masses instead of their differences. 
Note also that the ${\cal B}^-$s in some terms appear with negative prefactor. 
Importantly, thanks to partial
fractioning, the counterterm results can be expressed in terms of the
generalized one-loop
result of Eq.~\eqref{eq:genericaneu} and therefore of the standard one-loop
functions which depend only on a single mass ratio.  

The chargino vertex counterterm diagram, i.\,e.~the
diagram  with counterterm insertion at the photon--chargino--chargino
vertex, has a slightly more
complicated structure. 
The chargino vertex counterterm diagram of Fig.~\ref{fig:ctcases}(\CV) with
the generic counterterm Feynman rule $\gamma^\mu(v-a\gamma_5)$ leads to the expression  
\begin{align}
\begin{split}
\amuCTclass{\CV}{ij} =& 
\begin{aligned}[t]
  \frac{1}{16 \pi^2}\frac{m_{\mu}^2}{ m_{{\tilde \nu}_{\mu}}^2}
  \Bigg\{
  & \frac{1}{12}\left(v {\cal A}_{ji {\tilde \nu}_{\mu} }^{c +}-a {\cal A}_{ji {\tilde \nu}_{\mu}}^{c -}\right) {\cal F}_1^{C}(x_j,x_i) \\*
  &+v\frac{{\cal B}_{ji {\tilde \nu}_{\mu}}^{c +}}{6m_{\mu}} \left[{(m_{{\tilde \chi}^{-}_j} + m_{{\tilde \chi}^{-}_i}) {\cal F}_2^{C}(x_j,x_i)}{} + {6(m_{{\tilde \chi}^{-}_j}-m_{{\tilde \chi}^{-}_i}) {\cal F}_3^{C}(x_j,x_i)}{}\right]\\
  &+a\frac{{\cal B}_{ji {\tilde \nu}_{\mu}}^{c -}}{6m_{\mu}} \left[{(m_{{\tilde \chi}^{-}_j}-m_{{\tilde \chi}^{-}_i}) {\cal F}_2^{C}(x_j,x_i)}{} + {6(m_{{\tilde \chi}^{-}_j}+m_{{\tilde \chi}^{-}_i}) {\cal F}_3^{C}(x_j,x_i)}{}\right]
  \Bigg\}. 
\end{aligned}
\label{eq:countertermCV}
\end{split}
\end{align}
It cannot be written in terms of the
generalized one-loop result. Instead, and as an additional
complication, Eq.~\eqref{eq:countertermCV} contains new loop functions 
depending on two mass ratios, given in the appendix. Consistency with
the standard one-loop result is reflected in the relations
\begin{align}
{\cal F}_{1,2}^C(x_{i},x_{i})&={\cal F}_{1,2}^C(x_{i}),
&
{\cal F}_{3}^C(x_{i},x_{i})&=0.
\end{align}
The generic constants $v$ and $a$ have to be replaced by
\begin{subequations}
\begin{align}
  v &= \frac{1}{4}\left(\delta Z^{R*}_{\tilde{\chi}^{-},ji} + \delta Z^{L*}_{\tilde{\chi}^{-},ji} + \delta Z^{R}_{\tilde{\chi}^{-},ij} + \delta Z^{L}_{\tilde{\chi}^{-},ij}\right),\\
  a &= \frac{1}{4}\left(\delta Z^{R*}_{\tilde{\chi}^{-},ji} - \delta Z^{L*}_{\tilde{\chi}^{-},ji} + \delta Z^{R}_{\tilde{\chi}^{-},ij} - \delta Z^{L}_{\tilde{\chi}^{-},ij}\right).
\end{align}
\end{subequations}

Finally, the chargino self-energy counterterm contribution is obtained by
summing up the two chargino self-energy counterterm diagrams of
Fig.~\ref{fig:ctcases}(\CS). The result combines all
complications of the previous terms and reads
\begin{align}
\begin{split}
\amuCTclass{\CS}{ij} =& 
\frac{v}{m_{\tilde{\chi}^-_i}-m_{\tilde{\chi}^-_j}}
\Bigg[m_{\tilde{\chi}^-_j} \aCmPjij-m_{\tilde{\chi}^-_i}
   \aCmPjii\Bigg]
\\
&+
\frac{a}{m_{\tilde{\chi}^-_i}+m_{\tilde{\chi}^-_j}}
\Bigg[m_{\tilde{\chi}^-_j} \aCmMccijj+m_{\tilde{\chi}^-_i}
   \aCmMjii\Bigg]
\\
&+
\frac{s}{m_{\tilde{\chi}^-_i}-m_{\tilde{\chi}^-_j}}
\Bigg[\aCmPjij-\aCmPjii\Bigg]
\\
&+
\frac{p}{m_{\tilde{\chi}^-_i}+m_{\tilde{\chi}^-_j}}
\Bigg[\aCmMjii-\aCmMccijj\Bigg]
\\
&
\begin{aligned}
-\frac{1}{16 \pi^2}\frac{m_{\mu}^2}{ m_{{\tilde \nu}_{\mu}}^2}
 \Bigg\{
 &\frac{1}{12}\left(v   {\cal A}_{ji {\tilde \nu}_{\mu} }^{c +}-a {\cal A}_{ji {\tilde \nu}_{\mu}}^{c -}\right) {\cal F}_1^{C}(x_j,x_i) \\
 &+v\frac{{\cal B}_{ji {\tilde \nu}_{\mu}}^{c +}}{6m_{\mu}} \left[{(m_{{\tilde \chi}^{-}_j} + m_{{\tilde \chi}^{-}_i}) {\cal F}_2^{C}(x_j,x_i)}{} + {6(m_{{\tilde \chi}^{-}_j}-m_{{\tilde \chi}^{-}_i}) {\cal F}_3^{C}(x_j,x_i)}{}\right]\\
 &+a\frac{{\cal B}_{ji {\tilde \nu}_{\mu}}^{c -}}{6m_{\mu}} \left[{(m_{{\tilde \chi}^{-}_j}-m_{{\tilde \chi}^{-}_i}) {\cal F}_2^{C}(x_j,x_i)}{} + {6(m_{{\tilde \chi}^{-}_j}+m_{{\tilde \chi}^{-}_i}) {\cal F}_3^{C}(x_j,x_i)}{}\right]
 \Bigg\}. 
\end{aligned}
\label{eq:countertermCS}
\end{split}
\end{align} 
The generic constants $v,a,s,p$ have to be replaced by
\begin{subequations}
\begin{align}
  v &= \frac{1}{4}\left(\delta Z^{L*}_{\tilde{\chi}^{-},ji} + \delta Z^{R*}_{\tilde{\chi}^{-},ji} + \delta Z^{L}_{\tilde{\chi}^{-},ij} + \delta Z^{R}_{\tilde{\chi}^{-},ij}\right),\\
  a &= \frac{1}{4}\left(\delta Z^{L*}_{\tilde{\chi}^{-},ji} - \delta Z^{R*}_{\tilde{\chi}^{-},ji} + \delta Z^{L}_{\tilde{\chi}^{-},ij} - \delta Z^{R}_{\tilde{\chi}^{-},ij}\right),\\
  \begin{split}
    s &= - \frac{1}{4} m_{\tilde{\chi}^{-}_{i}} \left(\delta Z^{L}_{\tilde{\chi}^{-},ij} + \delta Z^{R}_{\tilde{\chi}^{-},ij}\right)
         - \frac{1}{4} m_{\tilde{\chi}^{-}_{j}} \left(\delta Z^{L*}_{\tilde{\chi}^{-},ji} + \delta Z^{R*}_{\tilde{\chi}^{0},ji}\right)\\
      &\quad - \frac{1}{2}\left(\delta m_{\tilde{\chi}^{-},ji} + \delta m_{\tilde{\chi}^{-},ij}^{*}\right),
  \end{split}\\
  \begin{split}
    p &= - \frac{1}{4} m_{\tilde{\chi}^{-}_{i}} \left(\delta Z^{L}_{\tilde{\chi}^{-},ij} - \delta Z^{R}_{\tilde{\chi}^{-},ij}\right)
         - \frac{1}{4} m_{\tilde{\chi}^{-}_{j}} \left(\delta Z^{R*}_{\tilde{\chi}^{-},ji} - \delta Z^{L*}_{\tilde{\chi}^{-},ji}\right)\\
      &\quad - \frac{1}{2}\left(\delta m_{\tilde{\chi}^{-},ji} - \delta m_{\tilde{\chi}^{-},ij}^{*}\right).
  \end{split}
\end{align}
\end{subequations}

We highlight several characteristic properties which are
shared by the more complicated two-loop contributions discussed below:
off-diagonal coupling combinations with indices~$ji$ and the
``minus''-coupling combinations~${\cal A}^-$ and~${\cal B}^-$
appear. Some parts of the counterterm results can be reduced by
partial fractioning to loop functions that appear already in the
standard one-loop results, some parts involve more complicated loop
functions that depend on two mass ratios. Results at ${\cal
  O}(\epsilon^0)$ for generic diagrams similar to these counterterm 
diagrams have already been computed in Ref.~\cite{HIRS1}.
\newpage
\section{Two-loop contributions}
\label{sec:twoloop}

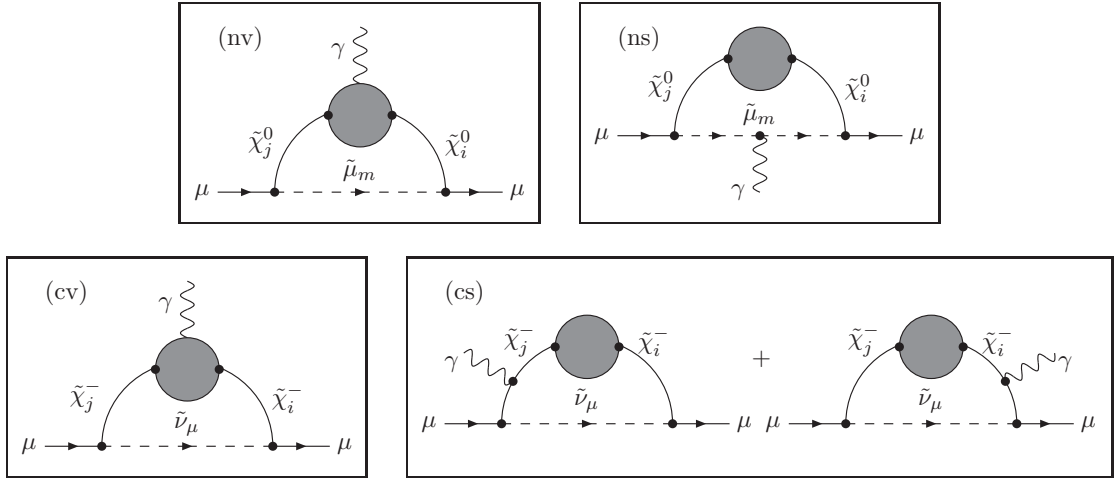
\begin{figure}[t]
\begin{center}
\null\hfill
\fbox{
\scalebox{.85}{
\begin{picture}(135,90)(-5,-5)
\Text(0,67.5)[lb]{(\NV)}
\Vertex(25,5){2}
\Vertex(100,5){2}
\CArc(62.5,5)(37.5,0,180)
\ArrowLine(0,5)(25,5)
\ArrowLine(100,5)(125,5)
\DashArrowLine(25,5)(100,5){4}
\Photon(62.5,53)(62.5,78){3}{3}
\CCirc(62.5,39){14}{Black}{Gray}
\Vertex(76.5,39){2}
\Vertex(48.5,39){2}
\Text(0,5)[r]{$\mu\ $}
\Text(125,5)[l]{$\ {\mu}$}
\Text(62.5,10)[b]{$\tilde{\mu}_m$}
\Text(25,20)[br]{$\tilde{\chi}^0_j$}
\Text(100,20)[bl]{$\tilde{\chi}^0_i$}
\Text(60,67.5)[r]{$\gamma\ $}
\end{picture}
}
}
\quad
\fbox{
\scalebox{.85}{
\begin{picture}(135,90)(-5,-30)
\Text(0,42.5)[lb]{(\NS)}
\Vertex(25,5){2}
\Vertex(100,5){2}
\Vertex(62.5,5){2}
\CArc(62.5,5)(37.5,0,180)
\ArrowLine(0,5)(25,5)
\ArrowLine(100,5)(125,5)
\DashArrowLine(25,5)(62.5,5){4}
\DashArrowLine(62.5,5)(100,5){4}
\Photon(62.5,5)(62.5,-20){3}{3}
\CCirc(62.5,39){14}{Black}{Gray}
\Vertex(76.5,39){2}
\Vertex(48.5,39){2}
\Text(0,5)[r]{$\mu\ $}
\Text(125,5)[l]{$\ {\mu}$}
\Text(62.5,10)[b]{$\tilde{\mu}_m$}
\Text(25,20)[br]{$\tilde{\chi}^0_j$}
\Text(100,20)[bl]{$\tilde{\chi}^0_i$}
\Text(60,-20)[r]{$\gamma\ $}
\end{picture}
}
}
\hfill\null
\\
\null\hfill
\fbox{
\scalebox{.85}{
\begin{picture}(135,90)(-5,-5)
\Text(0,67.5)[lb]{(\CV)}
\Vertex(25,5){2}
\Vertex(100,5){2}
\CArc(62.5,5)(37.5,0,180)
\ArrowLine(0,5)(25,5)
\ArrowLine(100,5)(125,5)
\DashArrowLine(25,5)(100,5){4}
\Photon(62.5,53)(62.5,78){3}{3}
\CCirc(62.5,39){14}{Black}{Gray}
\Vertex(76.5,39){2}
\Vertex(48.5,39){2}
\Text(0,5)[r]{$\mu\ $}
\Text(125,5)[l]{$\ {\mu}$}
\Text(62.5,10)[b]{$\tilde{\nu}_\mu$}
\Text(25,20)[br]{$\tilde{\chi}^-_j$}
\Text(100,20)[bl]{$\tilde{\chi}^-_i$}
\Text(60,67.5)[r]{$\gamma\ $}
\end{picture}
}
}
\quad
\fbox{
\scalebox{.85}{
\begin{picture}(125,90)(-5,-15)
\Text(0,57.5)[lb]{(\CS)}
\Vertex(25,5){2}
\Vertex(100,5){2}
\CArc(62.5,5)(37.5,0,180)
\ArrowLine(0,5)(25,5)
\ArrowLine(100,5)(125,5)
\DashArrowLine(25,5)(100,5){4}
\CCirc(62.5,39){14}{Black}{Gray}
\Vertex(76.5,39){2}
\Vertex(48.5,39){2}
\Vertex(30,23.75){2}
\Photon(30,23.75)(8.4,36.25){3}{3}
\Text(0,5)[r]{$\mu\ $}
\Text(125,5)[l]{$\ {\mu}$}
\Text(62.5,10)[b]{$\tilde{\nu}_\mu$}
\Text(85,35)[bl]{$\tilde{\chi}^-_i$}
\Text(40,35)[br]{$\tilde{\chi}^-_j$}
\Text(5,35)[t]{$\gamma\ $}
\end{picture}
\quad
\scalebox{1}{
\begin{picture}(5,90)(3,0)
\Text(0,45)[lb]{+}
\end{picture}
\quad
}
\begin{picture}(125,90)(10,-15)
\Vertex(25,5){2}
\Vertex(100,5){2}
\CArc(62.5,5)(37.5,0,180)
\ArrowLine(0,5)(25,5)
\ArrowLine(100,5)(125,5)
\DashArrowLine(25,5)(100,5){4}
\CCirc(62.5,39){14}{Black}{Gray}
\Vertex(76.5,39){2}
\Vertex(48.5,39){2}
\Vertex(95,23.75){2}
\Photon(95,23.75)(116.6,36.25){3}{3}
\Text(0,5)[r]{$\mu\ $}\Text(125,5)[l]{$\ {\mu}$}
\Text(62.5,10)[b]{$\tilde{\nu}_\mu$}
\Text(85,35)[bl]{$\tilde{\chi}^-_i$}
\Text(40,35)[br]{$\tilde{\chi}^-_j$}
\Text(120,35)[t]{$\ \gamma$}
\end{picture}
}}
\hfill\null
\caption{\label{fig:tlcases}
The four classes of two-loop diagrams with fermion/sfermion loop
insertions into the neutralino vertex~(\NV), neutralino self-energy~(\NS), 
chargino vertex~(\CV), and chargino self-energy~(\CS).
In the (\CS)~case always the sum of the two
contributing diagrams is considered. The dark circles
denote fermion/sfermion-loop insertions. 
}
\end{center}
\end{figure}

The two-loop diagrams with a closed fermion/sfermion loop can be
classified according to their topologies, as shown in Fig.~\ref{fig:tlcases},
where the fermion/sfermion loops are denoted by dark circles.
There are four classes:
\begin{itemize}
\item neutralino diagrams with an inner fermion/sfermion loop
  generating either an effective three-point
  neutralino--neutralino--photon vertex (\NV), or a neutralino
  self-energy (\NS). The inner loops are shown in more detail in 
  Figs.~\ref{fig:neutralinovertex} and \ref{fig:neutralinoselfenergy}.
\item chargino diagrams with an inner fermion/sfermion loop generating
  either an effective three-point chargino--chargino--photon vertex~(\CV),
  or a chargino self-energy~(\CS). The inner loops are shown in
  more detail in Figs.~\ref{fig:charvertex} and \ref{fig:charselfenergy}.
\end{itemize}
There are no fermion/sfermion-loop corrections to the external
vertices involving the muons, which is the reason why the counterterm
diagrams in Fig.~\ref{fig:finctcases}(\MuNV, \MuCV) are finite.

%
%
%
The computation of the diagrams has been carried out in two different
ways. The first way uses the procedure described in
Refs.~\cite{HSW03,HSW04} and is based on standard techniques for
evaluating two-loop integrals, reduction to master integrals, large
mass expansion and automated analytical simplification.


The second computation is completely different and uses the technique
of Barr-Zee diagrams~\cite{BarrZee}. In the simplest Barr-Zee diagrams
a closed loop generates an effective $\gamma$--$\gamma$--Higgs
vertex. The computational strategy is to first compute the inner
one-loop diagram alone using a Feynman parametrization, simplify it,
and then insert  
it into the second loop diagram. By performing the second loop
integration one obtains an integral representation of the full
two-loop diagram. This Barr-Zee technique has been employed to compute
several classes of contributions to electric~\cite{BarrZee,ChangKP98,Pilaftsis98} 
and magnetic dipole moments~\cite{ChangCCK00,CheungCK01,ArhribBaek,ChenGeng}. In the latter
references either fermion or sfermion loops generate a
$\gamma$--vector--Higgs interaction.
The diagrams considered in this paper can be regarded as
supersymmetric counterparts to this, since the fermion/sfermion loops
of Figs.~\ref{fig:neutralinovertex} and \ref{fig:charvertex} effectively
generate $\gamma$--gaugino--higgsino interactions. 

Compared to the previous applications of the method used by Barr and Zee the
diagrams considered here are more complicated for three reasons:
\begin{itemize}
\item There is one more heavy mass scale in the diagram.
  As a consequence the two-loop results depend on four dimensionless
  mass ratios instead of three.
\item The QED Ward identity constrains the results for the inner
  loops. As a consequence the inner loops in the references quoted above 
  can be simplified to expressions which depend only on a single
  covariant and a single scalar function. However, in our case the
  Ward identity allows four such covariants already in the simplest
  case.  
\item The inner loops can be ultraviolet divergent. Apart from the
  requirement of renormalization this implies that the outer loop has to
  be computed to higher orders in the dimensional
  regularization parameter~$\epsilon=(4-D)/2$. 
\end{itemize}

In the following the calculation and the results of the
four classes of Fig.~\ref{fig:tlcases} are described in detail.
The results are expressed in terms of generic neutralino and chargino
couplings, defined in Sec.~\ref{sec:couplings}.
All results will be given only for $i\ne j$; for the case $i = j$ a
limit can be performed. 

In the neutralino cases we will close with some remarks on the
structure of the results, similar to the remarks at the end of Sec.~\ref{sec:cts}.
We will be briefer in the chargino cases, where the
structure of the results is similar.

\subsection{Notations for the neutralino results}
\label{sec:neutralinonotations}

We first introduce abbreviations which help us to write
the neutralino two-loop results in a compact way. 

The Feynman parametrization of the inner loops leads to a denominator
of the form
\mbox{$\fx(1-\fx)\,\ell^2-(1-\fx)\,m_f^2-\fx\,m_{\tilde{f}}^2$}, which
depends on the Feynman parameter $\fx$. This defines a
propagator denominator
\begin{align}
{\cal D}_{f {\tilde f}_{k}}(\ell) &\equiv \ell^2 - m_{\BZNeut}^2(\fx)
\label{NeutralinoDff}
\end{align}
with momentum $\ell$ and an effective mass 
\begin{align}
m_{\BZ}^2(\fx)\equiv\frac{m_{f}^{2}}{\fx} + \frac{m_{\tilde{f}_k}^2}{1 - \fx}.
\label{EffectiveMass}
\end{align}
The quantities $m_{f}$ and $m_{\tilde f}$ denote fermion and sfermion masses,
respectively. 
Due to partial integration also the derivative of the denominator
appears, so it is convenient to introduce the abbreviation 
\begin{align}
\abbroneloop(\ell^2, m_f^2, m_{{\tilde f}_{k}}^2)\equiv
\frac{(1-2\fx)\,\ell^2 + m_{f}^{2} - m_{{\tilde f}_{k}}^{2}}{1 -\fx},
\label{DerivDff}
\end{align}
which is related to this derivative.
In the case of neutralinos the results depend on the electric 
charge $Q_f$ of the inner fermion, which equals \mbox{$+\frac{2}{3},-\frac{1}{3},-1$} for up-type quarks,
down-type quarks and charged leptons, respectively. The color factor
$\colorNumber$ is 1 for leptons and 3 for quarks.
The results further depend on the dimensionless mass ratios defined as 
\begin{align}
\xNeuti \equiv \frac{m_{i}^2}{\MssmsNeutTWO},\quad
\xNeutj \equiv \frac{m_{j}^2}{\MssmsNeutTWO},&\quad
\xfracNeut_{f} \equiv \frac{m_{f}^2}{\MssmsNeutTWO},\quad
\xfracNeut_{\tilde{f}_k} \equiv \frac{m_{\tilde{f}_k}^2}{\MssmsNeutTWO},\quad
\xfracNeut_{\BZNeut} \equiv \frac{m_{\BZNeut}^2(\fx)}{\MssmsNeutTWO},
\label{dimensionlessmassratio}
\end{align}
where the neutralino masses are defined as $m_{i,j}\equiv m_{\tilde{\chi}^0_{i,j}}$ and $m_{{\tilde \mu}_{m}}$ is the smuon mass. 
It is also useful to introduce the following abbreviations for logarithms 
\begin{align}
l_z &\equiv \log \xfracNeut_z,\quad L(m^2) \equiv \log\frac{m^2}{\muDR^2}. 
\label{logarithms}
\end{align}
Here $\muDR$ is the scale of dimensional regularization/dimensional
reduction (there is no difference between the two schemes for the
considered class of contributions). This scale does not drop out of
the final, renormalized two-loop result because of the  
$\overline{\text{DR}}$~renormalization of $\tan\beta$.

\subsection{Neutralino vertex contributions}
\label{sec:NV}

\setlength{\abovecaptionskip}{15pt plus 3pt minus 2pt}

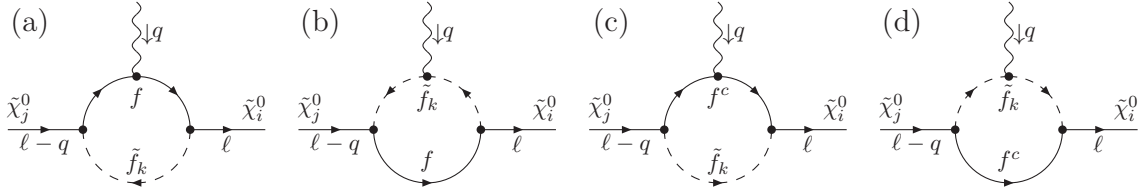
\begin{figure}
\begin{center}
\begin{tabular}{cccc}
\scalebox{.80}{
\begin{picture}(100,60)(-45,-5)
\ArrowLine(-60,10)(-25,10)
\Vertex(-25,10){2}
\Vertex(25,10){2}
\ArrowArcn(0,10)(25,180,90)
\ArrowArcn(0,10)(25,90,0)
\DashArrowArcn(0,10)(25,0,180){4}
\ArrowLine(25,10)(60,10)
\Photon(0,70)(0,35){2}{3}
\Vertex(0,35){2}
\Text(10,55)[]{$q$}
\Text(5,55)[]{$\downarrow$}
\Text(0,25)[]{$f$}
\Text(0,-5)[]{${\tilde f}_{k}$}
\Text(-55,20)[]{${ \tilde{\chi}}_{j}^{0}$}
\Text(42,3)[]{$\ell$}
\Text(-43,3)[]{$\ell-q$}
\Text(55,20)[]{${ \tilde{\chi}}_{i}^{0}$}
\Text(-50,60)[]{\Large(a)}
\end{picture}
}
& \quad
\scalebox{.80}{
\begin{picture}(100,60)(-45,-5)
\ArrowLine(-60,10)(-25,10)
\Vertex(-25,10){2}
\Vertex(25,10){2}
\ArrowArc(0,10)(25,180,360)
\DashArrowArc(0,10)(25,0,90){4}
\DashArrowArc(0,10)(25,90,180){4}
\ArrowLine(25,10)(60,10)
\Photon(0,70)(0,35){2}{3}
\Vertex(0,35){2}
\Text(10,55)[]{$q$}
\Text(5,55)[]{$\downarrow$}
\Text(0,-5)[]{$f$}
\Text(0,25)[]{${\tilde f}_{k}$}
\Text(-55,20)[]{${ \tilde{\chi}}_{j}^{0}$}
\Text(42,3)[]{$\ell$}
\Text(-43,3)[]{$\ell-q$}
\Text(55,20)[]{${ \tilde{\chi}}_{i}^{0}$}
\Text(-50,60)[]{\Large(b)}
\end{picture}
}
& \quad
\scalebox{.80}{
\begin{picture}(100,60)(-45,-5)
\ArrowLine(-60,10)(-25,10)
\Vertex(-25,10){2}
\Vertex(25,10){2}
\ArrowArcn(0,10)(25,180,90)
\ArrowArcn(0,10)(25,90,0)
\DashArrowArc(0,10)(25,180,360){4}
\ArrowLine(25,10)(60,10)
\Photon(0,70)(0,35){2}{3}
\Vertex(0,35){2}
\Text(10,55)[]{$q$}
\Text(5,55)[]{$\downarrow$}
\Text(0,25)[]{$f ^{c}$}
\Text(0,-5)[]{${\tilde f}_{k}$}
\Text(-55,20)[]{${ \tilde{\chi}}_{j}^{0}$}
\Text(42,3)[]{$\ell$}
\Text(-43,3)[]{$\ell-q$}
\Text(55,20)[]{${ \tilde{\chi}}_{i}^{0}$}
\Text(-50,60)[]{\Large(c)}
\end{picture}
}
& \quad
\scalebox{.80}{
\begin{picture}(100,60)(-45,-5)
\ArrowLine(-60,10)(-25,10)
\Vertex(-25,10){2}
\Vertex(25,10){2}
\ArrowArc(0,10)(25,180,360)
\DashArrowArcn(0,10)(25,180,90){4}
\DashArrowArcn(0,10)(25,90,0){4}
\ArrowLine(25,10)(60,10)
\Photon(0,70)(0,35){2}{3}
\Vertex(0,35){2}
\Text(10,55)[]{$q$}
\Text(5,55)[]{$\downarrow$}
\Text(0,-5)[]{$f ^{c}$}
\Text(0,25)[]{${\tilde f}_{k}$}
\Text(-55,20)[]{${ \tilde{\chi}}_{j}^{0}$}
\Text(42,3)[]{$\ell$}
\Text(-43,3)[]{$\ell-q$}
\Text(55,20)[]{${ \tilde{\chi}}_{i}^{0}$}
\Text(-50,60)[]{\Large(d)}
\end{picture}
}
\end{tabular}
\caption{\label{fig:neutralinovertex}
Feynman diagrams contributing to the neutralino vertex insertion. The sum of these diagrams is denoted as 
$i\Gamma _{i j {\tilde f}_{k}}^{0\mu}(\ell)$.
The momenta flow in the directions indicated by the arrows.}
\end{center}
\end{figure}

The neutralino vertex diagrams, shown in Fig.~\ref{fig:tlcases}(\NV),
constitute the only ultraviolet finite class of two-loop diagrams containing
a closed fermion/sfermion loop.
They show the closest similarity to the Barr-Zee type diagrams calculated in
Refs.~\cite{ChangCCK00,CheungCK01, ArhribBaek,ChenGeng}.
For each fermion/sfermion pair, there are four different inner loop
diagrams, shown in Fig.~\ref{fig:neutralinovertex}, which
generate an effective 
${\tilde{\chi}}^0{\tilde{\chi}}^0\gamma$ interaction.
This three-point function is calculated to first order in the photon
momentum $q$. 
The use of Feynman parameter representation then
yields the form of a Feynman propagator with an effective mass defined
in Eq.~\eqref{EffectiveMass} and simplifies the subsequent
calculation. 

For a generic fermion/sfermion \mbox{pair $(f,\tilde{f}_{k})$}, the sum of
the four inner loop diagrams of Fig.~\ref{fig:neutralinovertex} results in 
%
\begin{align}
\begin{split}
\Gamma _{i j \tilde{f}_{k}} ^{0 \mu} (\ell)=
\frac{e Q _{f}}{16 \pi ^{2}}\int _{0} ^{1}\frac{{\rm d}\fx}{2} &
\left[ \
\left( \atildepm - \atildemp \gamma ^{5} \right)
\frac{\slashed{\ell}\slashed{q}\gamma^{\mu}-\slashed{\ell}q^{\mu} + \slashed{q}\ell^{\mu}-(\ell\cdot q)\gamma^{\mu}}
{{\cal D}_{f {\tilde f}_{k}}(\ell)}\right.\\
&\left.
+\left( \btildepm - \btildemp \gamma ^{5} \right)
\frac{m_{f}}{\fx}\frac{\slashed{q} \gamma ^{\mu} - q^{\mu}}{{\cal D}_{f {\tilde f}_{k}}(\ell) } \
\right].
\end{split}
\label{NeutralinoVertexOneloop}
\end{align}

The appearing coupling combinations correspond to the imaginary and real parts
of the coupling constants involved:
\begin{subequations}
\label{ABbarDefs}
\begin{align}
\overline{{\cal A}}^{n \, \pm}_{i j {\tilde f}_{k}}
&\equiv {\cal A}^{n\,\pm}_{i j {\tilde f}_{k}}\mp{\cal A}^{n\,\pm *}_{i j {\tilde f}_{k}}, \\*
\overline{{\cal B}}^{n \, \pm}_{i j {\tilde f}_{k}}
&\equiv {\cal B}^{n\,\pm}_{i j {\tilde f}_{k}\,}\mp{\cal B}^{n\,\pm *}_{i j {\tilde f}_{k}}.
\end{align}
\end{subequations}
The result of Eq.~\eqref{NeutralinoVertexOneloop} can be divided into four parts 
according to these four different coupling combinations. 

Apparently, the QED Ward identity $q_\mu\Gamma _{i j \tilde{f}_{k}} ^{0\mu} (\ell)=0$  is manifestly valid
for each of the four parts separately.
In line with the interpretation from
Sec.~\ref{sec:couplings}, the chirality-conserving
${\cal A}$ combinations appear in terms with odd powers of $\gamma$-matrices;
the chirality-flipping ${\cal B}$ combinations in the ones with even powers.
Further, the ``plus''-combinations appear without $\gamma_5$;
the ``minus''-combinations with $\gamma_5$.

Inserting Eq.~\eqref{NeutralinoVertexOneloop} into the outer loop, performing the loop
integration and extracting $\amu$ yields
\begin{align}
\amuFSfclass{\NV}{ijm\tilde{f}_k}
=\int _{0} ^{1}{\rm d} \fx \Bigg[\ &
{{\cal A}^{n+}_{ji\tilde{\mu}_m}}\left(
{\overline{{\cal A}}^{n+}_{ij\tilde{f}_{\tp}}}{\FF{\NV+}{AA}}
+{\overline{{\cal B}}^{n+}_{ij\tilde{f}_{\tp}}}{\FF{\NV+}{AB}}
\right)
+{{\cal B}^{n+}_{ji\tilde{\mu}_m}}\left(
{\overline{{\cal A}}^{n+}_{ij\tilde{f}_{\tp}}}{\FF{\NV+}{BA}}
+{\overline{{\cal B}}^{n+}_{ij\tilde{f}_{\tp}}}{\FF{\NV+}{BB}}
\right)
\nonumber\\*
+&
{{\cal A}^{n-}_{ji\tilde{\mu}_m}}\left(
{\overline{{\cal A}}^{n-}_{ij\tilde{f}_{\tp}}}{\FF{\NV-}{AA}}
+{\overline{{\cal B}}^{n-}_{ij\tilde{f}_{\tp}}}{\FF{\NV-}{AB}}
\right)
+{{\cal B}^{n-}_{ji\tilde{\mu}_m}}\left(
{\overline{{\cal A}}^{n-}_{ij\tilde{f}_{\tp}}}{\FF{\NV-}{BA}}
+{\overline{{\cal B}}^{n-}_{ij\tilde{f}_{\tp}}}{\FF{\NV-}{BB}}
\right) \Bigg].
\label{eq:twoloopNV}
\end{align}

The loop functions $\FF{}{}$  are either symmetric or antisymmetric in~$i,j$ and read
\begin{subequations}
\label{loopfunctionsNV}
\begin{align}
\begin{split}
{\FF{\NV\pm}{AA}}
 =&\left(\frac{1}{16\pi^2}\right)^2\frac{\colorNumber Q_f}{4}\frac{m_{\mu }^2}{\MssmsNeutTWO}\frac{m_i}{m_i{\mp} m_j} \\*
& \times\left[\frac{1-\xfracNeut_{\BZNeut}+l_{{\BZNeut}} \xfracNeut_{{\BZNeut}}^2 }{(1-\xfracNeut_{{\BZNeut}}){}^2 (\xfracNeut_{{\BZNeut}}-\xNeuti) }
-\frac{1-\xNeuti+l_i \xNeutiTWO}{(1-\xNeuti){}^2 (\xfracNeut_{{\BZNeut}}-\xNeuti)}\right]+\plusswapij,
\end{split}
\\
\begin{split}
{\FF{\NV\pm}{AB}}
=&\left(\frac{1}{16\pi ^2}\right)^2\frac{{\colorNumber} Q_f}{4{\fx}}\frac{m_{\mu }^2m_{f}}{\MssmsNeutTWO}\frac{1}{\pm  m_i-m_j}\\*
&\times\left[\frac{1-\xfracNeut_{\BZNeut}+l_{{\BZNeut}}\xfracNeut_{{\BZNeut}}^2}{(1-\xfracNeut_{{\BZNeut}}){}^2 (\xfracNeut_{{\BZNeut}}-\xNeuti) }
-\frac{1-\xNeuti+l_i \xNeutiTWO}{(1-\xNeuti){}^2 (\xfracNeut_{{\BZNeut}}-\xNeuti)}\right]\pm\plusswapij,
\end{split}
\\
\begin{split}
{\FF{\NV\pm}{BA}}
=&\left(\frac{1}{16\pi ^2}\right)^2\frac{{\colorNumber} Q_f}{4}m_{\mu }\frac{1}{m_i{\mp} m_j}\\*
&\times\left[- \frac{ l_{{\BZNeut}}\xfracNeut_{{\BZNeut}}^2}{(1-\xfracNeut_{{\BZNeut}}) (\xfracNeut_{{\BZNeut}}-\xNeuti) }
+\frac{l_i\xNeutiTWO}{(1-\xNeuti) (\xfracNeut_{{\BZNeut}}-\xNeuti)}\right]\pm\plusswapij,
\end{split}
\\
\begin{split}
{\FF{\NV\pm}{BB}}
=&\left(\frac{1}{16\pi ^2}\right)^2\frac{{\colorNumber} Q_f}{2{\fx}}\frac{m_{\mu }m_{f}}{\MssmsNeutTWO}\frac{m_im_j}{m_i^2-m_j^2}\\*
&\times\left[-\frac{l_{{\BZNeut}} \xfracNeut_{{\BZNeut}}}{(1-\xfracNeut_{{\BZNeut}}) (\xfracNeut_{{\BZNeut}}-\xNeuti) }
+\frac{l_i \xNeuti}{(1-\xNeuti)(\xfracNeut_{{\BZNeut}}-\xNeuti){}}\right]+\plusswapij.
\end{split}
\end{align}
\end{subequations}

The notation~$\pm(i\leftrightarrow j)$ indicates adding or
subtracting the preceding expression with~$i$~and~$j$ exchanged.

We close with some comments on the structure of the results. The
${\cal A}^+, {\cal B}^+$-terms and the ${\cal A}^-, {\cal B}^-$-terms
in the inner loop generate the vector and axial vector parts of the
${\tilde{\chi}}^0{\tilde{\chi}}^0\gamma$ interaction and further terms
without/with $\gamma_5$, respectively. They combine with the coupling
combinations of the outer loop similarly to $v$ and $a$ in
 Eq.~\eqref{eq:countertermCV}, such that only ``$++$'' and ``$--$''
coupling combinations exist in Eq.~\eqref{eq:twoloopNV}. The index
structure of the outer loop, $ji\smum$, matches the index structure
$ij\tilde{f}_k$ of the inner loop.

An important property of the ${\cal B}$-terms of the inner loop is
that they are all proportional to the inner fermion mass $m_f$. This
factor is needed in the inner loop because of the chirality-flipping
nature of the ${\cal B}$ coupling combinations. As a result, the
${\cal B}$-terms of the inner loop can be sizeable only for
third-generation insertions. In contrast, the ${\cal B}$-terms of the
outer loop are always important; as already at one-loop, they are
$\tan\beta$-enhanced compared to the ${\cal A}$-terms of the outer
loop. 

The loop functions appearing in the two-loop results of
Eqs.~\eqref{loopfunctionsNV} depend on three mass ratios,
$\xNeuti,\xNeutj,\xfracNeut_{{\BZNeut}}$. The neutralino vertex
contributions are particularly simple in this respect, since the
masses $m_f$ and $m_{\tilde{f}_k}$ of the inner loop do not appear
individually but only in the combination $\xfracNeut_{{\BZNeut}}$.

\subsection{Neutralino self-energy contributions}

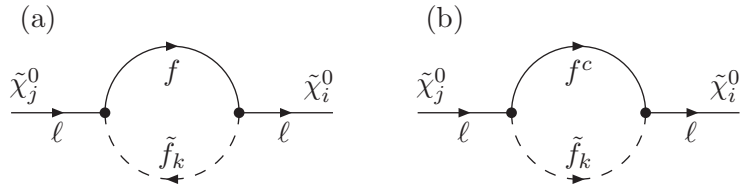
\begin{figure}
\begin{center}
\begin{tabular}{cc}
\scalebox{1}{
\begin{picture}(100,60)(-60,0)
\ArrowLine(-60,10)(-25,10)
\Vertex(-25,10){2}
\Vertex(25,10){2}
\ArrowArcn(0,10)(25,180,0)
\DashArrowArcn(0,10)(25,0,180){4}
\ArrowLine(25,10)(60,10)
\Text(0,25)[]{$f$}
\Text(0,-5)[]{${\tilde f}_{k}$}
\Text(-55,20)[]{${ \tilde{\chi}}_{j}^{0}$}
\Text(42,3)[]{$\ell$}
\Text(-43,3)[]{$\ell$}
\Text(55,20)[]{${ \tilde{\chi}}_{i}^{0}$}
\Text(-50,45)[]{(a)}
\end{picture}
}
& \quad\quad\quad
\scalebox{1}{
\begin{picture}(100,60)(-60,0)
\ArrowLine(-60,10)(-25,10)
\Vertex(-25,10){2}
\Vertex(25,10){2}
\ArrowArcn(0,10)(25,180,00)
\DashArrowArc(0,10)(25,180,360){4}
\ArrowLine(25,10)(60,10)
\Text(0,25)[]{$f ^{c}$}
\Text(0,-5)[]{${\tilde f}_{k}$}
\Text(-55,20)[]{${ \tilde{\chi}}_{j}^{0}$}
\Text(42,3)[]{$\ell$}
\Text(-43,3)[]{$\ell$}
\Text(55,20)[]{${ \tilde{\chi}}_{i}^{0}$}
\Text(-50,45)[]{(b)}
\end{picture}
}
\end{tabular}
\null
\caption{\label{fig:neutralinoselfenergy}
Feynman diagrams contributing to the neutralino self-energy insertion 
$i\Sigma^{0} _{i j {\tilde f}_{k}}({\ell})$.}
\end{center}
\end{figure}

The next diagram class in our consideration is the one corresponding
to Fig.~\ref{fig:tlcases}(\NS), 
where the inner loop generates a neutralino self-energy.
For each fermion/sfermion pair, the corresponding one-loop insertions
are given by the two Feynman diagrams shown in
Fig.~\ref{fig:neutralinoselfenergy}. 
The sum of these diagrams yields 
the effective self-energy insertion $i\Sigma^{0} _{i j {\tilde
    f}_{k}}({\ell})$, given by
%
\begin{align}
\begin{split}
\Sigma^{0}_{i j {\tilde f}_k}({\ell}) = 
\frac{1}{16 \pi ^{2}} \int _{0} ^{1} \frac{{\rm d} \fx}{2} & \left[ \
\Big(\atildepp - \atildemm \gamma ^{5}\Big) \fx \slashed\ell
\Bigg(\frac{1}{\epsilon} - L(m_{{\tilde f}_{k}}^2 )+
\frac{\abbroneloop(\ell^2, m_f^2, m_{{\tilde f}_{k}}^2)/2}{{\cal D}_{f {\tilde f}_{k}}(\ell)}\Bigg)\right. \\
&+\left.\Big(\btildepp - \btildemm \gamma ^{5}\Big) m_{f}
\Bigg(\frac{1}{\epsilon} - L(m_{{\tilde f}_{k}}^2)+
\frac{\abbroneloop(\ell^2, m_f^2, m_{{\tilde f}_{k}}^2)}{\phantom{2}{\cal D}_{f {\tilde f}_{k}}(\ell)}\Bigg)
\right].
\end{split}
\label{neutself}
\end{align}

Compared to the neutralino vertex contributions in Eqs.~\eqref{ABbarDefs}, the coupling 
combinations appear with opposite signs:
\begin{subequations}
\label{ABtildeDefs}
\begin{align}
\widetilde{{\cal A}}^{n \, \pm}_{i j {\tilde f}_{k}} &\equiv
{\cal A}^{n\,\pm}_{i j {\tilde f}_{k}}\pm{\cal A}^{n\,\pm *}_{i j {\tilde f}_{k}}, \\*
\widetilde{{\cal B}}^{n \, \pm}_{i j {\tilde f}_{k}} &\equiv
{\cal B}^{n\,\pm}_{i j {\tilde f}_{k}\,}\pm{\cal B}^{n\,\pm *}_{i j {\tilde f}_{k}}. 
\end{align}
\end{subequations}
Similar to the previous case, the chirality-conserving ${\cal A}$-combinations
appear in the terms with odd powers of $\gamma$-matrices,
the chirality-flipping~${\cal B}$-couplings in those without $\gamma$-matrix.
Also the ``plus''(``minus'')-combinations appear without(with) $\gamma_5$.

Inserting Eq.~\eqref{neutself} into the two-loop diagrams of
Fig.~\ref{fig:tlcases}(\NS) and performing the loop integration
yields a rather compact expression for $a_\mu$.
The divergent part can be trivially read off from combining
Eqs.~\eqref{eq:countertermNS} and~\eqref{neutself}. 
The finite part is given by:
\begin{align}
\amuFSfclass{\NS}{ijm\tilde{f}_k}=\int _{0} ^{1}{\rm d} \fx
\Bigg[& {{\cal A}^{n+}_{ji\tilde{\mu}_m}}\left(
{\widetilde{{\cal A}}^{n+}_{ij\tilde{f}_{\tp}}}{\FF{\NS+}{AA}}+
{\widetilde{{\cal B}}^{n+}_{ij\tilde{f}_{\tp}}}{\FF{\NS+}{AB}}
\right)+{{\cal B}^{n+}_{ji\tilde{\mu}_m}}\left(
{\widetilde{{\cal A}}^{n+}_{ij\tilde{f}_{\tp}}}{\FF{\NS+}{BA}}+
{\widetilde{{\cal B}}^{n+}_{ij\tilde{f}_{\tp}}}{\FF{\NS+}{BB}}
\right)
\nonumber \\*
+&
{{\cal A}^{n-}_{ji\tilde{\mu}_m}}\left(
{\widetilde{{\cal A}}^{n-}_{ij\tilde{f}_{\tp}}}{\FF{\NS-}{AA}}+
{\widetilde{{\cal B}}^{n-}_{ij\tilde{f}_{\tp}}}{\FF{\NS-}{AB}}
\right)+{{\cal B}^{n-}_{ji\tilde{\mu}_m}}\left(
{\widetilde{{\cal A}}^{n-}_{ij\tilde{f}_{\tp}}}{\FF{\NS-}{BA}}+
{\widetilde{{\cal B}}^{n-}_{ij\tilde{f}_{\tp}}}{\FF{\NS-}{BB}}
\right) \Bigg].
\end{align}

Each of the loop functions $\FF{\NS\pm}{XY}$ can be expressed in terms
of two simpler functions $\FF{\NS1}{XY}$ and $\FF{\NS2}{XY}$, such
that the dependence on $\xfracNeut_{\BZNeut}$ and $\xNeuti/\xNeutj$
is essentially separated:
\begin{subequations}
\label{loopfunctionsNS}
\begin{align}
{\FF{\NS\pm}{AA}}&=\left(\frac{1}{16\pi ^2}\right)^2\colorNumber \frac{ m_{\mu }^2}{\MssmsNeutTWO}
\frac{m_i}{m_i{\mp} m_j}\bigg[{\FF{\NS1}{AA}}+{\FF{\NS2}{AA}}\bigg]+\plusswapij,\\
{\FF{\NS\pm}{AB}}&=\left(\frac{1}{16\pi ^2}\right)^2\colorNumber \frac{m_{\mu }^2m_{f'}}{\MssmsNeutTWO}
\frac{1}{\pm  m_i- m_j}\bigg[{\FF{\NS1}{AB}}+{\FF{\NS2}{AB}}\bigg]\pm\plusswapij,\\
{\FF{\NS\pm}{BA}}&=\left(\frac{1}{16\pi ^2}\right)^2\colorNumber m_{\mu }\frac{1}{m_i{\mp} m_j}{}
\bigg[{\FF{\NS1}{BA}}+{\FF{\NS2}{BA}}\bigg]\pm\plusswapij,\\
{\FF{\NS\pm}{BB}}&=\left(\frac{1}{16\pi ^2}\right)^2{\colorNumber} \frac{m_{\mu }m_{f'}}{\MssmsNeutTWO}
\frac{m_i}{\pm  m_i- m_j}\bigg[{\FF{\NS1}{BB}}+{\FF{\NS2}{BB}}\bigg]+\plusswapij,
\end{align}
\end{subequations}
with
\begin{subequations}
\begin{align}
{\FF{\NS1}{AA}}=&\,
{\fx}\,\frac{\abbroneloop(\xfracNeut_{\BZNeut},\xfracNeut_f,\xfracNeut_{\tilde{f}_k})}{\xfracNeut_{{\BZNeut}}-\xNeuti}
\frac{ F_1^N(\xfracNeut_{{\BZNeut}})}{48},\\*
\begin{split}
{\FF{\NS2}{AA}}=&\,\bigg(-4{}\LLNeut- l_i-2{}l_{\tilde{f}_k}\\*
&\ \ \,-\frac{1}{6 \xNeutiTWO}+\frac{1}{\xNeuti}+\frac{5}{2}+\frac{\xNeuti}{3}
-2{\fx}\,\frac{\abbroneloop(\xNeuti,\xfracNeut_f,\xfracNeut_{\tilde{f}_k})}{\xfracNeut_{{\BZNeut}}-\xNeuti}\bigg)
\frac{F_1^N(\xNeuti)}{96}+\frac{1-8\xNeuti-4 \xNeutiTWO}{288\xNeutiTWO},
\end{split}\\
{\FF{\NS1}{AB}}=&\,
\frac{\abbroneloop(\xfracNeut_{\BZNeut},\xfracNeut_f,\xfracNeut_{\tilde{f}_k})}{\xfracNeut_{{\BZNeut}}-\xNeuti}
\frac{ F_1^N(\xfracNeut_{{\BZNeut}})}{24},\\*
\begin{split}
{\FF{\NS2}{AB}}=&\,\bigg(-4 \LLNeut- l_i-2l_{\tilde{f}_k}\\*
&\ \ \,-\frac{1}{6 \xNeutiTWO}+\frac{1}{\xNeuti}+\frac{5}{2}+\frac{\xNeuti}{3}
-2\,\frac{\abbroneloop(\xNeuti,\xfracNeut_f,\xfracNeut_{\tilde{f}_k})}{\xfracNeut_{{\BZNeut}}-\xNeuti}\bigg)
\frac{F_1^N(\xNeuti)}{48}+\frac{1-8\xNeuti}{144\xNeutiTWO},
\end{split}\\
{\FF{\NS1}{BA}}=&\,
{\fx}\,\frac{\abbroneloop(\xfracNeut_{\BZNeut},\xfracNeut_f,\xfracNeut_{\tilde{f}_k})}{\xfracNeut_{{\BZNeut}}-\xNeuti}
\frac{\xfracNeut_{\BZNeut} \, F_2^N(\xfracNeut_{{\BZNeut}})-3}{24},\\*
\begin{split}
{\FF{\NS2}{BA}}=&\,\bigg(-4 \LLNeut-l_i-2 l_{\tilde{f}_k}\\*
&\ \ \,-\frac{1}{2 \xNeutiTWO}+\frac{2}{\xNeuti}+\frac{1}{2}+\xNeuti
-2{\fx}\,\frac{\abbroneloop(\xNeuti,\xfracNeut_f,\xfracNeut_{\tilde{f}_k})}{\xfracNeut_{{\BZNeut}}-\xNeuti}\bigg)
\frac{\xNeuti \, F_2^N(\xNeuti)-3}{48}-\frac{3-15\xNeuti}{96\xNeutiTWO},
\end{split}\\
{\FF{\NS1}{BB}}=&\,
\frac{\abbroneloop(\xfracNeut_{\BZNeut},\xfracNeut_f,\xfracNeut_{\tilde{f}_k})}{\xfracNeut_{{\BZNeut}}-\xNeuti}
\frac{ F_2^N(\xfracNeut_{{\BZNeut}})}{12},\\*
\begin{split}
{\FF{\NS2}{BB}}=&\,\bigg(-4 \LLNeut-l_i-2 l_{\tilde{f}_k}\\*
&\ \ \,+\frac{1}{2 \xNeutiTWO}+2+\frac{\xNeuti}{2}
-2\,\frac{\abbroneloop(\xNeuti,\xfracNeut_f,\xfracNeut_{\tilde{f}_k})}{\xfracNeut_{{\BZNeut}}-\xNeuti}\bigg)
\frac{F_2^N(\xNeuti)}{24}-\frac{1+\xNeuti}{16\xNeuti},
\end{split}
\end{align}
\end{subequations}

The comments at the end of Sec.~\ref{sec:NV} apply here as well.
However, the loop functions are now more complicated. The masses $m_f$ and
$m_{\tilde{f}_k}$ appear not only via the combination
$\xfracNeut_{{\BZNeut}}$ but also explicitly via
$\xfracNeut_f$ and $\xfracNeut_{\tilde{f}_k}$. Their dependence is
localized to the abbreviation $\abbroneloop$, which already appears in
the inner loop, i.\,e.~the self energy Eq.~\eqref{neutself}. In spite of
this complication, the loop functions in Eq.~\eqref{loopfunctionsNS}
could be partially expressed in terms of the one-loop
functions~$F_{1,2}^N$.  This is similar to the corresponding
counterterm diagram in Fig.~\ref{fig:ctcases}(\NS), Eq.~\eqref{eq:countertermNS}.

\subsection{Notations for the chargino results}

We now turn to the chargino results. Since the chargino is negatively
charged, the inner loop contains the fermion $f'$, which is the SU(2) doublet partner
of the fermion $f$. In analogy to the neutralino case we introduce an effective mass 
\begin{align}
m_{\BZChar}^2(\fx)\equiv\frac{m_{f'}^{2}}{\fx} + \frac{m_{\tilde{f}_k}^2}{1 - \fx},
\end{align}
and its corresponding propagator denominator 
\begin{align}
{\cal D}_{f' {\tilde f}_{k}}(\ell) &\equiv \ell^2 - m_{\BZChar}^2(\fx).
\label{CharginoDff}
\end{align}

As mentioned in Sec.~\ref{sec:couplings}, the chargino
interactions in the Lagrangian~\eqref{unmodifiedLagrangian} can be
rewritten using flipping rules for the anti-up-type fermions. 
Recalling the $(f' , {\tilde f})$ combinations in Eq.~\eqref{sumsextent} 
we apply $Q_{u^c}=-\frac{2}{3}$ for anti-up-type quarks, $Q_{d}=-\frac{1}{3}$
for down-type quarks, and $Q_{l}=-1$ for leptons.  
In this way, the relation for charge conservation
$Q_{f'}-Q_{\tilde{f}_k}=Q_{\tilde{\chi}^{-}}=-1$  is always valid.
The chargino results further depend on the color factor 
$\colorNumber$ of the inner fermion, the chargino masses 
$m_{i,j}\equiv m_{\tilde{\chi}^-_{i,j}}$ and the
dimensionless mass ratios
\begin{align}
\xChari \equiv  \frac{m_{i}^2}{\MssmsCharTWO},\quad
\xCharj \equiv  \frac{m_{j}^2}{\MssmsCharTWO},&\quad
\xfracChar_{f'} \equiv  \frac{m_{f'}^2}{\MssmsCharTWO},\quad
\xfracChar_{\tilde{f}_k} \equiv  \frac{m_{\tilde{f}_k}^2}{\MssmsCharTWO},\quad
\xfracChar_{\BZChar} \equiv  \frac{m_{\BZChar}^2(\fx)}{\MssmsCharTWO}. 
\end{align}
The logarithms defined in Eq.~\eqref{logarithms} are modified to
\begin{align}
l_z &\equiv  \log \xfracChar_z,\quad L(m^2) \equiv  \log\frac{m^2}{\muDR^2}
\label{kinematicvariablesChar}
\end{align}
for the chargino case. 

\subsection{Chargino vertex contributions}

\begin{figure}
\begin{center}
\begin{tabular}{cc}
\begin{picture}(100,60)(-60,-10)
\ArrowLine(-60,10)(-25,10)
\Vertex(-25,10){2}
\Vertex(25,10){2}
\ArrowArcn(0,10)(25,90,0)
\ArrowArcn(0,10)(25,180,90)
\DashArrowArcn(0,10)(25,360,180){4}
\ArrowLine(25,10)(60,10)
\Photon(0,70)(0,35){2}{3}
\Vertex(0,35){2}
\Text(10,55)[]{$q$}
\Text(5,55)[]{$\downarrow$}
\Text(0,25)[]{$f'$}
\Text(0,-5)[]{${\tilde f}_{k}$}
\Text(-55,20)[]{${ \tilde{\chi}_{j}}^{-}$}
\Text(42,3)[]{$\ell$}
\Text(-43,3)[]{$\ell-q$}
\Text(55,20)[]{${ \tilde{\chi}_{i}}^{-}$}
\Text(-50,45)[]{$(a)$}
\end{picture}
&\quad\quad\quad
\begin{picture}(100,60)(-60,-10)
\ArrowLine(-60,10)(-25,10)
\Vertex(-25,10){2}
\Vertex(25,10){2}
\ArrowArc(0,10)(25,180,360)
\DashArrowArc(0,10)(25,90,180){4}
\DashArrowArc(0,10)(25,0,90){4}
\ArrowLine(25,10)(60,10)
\Photon(0,70)(0,35){2}{3}
\Vertex(0,35){2}
\Text(10,55)[]{$q$}
\Text(5,55)[]{$\downarrow$}
\Text(0,-5)[]{$f'$}
\Text(0,25)[]{${\tilde f} _{k}$}
\Text(-55,20)[]{${ \tilde{\chi}_{j}}^{-}$}
\Text(42,3)[]{$\ell$}
\Text(-43,3)[]{$\ell-q$}
\Text(55,20)[]{${ \tilde{\chi}_{i}}^{-}$}
\Text(-50,45)[]{$(b)$}
\end{picture}
\end{tabular}
\end{center}
\caption{\label{fig:charvertex}%
Feynman diagrams for the chargino vertex insertion~$i\Gamma _{i j {\tilde f}_{k}} ^{- \, \mu} (\ell)$.
The possible insertions are given in Eq.~\eqref{sumsextent} and involve only negative fermions
and positive sfermions.}
\end{figure}
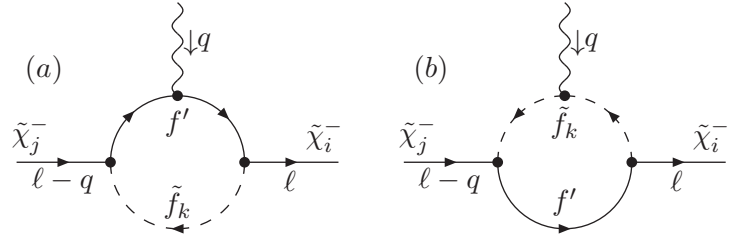
%

The chargino vertex contributions are shown in
Fig.~\ref{fig:tlcases}(\CV).
The corresponding one-loop insertions are shown in
Fig.~\ref{fig:charvertex} and can be written as:
\begin{align}
\begin{split}
\Gamma _{i j {\tilde f}_{k}} ^{- \mu} (\ell) =
\frac{e Q_{f'}}{16\pi^2}\int_{0}^{1} \frac{{\rm d} \fx}{2}
\Bigg\{
&\left(\caplus - \caminu \gamma_{5} \right)
\frac{\slashed{\ell}\slashed{q}\gamma^{\mu} -\slashed{\ell} q^{\mu} + 
\slashed{q} \ell^{\mu}- (\ell\cdot q)\gamma^{\mu}}{\Dfpf}\\
+&\left(\cbplus - \cbminu \gamma_{5} \right)
\frac{m_{f'}}{\fx} \frac{\slashed{q} \gamma ^{\mu} - q^{\mu}}{\Dfpf} 
\Bigg\}
\\
-\frac{e Q_{{\tilde \chi}^{-}}}{16 \pi^2}\int_{0}^{1}\frac{{\rm d}\fx}{2}
\Bigg\{
&\left(\caplus - \caminu \gamma_{5} \right)\fx \\
&\times\Bigg[
\left(\frac{1}{\epsilon} - L(m_{{\tilde f}_{k}}^2) \right)\gamma^{\mu} 
-2\frac{(\ell \cdot q)\,{\slashed \ell}\,\ell^{\mu}}{\DfpfTWO}\\
&\quad\quad+\frac{
\slashed{q}\ell^{\mu} + \slashed{\ell}q^{\mu} +
(\ell \cdot q) \gamma^{\mu} - 2\,\slashed{\ell}\,\ell^{\mu}
+\abbroneloop(\ell^2, m_{f'}^2, m_{{\tilde f}_{k}}^2)\gamma^{\mu}/2
}{\Dfpf}
\Bigg]\\
+&\left(\cbplus - \cbminu \gamma_{5} \right)m_{f'}
\Bigg[-2 \frac{(\ell \cdot q) \ell^{\mu}}{\DfpfTWO}+
\frac{q^{\mu} - 2 \ell ^{\mu}}{\Dfpf}
\Bigg]
\Bigg\}.
\end{split}
\label{CharginoVertexOneLoop}
\end{align}
The part proportional to the inner fermion charge $Q_{f'}$ can
be written in exactly the same way as the neutralino vertex
corrections. The remaining terms proportional
to $Q_{\tilde{\chi}^{-}}$ have no neutralino counterpart.
They are divergent and have a far more involved structure.

Inserting Eq.~\eqref{CharginoVertexOneLoop} into the outer
loop and integrating over the loop momentum yields for the finite part
%
\begin{align}
\amuFSfclass{\CV}{ij\tilde{f}_k}=\int _{0} ^{1}{\rm d} \fx
\Bigg[&
{{\cal A}^{c+}_{j i\tilde{\nu}_\mu}}\left(
{\caplus}{\FF{\CV+}{AA}}+
{\cbplus}{\FF{\CV+}{AB}}
\right)+
{{\cal B}^{c+}_{j i\tilde{\nu}_\mu}}\left(
{\caplus}{\FF{\CV+}{BA}}+
{\cbplus}{\FF{\CV+}{BB}}
\right)
\nonumber\\*
+&
{{\cal A}^{c-}_{j i\tilde{\nu}_\mu}}\left(
{\caminu}{\FF{\CV-}{AA}}+
{\cbminu}{\FF{\CV-}{AB}}
\right)+
{{\cal B}^{c-}_{j i\tilde{\nu}_\mu}} \left(
{\caminu}{\FF{\CV-}{BA}}+
{\cbminu}{\FF{\CV-}{BB}}
\right)
\Bigg].
\end{align}

\newcommand{\kindex}{x}
Now the loop functions have a significantly more involved
structure. On the one hand this is due to the additional terms in
the inner loop, and on the other hand it is caused by the outer loop, for which
the counterterm result of Eq.~\eqref{eq:countertermCV} already provides
an illustration. 
For the loop functions we find the following expressions
\begin{subequations}
\label{loopfunctionsCV}
\begin{align}
\begin{split}
{\FF{\CV\pm}{AA}}=&\left(\frac{1}{16\pi ^2}\right)^2
\colorNumber\frac{m_{\mu }^2 }{{\MssmsChar}^2}
\bigg[\frac{1}{\xfracChar_i-\xfracChar_j}
\bigg(\frac{{\FF{\CV 1a}{AA}}(\xfracChar_\BZChar)-{\FF{\CV 1a}{AA}}(\xfracChar_i)}
  {\xfracChar_\BZChar-\xfracChar_i}+{\FF{\CV 2}{AA}}(\xfracChar_i)\bigg)\\*
& \pm\frac{m_im_j}{m_i^2-m_j^2}
\frac{{\FF{\CV 1b}{AA}}(\xfracChar_\BZChar)-{\FF{\CV 1b}{AA}}(\xfracChar_i)}
  {\xfracChar_\BZChar-\xfracChar_i}
\bigg]+\plusswapij,
\end{split}\\
{\FF{\CV\pm}{AB}}=&\left(\frac{1}{16\pi ^2}\right)^2
\colorNumber\frac{m_{\mu }^2m_{f'} }{{\MssmsChar}^2}\frac{1}{{\pm} m_i-m_j}\frac{1}{{\fx}}
\bigg[\frac{{\FF{\CV 1}{AB}}(\xfracChar_\BZChar)-{\FF{\CV 1}{AB}}(\xfracChar_i)}
  {\xfracChar_\BZChar-\xfracChar_i}
\bigg]\pm\plusswapij,\\
\begin{split}
{\FF{\CV\pm}{BA}}=&\left(\frac{1}{16\pi ^2}\right)^2
{\colorNumber} m_{\mu }
\bigg[\frac{1}{m_i{\mp}\text{   }m_j}
\bigg(\frac{{\FF{\CV1a}{BA}}(\xfracChar_\BZChar)-{\FF{\CV1a}{BA}}(\xfracChar_i)}
  {\xfracChar_\BZChar-\xfracChar_i}
+{\FF{\CV2a}{BA}}\bigg)\\
& +\frac{1}{m_i{\pm} m_j}
\bigg(\frac{{\FF{\CV1b}{BA}}}{\xfracChar_\BZChar-\xfracChar_i}
+{\FF{\CV2b}{BA}}\bigg)\bigg]\pm\plusswapij,
\end{split}\\
\begin{split}
{\FF{\CV\pm}{BB}}=&\left(\frac{1}{16\pi ^2}\right)^2
\colorNumber\frac{{}m_{\mu }m_{f'} }{{\MssmsChar}^2}
\bigg[\frac{{\pm1}}{\xfracChar_i-\xfracChar_j}
\frac{{\FF{\CV 1}{BB}}(\xfracChar_\BZChar)-{\FF{\CV 1}{BB}}(\xfracChar_i)}
  {\xfracChar_\BZChar-\xfracChar_i} \\*
& +\frac{m_im_j}{m_i^2-m_j^2}
\frac{{\FF{\CV 1b}{BB}}(\xfracChar_\BZChar)-{\FF{\CV 1b}{BB}}(\xfracChar_i)}
  {\xfracChar_\BZChar-\xfracChar_i}\bigg]+\plusswapij,
\end{split}
\end{align}
\end{subequations}
with
\begin{subequations}
\begin{align}
\begin{split}
{\FF{\CV 1a}{AA}}(\xfracChar_{\kindex})=\ &
\frac{Q_{f'}}{4}
\frac{\xfracChar_{\kindex} (1-\xfracChar_{\kindex}+l_\kindex \xfracChar_{\kindex}^2)}{(1-\xfracChar_{\kindex}){}^2}+
\frac{{\fx}}{4}
\frac{1-\xfracChar_{\kindex}+l_\kindex \xfracChar_{\kindex}^3}{(1-\xfracChar_{\kindex}){}^2}+
\frac{{\fx}}{24 (1-{\fx})}
\frac{1}{(1-\xfracChar_{\kindex}){}^3} \\*
& \times
\bigg[ 2-6 \xfracChar_{\kindex}+4 \xfracChar_{\kindex}^2-3 l_\kindex \xfracChar_{\kindex}^3+l_\kindex \xfracChar_{\kindex}^4+
  (\xfracChar_{f'}-\xfracChar_{\tilde{f}_k})\xfracChar_{\kindex}
  (2-2 \xfracChar_{\kindex}+3l_\kindex{}\xfracChar_{\kindex}-l_\kindex\xfracChar_{\kindex}^2)\bigg],
\end{split}\\*
{\FF{\CV 1b}{AA}}\left(\xfracChar_{\kindex}\right)=\ &
\frac{Q_{f'}}{4}
\frac{1-\xfracChar_{\kindex}+l_\kindex \xfracChar_{\kindex}^2}{\left(1-\xfracChar_{\kindex}\right){}^2}+
\frac{{\fx} }{12}
\frac{3-8 \xfracChar_{\kindex}+5 \xfracChar_{\kindex}^2-2 l_\kindex \xfracChar_{\kindex}^3}
  {\left(1-\xfracChar_{\kindex}\right){}^3},\\
\begin{split}
{\FF{\CV 2}{AA}}\left(\xfracChar_{\kindex}\right)=\ &
\bigg[-12 \LLChar-6 l_{\tilde{f}_k}-3 l_\kindex+5-\frac{6}{\xfracChar_{\kindex}}\bigg]
\frac{\xfracChar_{\kindex} \left(2-2 \xfracChar_{\kindex}+3 l_\kindex \xfracChar_{\kindex}-l_\kindex \xfracChar_{\kindex}^2\right)}
  {144\left(1-\xfracChar_{\kindex}\right){}^3} \\*
& +\frac{1+2 l_\kindex \xfracChar_{\kindex}-\xfracChar_{\kindex}^2}
  {12\left(1-\xfracChar_{\kindex}\right){}^3},
\end{split}\\
{\FF{\CV 1}{AB}}\left(\xfracChar_{\kindex}\right)=\ &
{\FF{\CV 1b}{AA}}\left(\xfracChar_{\kindex}\right),\\
\begin{split}
{\FF{\CV 1a}{BA}}\left(\xfracChar_{\kindex}\right)=\ &
-\frac{Q_{f'}}{4}
\frac{l_\kindex \xfracChar_{\kindex}^2}{ 1-\xfracChar_{\kindex} }+
\frac{{\fx}}{8}
\bigg[\frac{2}{1-\xfracChar_{\kindex}}+
  \frac{ l_\kindex \xfracChar_{\kindex}^2\left(1+\xfracChar_{\kindex}\right)}
  {\left(1-\xfracChar_{\kindex}\right){}^2 }\bigg] \\*
& -
\frac{1}{16}
\fx\,\abbroneloop(\xfracChar_\BZChar,\xfracChar_{f'},\xfracChar_{\tilde{f}_k})
\bigg[\frac{1}{1-\xfracChar_{\kindex} }+
\frac{ l_\kindex \xfracChar_{\kindex}\left(2-\xfracChar_{\kindex}\right)}
  {\left(1-\xfracChar_{\kindex}\right){}^2}
\bigg],
\end{split}\\
\begin{split}
{\FF{\CV 1b}{BA}}=\ &
\frac{1}{16}
\fx\,\abbroneloop(\xfracChar_\BZChar,\xfracChar_{f'},\xfracChar_{\tilde{f}_k})
\bigg[\frac{l_i\xfracChar_i(\xfracChar_i^2-2 \xfracChar_j+\xfracChar_i \xfracChar_j)}
  {\left(\xfracChar_i-\xfracChar_j\right)\left(1-\xfracChar_i\right){}^2}+
\frac{\xfracChar_i}{1-\xfracChar_i}+ \\*
& +\frac{ l_{\BZChar} \xfracChar_\BZChar^2 }{1-\xfracChar_\BZChar}
\bigg(\frac{2}{\xfracChar_i-\xfracChar_j}-\frac{1}{\xfracChar_\BZChar-\xfracChar_i}\bigg)\bigg] 
-\frac{{\fx}}{8}\frac{ l_i \xfracChar_i^2\text{   }}{1-\xfracChar_i},
\end{split}\\
\begin{split}
{\FF{\CV 2a}{BA}}=\ &
\frac{\xfracChar_i\left(-3+3 \xfracChar_i-4 l_i+l_i \xfracChar_i+2 l_i^2-l_i^2 \xfracChar_i\right)}
  {32\left(1-\xfracChar_i\right){}^2}\\*
& +
\frac{1}{16}\bigg[2 \LLChar+l_{\tilde{f}_k}-\fx\frac{1-2 {\fx}}{1-{\fx}}\bigg]\times
\bigg[\frac{l_i \left(2-\xfracChar_i\right)\xfracChar_i}
  {\left(1-\xfracChar_i\right){}^2}
+\frac{\xfracChar_i}{1-\xfracChar_i}\bigg],
\end{split}\\
\begin{split}
{\FF{\CV 2b}{BA}}=\ &
{\FF{\CV 2a}{BA}}+
\frac{\left(3 l_i-l_i^2\right) \xfracChar_i^2}
  {16 \left(1-\xfracChar_i\right) \left(\xfracChar_i-\xfracChar_j\right)} \\*
& -\frac{1}{8}\ \bigg[2 \LLChar+l_{\tilde{f}_k}-\fx\frac{1-2{\fx}}{1-{\fx}}\bigg]\times
\bigg[\frac{l_i \xfracChar_i^2}
  {\left(1-\xfracChar_i\right) \left(\xfracChar_i-\xfracChar_j\right)}
\bigg],
\end{split}\\
{\FF{\CV 1}{BB}}\left(\xfracChar_{\kindex}\right)=\ &
\frac{1-\xfracChar_{\kindex}+l_\kindex \xfracChar_{\kindex}^2}{4\left(1-\xfracChar_{\kindex}\right){}^2} ,\\
{\FF{\CV 1b}{BB}}\left(\xfracChar_{\kindex}\right)=\ &
-\frac{Q_{f'}}{2{\fx}}
\frac{l_\kindex \xfracChar_{\kindex}}{1-\xfracChar_{\kindex}}+
\frac{1-\xfracChar_{\kindex}+l_\kindex \xfracChar_{\kindex}^2}{4\left(1-\xfracChar_{\kindex}\right){}^2}
\end{align}
\end{subequations}
In some of these functions an explicit argument $\xfracChar_\kindex$ is
specified, which is specialized in 
Eqs.\ (\ref{loopfunctionsCV}) to  $\xfracChar_\kindex\in\{\xfracChar_i,\xfracChar_j,\xfracChar_\BZChar\}$. 

\subsection{Chargino self-energy contributions}

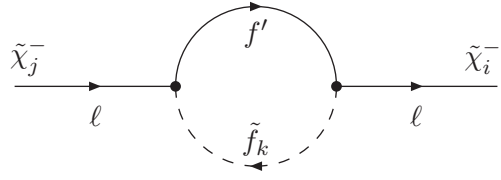
\begin{figure}
\begin{picture}(100,60)(-200,-40)
\ArrowLine(-90,0)(-30,0)
\Vertex(-30,0){2}
\ArrowArcn(0,0)(30,180,0)
\DashArrowArcn(0,0)(30,0,180){4}
\Vertex(30,0){2}
\ArrowLine(30,0)(90,0)
\Text(0,20)[]{$f'$}
\Text(0,-20)[]{${\tilde f}_{k}$}
\Text(-85,11)[]{${ \tilde{\chi}} _{j}^{-}$}
\Text(-60,-11)[]{$\ell$}
\Text(85,11)[]{${ \tilde{\chi}} _{i}^{-}$}
\Text(60,-11)[]{$\ell$}
\end{picture}
\caption{Feynman diagram for the chargino self-energy insertion
$i\Sigma^{-}_{i j {\tilde f}_k}({\ell})$.}
\label{fig:charselfenergy}
\end{figure}

The last class of diagrams is the one corresponding to
Fig.~\ref{fig:tlcases}(\CS), where the fermion/sfermion loop
generates a chargino self-energy.
Fig.~\ref{fig:charselfenergy} shows the chargino self-energy one-loop diagram.
%
The result has the same structure as the corresponding
neutralino self-energy result, see Eq.~\eqref{neutself},
%

\begin{align}
\begin{split}
\Sigma^{-}_{i j {\tilde f}_k}({\ell}) = 
\frac{1}{16 \pi ^{2}} \int _{0} ^{1} \frac{{\rm d} \fx}{2} & \left[ \
\Big(\acplusijsfermion - \acminusijsfermion \gamma ^{5}\Big) \fx \slashed\ell
\Bigg(\frac{1}{\epsilon} - L(m_{{\tilde f}_{k}}^2 )+\frac{\abbroneloop(\ell^2, m_{f'}^2, m_{{\tilde f}_{k}}^2)/2}{{\cal D}_{f' {\tilde f}_{k}}(\ell)}\Bigg)\right. \\
&+\left.\Big(\bcplusijsfermion - \bcminusijsfermion \gamma ^{5}\Big) m_{f}
\Bigg(\frac{1}{\epsilon} - L(m_{{\tilde f}_{k}}^2)+\frac{\abbroneloop(\ell^2, m_{f'}^2, m_{{\tilde f}_{k}}^2)}{{\cal D}_{f' {\tilde f}_{k}}(\ell)}\Bigg)
\right].
\end{split}
\label{eq,charself}
\end{align}
This self-energy vertex can be inserted in the outer loop
in two different ways, as shown in Fig.~\ref{fig:tlcases}(\CS). 
Inserting the self-energy result into the two-loop diagrams,
the following contribution to $a_\mu$ is obtained:
%
\begin{align}
\amuFSfclass{\CS}{ij\tilde{f}_k}=\int _{0} ^{1}{\rm d} \fx 
\Big[
&{{\cal A}^{c+}_{j i \tilde{\nu}_\mu}}
\left(
{\caplus}{\FF{\CS +}{AA}}+
{\cbplus}{\FF{\CS +}{AB}}
\right)+
{{\cal B}^{c+}_{j i \tilde{\nu}_\mu}}
\left(
{\caplus}{\FF{\CS +}{BA}}+
{\cbplus}{\FF{\CS +}{BB}}
\right)
\nonumber\\*+
&
{{\cal A}^{c-}_{ji\tilde{\nu}_\mu}}
\left(
{\caminu}{\FF{\CS -}{AA}}+
{\cbminu}{\FF{\CS -}{AB}}
\right)+{{\cal B}^{c-}_{ji\tilde{\nu}_\mu}}
\left(
{\caminu}{\FF{\CS -}{BA}}+
{\cbminu}{\FF{\CS -}{BB}}
\right)
\Big].
\end{align}
As is clear from the corresponding one-loop counterterm diagrams of
Fig.~\ref{fig:ctcases}(\CS), Eq.~\eqref{eq:countertermCS}, the
chargino self-energy corrections lead to the most complicated
two-loop expressions. 
It would result in very long formulas for the loop
functions~$\FF{\CS\pm}{XY}$ to express the ${\cal O}(\epsilon^0)$-result
like in the previous cases. 
Hence we employ the $\epsilon$-dependent one-loop
functions defined with two variables~$\FTIL_{1,2,3}^{C}(\xCharj,\xChari)$. 
The~${\cal O}(\epsilon^0)$-part can then be obtained by evaluating the following expressions explicitly. 

Furthermore, the particular structure of the one-loop self-energy result is used. 
All terms have the form of a counterterm insertion, possibly
multiplied with an additional propagator and~$\fx$-dependent
rational functions, which can be split off:
\begin{subequations}
\begin{align}
{\FF{\CS \pm}{XA}}&=\Bigg(\frac{1}{16\pi^2}\Bigg)^2
\Bigg[\Bigg(\frac{1}{{\epsilon}}-\LLChar-l_{\tilde{f}_k}-\frac{2{\fx}-1}{2(1-{\fx})}\Bigg)
{\FF{\CS 1\pm}{XA}}+\frac{\abbroneloop(\xfracChar_{\BZChar},\xfracChar_{f'},\xfracChar_{\tilde{f}_k})}{2}{\FF{\CS 2\pm}{XA}}\Bigg]\fx,\\*
{\FF{\CS \pm}{XB}}&=\Bigg(\frac{1}{16\pi^2}\Bigg)^2
\Bigg[\Bigg(\frac{1}{{\epsilon}}-\LLChar-l_{\tilde{f}_k}-\frac{2{\fx}-1}{\phantom{2(}1-{\fx}\phantom{)}}\Bigg)
{\FF{\CS 1\pm}{XB}}+\abbroneloop(\xfracChar_{\BZChar},\xfracChar_{f'},\xfracChar_{\tilde{f}_k})\, {\FF{\CS 2\pm}{XB}}\Bigg],
\end{align}
\end{subequations}
where $X\in\{A,B\}$. The individual results for the coefficients of
the $\fx$-dependent functions are
\begin{subequations}
\begin{align}
{\FF{\CS 1\pm}{AA}}=&
-\frac{\colorNumber}{24}\frac{m_{\mu }^2}{\MssmsCharTWO}\frac{m_i}{m_i{\mp} m_j}
\Big[{\FTIL_{\rm 1}^{\rm C}}(\xChari)+{\FTIL_{\rm 1}^{\rm C}}(\xCharj,\xChari)\Big]+\plusswapij, \\*
\begin{split}
{\FF{\CS 2\pm}{AA}}=&\
\frac{{\colorNumber}}{24}\frac{ m_{\mu }^2}{\MssmsCharTWO}\frac{m_i}{m_i{\mp}m_j}\frac{1}{\xfracChar_{{\BZChar}}-\xChari} \\*
& \times\Big[{\FTIL_{\rm 1}^{\rm C}}(\xChari)-{\FTIL_{\rm 1}^{\rm C}}(\xfracChar_{{\BZChar}},\xChari)
-{\FTIL_{\rm 1}^{\rm C}}(\xCharj,\xfracChar_{{\BZChar}})+{\FTIL_{\rm 1}^{\rm C}}(\xCharj,\xChari)\Big]+\plusswapij,
\end{split} \\
{\FF{\CS 1\pm}{AB}}=&
-\frac{{\colorNumber}}{24}\frac{m_{\mu }^2}{\MssmsCharTWO}m_{f'}\frac{1}{\pm  m_i-m_j}
{\FTIL_{\rm 1}^{\rm C}}(\xChari)\pm\plusswapij,\\*
\begin{split}
{\FF{\CS 2\pm}{AB}}=&\
\frac{{\colorNumber}}{24}\frac{m_{\mu }^2}{\MssmsCharTWO}m_{f'}\frac{1}{\pm m_i-m_j}\frac{1}{\xfracChar_{{\BZChar}}-\xChari} \\*
& \times\Big[{\FTIL_{\rm 1}^{\rm C}}(\xChari)-{\FTIL_{\rm 1}^{\rm C}}(\xfracChar_{{\BZChar}},\xChari)
-{\FTIL_{\rm 1}^{\rm C}}(\xCharj,\xfracChar_{{\BZChar}})+{\FTIL_{\rm 1}^{\rm C}}(\xCharj,\xChari)\Big] \pm\plusswapij,
\end{split} \\
\begin{split}
{\FF{\CS 1\pm}{BA}}=&
-\frac{{\colorNumber}}{12}\frac{m_{\mu }}{\MssmsCharTWO}\frac{m_i}{m_i{\mp} m_j} \\* 
&\times\Big[2{}m_i {\FTIL_{\rm 2}^{\rm C}}(\xChari)+( m_i{\pm}{}m_j) 
{\FTIL_{\rm 2}^{\rm C}}(\xCharj,\xChari)-6 ( m_i{\mp}{}m_j) {\FTIL_{\rm 3}^{\rm C}}(\xCharj,\xChari)\Big]\pm\plusswapij,
\end{split}\\*
\begin{split}
{\FF{\CS 2\pm}{BA}}=
&-\frac{{\colorNumber}}{12}
\frac{m_{\mu }}{\MssmsCharTWO}
\frac{m_i}{m_i \mp m_j}
\frac{1}{\xfracChar_{{\BZChar}}-\xChari} \\*
&\times\Big[2{}m_i \Big(-{\FTIL_{\rm 2}^{\rm C}}(\xChari)+{\FTIL_{\rm 2}^{\rm C}}(\xfracChar_{{\BZChar}},\xChari)\Big)
+(m_i{\pm}{}m_j) \Big({\FTIL_{\rm 2}^{\rm C}}(\xCharj,\xfracChar_{{\BZChar}})-{\FTIL_{\rm 2}^{\rm C}}(\xCharj,\xChari)\Big) \\*
&\quad +6( m_i{\mp}{}m_j) \Big({\FTIL_{\rm 3}^{\rm C}}(\xCharj,\xChari)-{\FTIL_{\rm 3}^{\rm C}}(\xCharj,\xfracChar_{{\BZChar}})\Big)\Big]+\plusswapij,
\end{split}\\
{\FF{\CS 1\pm}{BB}}=&
-\frac{{\colorNumber}}{6}\frac{ m_{\mu } m_{f'} }{\MssmsCharTWO}\frac{m_i }{\pm  m_i- m_j}
{\FTIL_{\rm 2}^{\rm C}}(\xChari)+\plusswapij,\\* \nonumber
\begin{split}
{\FF{\CS 2\pm}{BB}}=
&-\frac{{\colorNumber}}{12}
\frac{m_{\mu } m_{f'}}{\MssmsCharTWO}
\frac{1}{\pm  m_i- m_j}
\frac{1}{\xfracChar_{{\BZChar}}-\xChari} \\*
&\times\Big[2{}m_i \Big(-{\FTIL_{\rm 2}^{\rm C}}(\xChari)+{\FTIL_{\rm 2}^{\rm C}}(\xfracChar_{{\BZChar}},\xChari)\Big)
+(m_i{\pm}{}m_j) \Big({\FTIL_{\rm 2}^{\rm C}}(\xCharj,\xfracChar_{{\BZChar}})-{\FTIL_{\rm 2}^{\rm C}}(\xCharj,\xChari)\Big) \\*
&\quad +6( m_i{\mp}{}m_j) \Big({\FTIL_{\rm 3}^{\rm C}}(\xCharj,\xChari)-{\FTIL_{\rm 3}^{\rm C}}(\xCharj,\xfracChar_{{\BZChar}})\Big)\Big]+\plusswapij.
\end{split}
\end{align}
\end{subequations}

\newpage
\newcommand{\amuWHnu}{a_{\mu}^{1{\rm L}}({\tilde W}\text{--}{\tilde H},{\tilde \nu}_{\mu})} 
\newcommand{\amuWHmuL}{a_{\mu}^{1{\rm L}}({\tilde W}\text{--}{\tilde H},{\tilde \mu}_{L})} 
\newcommand{\amuBHmuL}{a_{\mu}^{1{\rm L}}({\tilde B}\text{--}{\tilde H},{\tilde \mu}_{L})}
\newcommand{\amuBHmuR}{a_{\mu}^{1{\rm L}}({\tilde B}\text{--}{\tilde H},{\tilde \mu}_{R})}
\newcommand{\amuBmuLmuR}{a_{\mu}^{1{\rm L}}({\tilde B},{\tilde \mu}_{L}\text{--}{\tilde \mu}_{R})} 

\newcommand{\amuWHnutwoL}{a_{\mu}^{2{\rm L}, f{\tilde f}\, {\rm LL}}({\tilde W}\text{--}{\tilde H},{\tilde \nu}_{\mu})} 
\newcommand{\amuWHmuLtwoL}{a_{\mu}^{2{\rm L},f{\tilde f}\, {\rm LL}}({\tilde W}\text{--}{\tilde H},{\tilde \mu}_{L})} 
\newcommand{\amuBHmuLtwoL}{a_{\mu}^{2{\rm L},f{\tilde f}\, {\rm LL}}({\tilde B}\text{--}{\tilde H},{\tilde \mu}_{L})}
\newcommand{\amuBHmuRtwoL}{a_{\mu}^{2{\rm L},f{\tilde f}\, {\rm LL}}({\tilde B}\text{--}{\tilde H},{\tilde \mu}_{R})}
\newcommand{\amuBmuLmuRtwoL}{a_{\mu}^{2{\rm L},f{\tilde f}\, {\rm LL}}({\tilde B},{\tilde \mu}_{L}\text{--}{\tilde \mu}_{R})} 
\newcommand{\logscale}{m_{\text{SUSY}}}

\newcommand{\Deltagone}{\Delta_{g_{1}}}
\newcommand{\Deltagtwo}{\Delta_{g_{2}}}
\newcommand{\DeltaYukHiggsino}{\Delta_{\tilde H}}
\newcommand{\DeltaYukBinoHiggsino}{\Delta_{{\tilde B}{\tilde H}}}
\newcommand{\DeltaYukWinoHiggsino}{\Delta_{{\tilde W}{\tilde H}}}
\newcommand{\Deltatanbe}{\Delta_{t_{\beta}}}

\newcommand{\constA}{0.015}
\newcommand{\constB}{0.015}
\newcommand{\constC}{0.015}
\newcommand{\constD}{0.04}
\newcommand{\constE}{0.03}

\section{Overview of input parameters and benchmark scenarios}
\label{sec:inputparameters}

In the remaining three sections the phenomenological
behaviour of the results is discussed. The present section gives an
overview of the input 
parameters and useful benchmark parameter scenarios; then a
compact approximation is provided, and finally the parameter
dependence of the fermion/sfermion-loop contributions to $a_\mu$ is
analyzed in detail.

The fermion/sfermion-loop contributions to $a_\mu$ depend on the
following fifteen parameters.
\begin{itemize}
\item One-loop parameters: 
\begin{align}
\mu, M_1, M_2, M_{E}, M_{L}, \tan\beta.
\end{align}
Of course, all parameters of the one-loop SUSY contributions appear
again.
\item Two-loop sfermion-mass parameters appearing in the inner loop:
\begin{align}
M_U, 
M_D, M_Q, M_{U3}, M_{D3}, M_{Q3}, M_{E3}, M_{L3}.
\end{align}
The additional sensitivity on these sfermion masses of all generations 
is one of the most important properties of the fermion/sfermion-loop
contributions. For simplicity, the sfermion-mass parameters of the
first two generations are set equal; hence the first-generation
slepton masses do not appear as free parameters here.
\item Stop $A$-parameter:
\begin{align}
A_t.
\end{align}
All other $A$-parameters appear only multiplied with small fermion
masses and are neglected.
\end{itemize}
In our analysis all parameters are considered to be real quantities,
and generation mixing is neglected. 
SM input parameters are defined as in Ref.~\cite{PDG2012}:
    \begin{align}
      \begin{alignedat}{2}
        M_{W} &= (80.385 \pm 0.015)~\text{GeV}, &\quad m_{t} &= (173.5 \pm 0.6 \pm 0.8)~\text{GeV},\\
        M_{Z} &= (91.1876 \pm 0.0021)~\text{GeV}, &\quad m_{\mu} &= (105.6583715 \pm 0.0000035)~\text{MeV}.\\
      \end{alignedat}
    \end{align}
Since we define $\tan\beta$ in the 
$\overline{\text{DR}}$~renormalization scheme, the final result also
depends on the scale $\muDR$, which we always set to the SPS1a value
$\muDR=454.7$~GeV \cite{SPSDef}. 

\begin{table}
\centering
\begin{tabular}{l c c c c c c}
					& BM1	& BM2	& BM3	& BM4	\\
\hline\hline
$\mu [\text{GeV}]$			& 350	& 1300	& 4000	& $-160$	\\
$\text{tan}\beta$			& 40	& 40	& 40	& 50	\\
$M_1 [\text{GeV}]$			& 150	& 150	& 150	& 140	\\
$M_2 [\text{GeV}]$			& 300	& 300	& 300	& 2000	\\
$M_{E} [\text{GeV}]$			& 400	& 400	& 400	& 200	\\
$M_{L} [\text{GeV}]$ 			& 400	& 400	& 400	& 2000	\\
$\amuSUOL[ 10^{-10}]$	& 44.02	& 26.95	& 46.78	& 15.98
\end{tabular}
\caption{\label{BMDefinition} Definition of the benchmark points; see
  also Ref.~\cite{fsf2loopA}. In BM1, all one-loop masses are similar;
  in BM2, the $\mu$-parameter is increased by a factor 4. In BM3, the
  $\mu$-parameter is 
  very large and the bino-exchange contribution dominates. In BM4, the
  contribution from the right-handed smuon dominates.
}
\end{table}

In the following numerical discussions, the benchmark points for the
one-loop parameters, introduced in Ref.~\cite{fsf2loopA} and defined
in Tab.~\ref{BMDefinition}, are used repeatedly.
They characterize qualitatively different regions of the one-loop
parameter space, where the one-loop result is dominated by different
mass-insertion diagrams (see Refs.~\cite{moroi, review,
  Cho:2011rk}).
\begin{itemize}
\item BM1: All one-loop masses are similar; the one-loop contribution
  to $a_\mu$ is dominated by the chargino mass-insertion diagram
  $\amuWHnu$ with wino--Higgsino exchange.
\item BM2: The $\mu$-parameter is increased by a factor $\sim4$. The
  one-loop chargino contribution $\amuWHnu$ and the bino-exchange contribution
  $\amuBmuLmuR$ are similar. The well-known benchmark point SPS1a
  \cite{SPSDef} has a similar characteristic.
\item BM3: The $\mu$-parameter is very large. All one-loop
  contributions involving
  higgsinos are suppressed, and the bino-exchange contribution
  $\amuBmuLmuR$ dominates. Parameter scenarios with this
  characteristic have been studied extensively also in
  Refs.~\cite{Endo:2013bba,Endo:2013lva} recently.
\item BM4: The parameters are chosen such that the right-handed smuon
  contribution $\amuBHmuR$ dominates: $M_2$ and $M_{L}$ are heavy, and
  the other three one-loop mass parameters are light. The
  $\mu$-parameter is negative to allow for a positive contribution to
  $a_\mu$. Parameter scenarios with this characteristic have been
  studied recently also in Ref.~\cite{Grothaus:2012js}.
\end{itemize}

\newpage

\section{Leading logarithmic approximation}
%
\label{sec:leadinglog}

As a first step of the numerical discussion, a very compact approximate formula is provided.  
It can be easily implemented, and it captures many features of the
qualitative and quantitative behaviour of the exact result.  
The approximation is based on the leading logarithms of the result. 

As discussed in Ref.~\cite{fsf2loopA}, the fermion/sfermion-loop contributions
are logarithmically enhanced if the sfermions in the inner loop become heavy.
This non-decoupling behaviour can be understood in an effective field theory. 
If heavy sfermions are integrated out, the effective field theory is not supersymmetric anymore,
and gaugino and higgsino couplings can differ from the corresponding gauge and Yukawa couplings. 

Based on this idea, we start from the one-loop result, approximated by
mass-insertion diagrams, see Refs.~\cite{moroi, review,
  Cho:2011rk}.\footnote{%
Note that we only use $\amuSUOLapprox$ and the definitions of
Eq.\ (\ref{eq:onelooplogapprox}) 
as  building blocks
in an approximation of the two-loop results.  They  should not be used in
a precision evaluation of the SUSY one-loop contributions, 
since the error can be significant.
}
In the form given in Ref.~\cite{Cho:2011rk}
the approximation reads
\begin{align}
\amuSUOLapprox &= \amuWHnu + \amuWHmuL + \amuBHmuL \nonumber\\*
&\quad + \amuBHmuR + \amuBmuLmuR,
\end{align}
with
\begin{subequations}
\label{eq:onelooplogapprox}
\begin{align}
\amuWHnu &= \frac{g_{2}^{2}}{8 \pi ^{2}} \frac{m_{\mu} ^{2} M_{2}}{m_{{\tilde \nu}_{\mu}}^{4}}\,\mu \tan\beta\, 
F_{a}\left(\frac{M_{2}^{2}}{m_{{\tilde \nu}_{\mu}}^{2}},\frac{\mu^{2}}{m_{{\tilde \nu}_{\mu}}^{2}}\right), \\* 
\amuWHmuL &= - \frac{g_{2}^{2}}{16 \pi ^{2}} \frac{m_{\mu}^{2} M_{2}}{M_{L 2}^{4}}\,\mu \tan\beta\,
F_{b}\left(\frac{M_{2}^{2}}{M_{L 2}^{2}},\frac{\mu^{2}}{M_{L 2}^{2}}\right),\\*
\amuBHmuL &= \frac{g_{1}^{2}}{16 \pi ^{2}}\frac{m_{\mu}^{2} M_{1}}{M_{L 2}^{4}}\,\mu \tan\beta\,
F_{b}\left(\frac{M_{1}^{2}}{M_{L 2}^{2}},\frac{\mu^{2}}{M_{L 2}^{2}}\right), \\* 
\amuBHmuR &= - \frac{g_{1}^{2}}{8 \pi ^{2}} \frac{m_{\mu}^{2} M_{1}}{M_{E 2}^{4}}\,\mu \tan\beta\, 
F_{b}\left(\frac{M_{1}^{2}}{M_{E 2}^{2}},\frac{\mu^{2}}{M_{E 2}^{2}}\right), \\* 
\amuBmuLmuR &= \frac{g_{1}^{2}}{8 \pi ^{2}} \frac{m_{\mu}^{2}}{M_{1}^{3}}\,\mu \tan\beta\, 
F_{b}\left(\frac{M_{L 2}^{2}}{M_{1}^{2}},\frac{M_{E 2}^{2}}{M_{1}^{2}}\right).
\end{align} 
\end{subequations}
The loop functions appearing here are defined as
\begin{subequations}
\begin{align}
F_{a}(x, y) &= -\frac{G_{3}(x) - G_{3}(y)}{x - y}, \\
F_{b}(x, y) &= -\frac{G_{4}(x) - G_{4}(y)}{x - y},
\end{align}
\end{subequations}
with
\begin{subequations}
\begin{align}
G_{3} (x) &= \frac{1}{2(x-1)^{3}}\Big[ (x-1)(x-3) + 2\log x\Big],\\
G_{4} (x) &= \frac{1}{2(x-1)^{3}}\Big[ (x-1)(x+1) - 2 x \log x \Big] .
\end{align}
\end{subequations}

In terms of these expressions, the
leading logarithmic approximation of the fermion/sfermion two-loop
contributions, $\amuFSfL$, is  given by
\begin{align}
\begin{split}
\amuFSfL =&\
\amuWHnu\,
\Big(\Deltagtwo + \DeltaYukHiggsino + \DeltaYukWinoHiggsino + \Deltatanbe + \constA\Big),\\*
&+\amuWHmuL\,
\Big(\Deltagtwo + \DeltaYukHiggsino + \DeltaYukWinoHiggsino + \Deltatanbe + \constB\Big),\\*
&+\amuBHmuL\,
\Big(\Deltagone + \DeltaYukHiggsino + \DeltaYukBinoHiggsino + \Deltatanbe + \constC\Big),\\*
&+\amuBHmuR\,
\Big(\Deltagone + \DeltaYukHiggsino + \DeltaYukBinoHiggsino + \Deltatanbe + \constD\Big),\\*
&+\amuBmuLmuR\,
\Big(\Deltagone + \Deltatanbe + \constE\Big).
\label{eq:twolooplogapprox}
\end{split}
\end{align}
The shifts $\Deltagone$, $\Deltagtwo$, $\DeltaYukHiggsino$, $\DeltaYukBinoHiggsino$, $\DeltaYukWinoHiggsino$ and $\Deltatanbe$ 
are defined as follows:
\begin{subequations}
\begin{align}
\begin{split}
  \Deltagone &=
  \begin{aligned}[t]
     \frac{g_{1}^{2}}{16 \pi ^2} \frac{4}{3}\biggl(
     & \frac{8}{3} \log\frac{M_{U }}{\logscale} + \frac{4}{3} \log\frac{M_{U 3}}{\logscale} + \frac{2}{3} \log\frac{M_{D }}{\logscale} + \frac{1}{3} \log\frac{M_{D 3}}{\logscale}\\
     &+ \frac{1}{3} \log\frac{M_{Q }}{\logscale} + \frac{1}{6} \log\frac{M_{Q 3}}{\logscale} + \log\frac{M_{E3}}{\logscale} + \frac{1}{2} \log\frac{M_{L 3}}{\logscale}\biggr),
   \end{aligned}
\end{split}\\
\Deltagtwo &= \frac{g_{2}^{2}}{16 \pi ^{2}} \frac{4}{3}
\left(3 \log\frac{M_{Q }}{\logscale} + \frac{3}{2}\log\frac{M_{Q 3}}{\logscale} 
+ \frac{1}{2}\log\frac{M_{L 3}}{\logscale}
\right),\\*
\begin{split}
  \DeltaYukHiggsino &=
  \begin{aligned}[t]
    \frac{1}{16 \pi ^2} \frac{1}{2}\biggl(
    & 3 y_{t}^{2} \log\frac{M_{U 3}}{\logscale} + 3 y_{b}^{2} \log\frac{M_{D 3}}{\logscale} + 3 (y_{t}^{2} + y_{b}^{2}) \log\frac{M_{Q 3}}{\logscale}\\
    &+ y_{\tau}^{2} \log\frac{M _{E 3}}{\logscale} + y_{\tau}^{2} \log\frac{M _{L 3}}{\logscale}\biggr),
  \end{aligned}
\end{split}\\*
\DeltaYukBinoHiggsino &= \frac{1}{16 \pi ^{2}} 
y_{t}^{2} \left(2 \log\frac{M_{Q 3}}{\logscale} - 8 \log\frac{M_{U 3}}{\logscale}\right), \\
\DeltaYukWinoHiggsino &= \frac{1}{16 \pi ^{2}} 
y_{t}^{2} \left(-6\log\frac{M_{Q 3}}{\logscale} \right),\\ 
\Deltatanbe &= \frac{1}{16 \pi ^{2}}(3 y_{b}^{2} - 3 y_{t}^{2} + y_{\tau}^{2})\log\frac{\muDR}{\logscale}, 
\end{align}
\end{subequations}
where $\logscale = \min (\lvert\mu\rvert, \lvert M_{1}\rvert, \lvert M_{2}\rvert, M_{L 2}, M_{E 2})$. 
The gauge and Yukawa coupling constants in the  coefficients are given
in Eq.~\eqref{DefGaugeYukawaCouplings}.

\begin{figure}
\subfiguretopcaptrue
\subfigure[]{\epsfxsize=0.32\textwidth\epsfbox{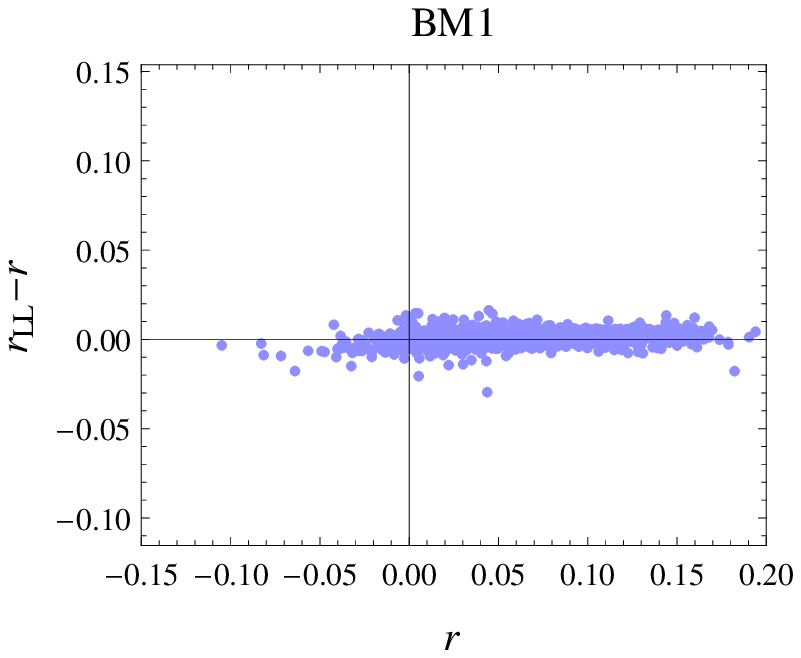}}
\subfigure[]{\epsfxsize=0.32\textwidth\epsfbox{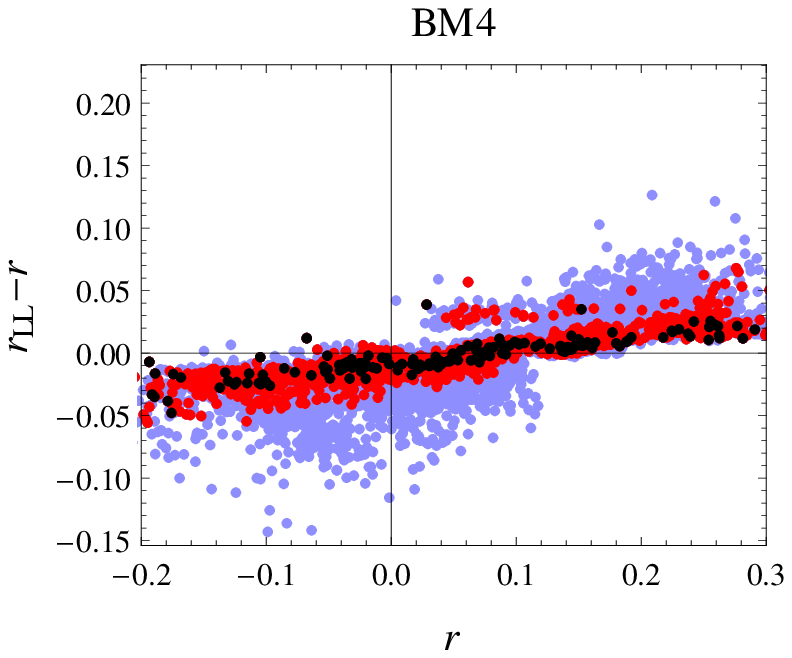}}
\subfigure[]{\epsfxsize=0.32\textwidth\epsfbox{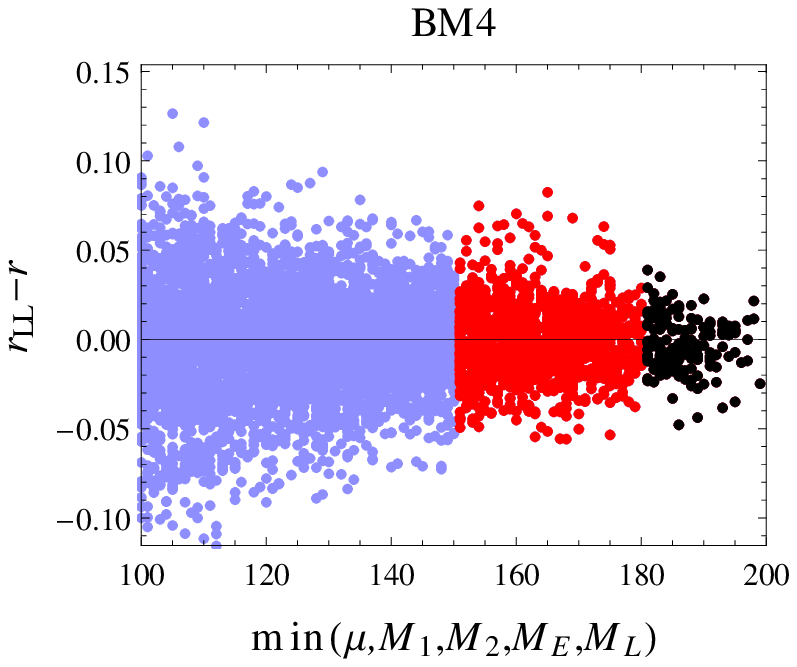}}
\vspace{-2ex}
\caption{\label{fig:logscatterplot}The exact result for 
\mbox{$r\equiv{\amuFSf}/{\amuSUOL}$}
compared with the approximation expressed as
\mbox{$r_{\rm LL}\equiv{\amuFSfL}/{\amuSUOL}$}. The mass parameters are chosen randomly 
around the benchmark points BM1 and BM4, with the ranges given in
Tab.~\ref{table:scatterregion}. 
The first and second figure show $r_{\rm LL}-r$ as a function of $r$, 
the third figure shows $r_{\rm LL}-r$ as a function of the minimum SUSY mass $\min(\lvert\mu\rvert, \lvert M_1\rvert, \lvert M_2\rvert, M_{E}, M_{L})$. 
The light blue points correspond to the same data points as the ones used for Fig.~5 of Ref.~\cite{fsf2loopA}, 
the red (black) points to a minimum SUSY mass bigger than $150\ (180)$~GeV.}
\end{figure}

Here, $\Deltagone$ and $\Deltagtwo$ are effective shifts to the gaugino couplings of the bino and wino, respectively. 
The logarithms of the inner sfermion masses appear weighted with the respective squared gauge couplings. 
$\DeltaYukHiggsino$ corresponds to the higgsino self energies and contains logarithms multiplied with squared Yukawa couplings. 
Here, the 1st and 2nd generation Yukawa couplings are neglected. 
$\DeltaYukBinoHiggsino$ and $\DeltaYukWinoHiggsino$ correspond to effective ${\tilde B}\text{--}{\tilde H}$ and
${\tilde W}\text{--}{\tilde H}$ transitions generated by fermion/sfermion loops. 
$\Deltatanbe$ arises from ${\overline{ \text{DR}}}$ renormalization of $\tan\beta$ and contains $\muDR$. 
The non-logarithmic numerical constants appearing in Eq.~\eqref{eq:twolooplogapprox}
approximate the typical magnitude of the additional non-logarithmic contributions. 
They have been obtained by fitting Eq.\ (\ref{eq:twolooplogapprox}) to
the exact result for the data points shown in
Fig.~\ref{fig:logscatterplot}. 
\pagebreak[4]

We briefly summarize the terms which are neglected by this approximation and state criteria when the approximation is expected to fail. 
\begin{itemize}
\item Already the one-loop approximation in Eqs.~\eqref{eq:onelooplogapprox} neglects
  terms with a relative suppression of the orders ${\cal O}(1/\tan\beta)$ and ${\cal O}(M_{Z}^{2}/M_{\text{SUSY}}^{2})$,
  where $M_{\text{SUSY}}$ denotes the relevant SUSY masses appearing at the one-loop level.
  Hence, the two-loop approximation becomes invalid if $\tan\beta$ or $M_{\text{SUSY}}$ become too small. 
\item The dependence on the inner sfermion masses beyond leading logarithms, in particular the behaviour
  for small inner sfermion masses and large mixing, is neglected. Furthermore,
  the dependence on all one-loop parameters $\mu, M_{1}, M_{2},
  M_{L},M_{E}$ and $\tan\beta$ (beyond the one-loop dependence) is neglected and replaced by the
  numerical constants in Eq.~\eqref{eq:twolooplogapprox}. 
\end{itemize}

\begin{table}
\centering
\begin{tabular}{l c c c c c c}
				& BM1	& BM2	& BM3	& BM4 \\
\hline\hline
$\amuSUOL[ 10^{-10}]$		& 44.02	& 26.95	& 46.78	& 15.98	\\
$\amuSUOLapprox[ 10^{-10}]$	& 44.69	& 27.25	& 47.41	& 17.59	\\
$r=\amuFSf/\amuSUOL$		& 0.041	& 0.045	& 0.047	& 0.049	\\
$r_{\rm LL}=\amuFSfL/\amuSUOL$	& 0.040	& 0.043	& 0.047	& 0.053 \\
\end{tabular}
\caption{\label{BMApprox} Comparison of exact results and corresponding approximation
formulas. The first and the third line are taken from \mbox{Ref.\ \cite{fsf2loopA}}.
}
\end{table}

%
\begin{table}
\centering
\begin{tabular}{l c c}
			& BM1		& BM4 \\
\hline\hline
$\mu [\text{GeV}]$	& $[100,200]$	& $[-200,-100]$ \\
$\tan\beta$		& $40$		& $50$ \\    
$M_1 [\text{GeV}]$	& $[100,200]$	& $[100,200]$ \\
$M_2 [\text{GeV}]$	& $[200,400]$	& $[1000,3000]$ \\
$M_{E} [\text{GeV}]$	& $[200,500]$	& $[100,300]$ \\
$M_{L} [\text{GeV}]$	& $[200,500]$	& $[1000,3000]$ \\
\end{tabular}
\caption{\label{table:scatterregion} Scan intervals for the least
  restrictive light blue parameter regions of Fig.~\ref{fig:logscatterplot}.}
\end{table}

We have verified that the approximation is in good agreement with
the results for the benchmark points BM1...BM4 from
Ref.~\cite{fsf2loopA}.\footnote{%
Some of the benchmark points involve equal mass parameters,
$M_{E}=M_{L}$, $M_2=M_{L}$. A direct evaluation of the approximation
formulas for these benchmark points
would suffer from the artificial singularities of the loop functions
$F_a(x,y)$ and $F_b(x,y)$  for $x=y$. These are avoided in the
numerical evaluation of the 
approximations in Tab.~\ref{BMApprox} by 
shifting the mass parameters of the benchmark points in a numerically
insignificant way, but such that all
mass parameters are different. Note that this problem is not present
in the exact result.
}
Tab.~\ref{BMApprox} shows a comparison of the results from Ref.~\cite{fsf2loopA}
(repeated in the first and third line) with the corresponding one-loop and leading log
approximations (second and fourth line).
Also, all entries of Tab.~3 of Ref.~\cite{fsf2loopA} are reproduced
well, except for the coefficient of $\log(M_{U 3})$ in the case of BM4. 
The reason is that BM4 has very small SUSY masses, so it is outside the region of approximation validity. 
Fig.~\ref{fig:logscatterplot} quantifies how well the approximation works. 
It compares the approximate results with the exact ones, normalized to the one-loop result,
for a random set of parameters.
The same data set as for Fig.~5 of Ref.~\cite{fsf2loopA} has been used,
see Tab.~\ref{table:scatterregion} for the one-loop masses;
the other eight sfermion-mass parameters appearing in the inner loop
are varied in the range $[10^3,10^6]$~GeV.

The scatter plot around BM1, Fig.~\ref{fig:logscatterplot}(a),
shows an almost perfect agreement between approximate and exact results. 
For almost all data points, the difference is at most $1\%$ of the one-loop result. 
Hence, the approximation represents a significant improvement compared to the
fit formula given in Ref.~\cite{fsf2loopA} with fixed coefficients of the logarithms.

For the scatter plot around BM4, Fig.~\ref{fig:logscatterplot}(b), the
improvement compared to the fit formula of Ref.~\cite{fsf2loopA} is only marginal. 
This is  due to the smallness of the SUSY masses in these data points. 
The blue/red/black points in Fig.~\ref{fig:logscatterplot}(b) are the points
for which the minimum SUSY mass is $\geq 100/150/180{\rm GeV}$, respectively. 
Fig.~\ref{fig:logscatterplot}(c) shows the same data points as a
function of the minimum SUSY mass. The figures confirm that the
approximation quickly improves as the ratio $M_{\text{SUSY}}/M_{Z}$
increases.
\newpage
\section{Numerical analysis}
\label{sec:numerics}


In this section the parameter dependence of the
fermion/sfermion-loop contributions to $a_\mu$ is analyzed in
detail. To begin with, we recall that the fifteen relevant parameters
can be devided into one-loop parameters, two-loop sfermion-mass
parameters for the inner loops, and the stop $A$-parameter, see
Sec.\ \ref{sec:inputparameters}:
\begin{center}
$\mu, M_1, M_2, M_{E}, M_{L}, \tan\beta,$
\\
$
M_U, 
M_D, M_Q, M_{U3}, M_{D3}, M_{Q3}, M_{E3}, M_{L3},
$
\\
$A_t.$
\end{center}
As stated above, the additional sensitivity on the sfermion masses of
all generations is a distinctive feature of the fermion/sfermion-loop
contributions. 

The dependence on all parameters is studied systematically,
starting with the region of large two-loop sfermion masses,
where the leading logarithmic approximation of Sec.~\ref{sec:leadinglog} is
valid. Then, the focus is set on smaller inner sfermion masses and the
influence of stop mixing.

We also briefly recall the benchmark points for the
one-loop parameters, introduced in Ref.~\cite{fsf2loopA} and
in Sec.\ \ref{sec:inputparameters}, Tab.~\ref{BMDefinition}.
They characterize qualitatively different regions of the one-loop
parameter space, in particular, BM1 is a point where all one-loop
masses are similar. BM3 is a point where the bino-exchange
contribution strongly dominates, and BM4 is a point where the
right-handed smuon contribution dominates.

\subsection{Parameter region of the leading logarithmic approximation}

The leading logarithmic behaviour has been studied extensively in
Ref.~\cite{fsf2loopA}, and it can be well understood from the
approximation $\amuFSfL$ in Sec.~\ref{sec:leadinglog}.
The approximation is expected to be valid if the hierarchy
$(\text{2-loop sfermion masses}) \gg (\text{1-loop masses}) \gg M_Z$
holds.
In practice it is already good for one-loop masses above around
200~GeV and two-loop masses around 1~TeV, as shown in
Fig.~\ref{fig:logscatterplot}.

The largest two-loop contributions can arise from the correction
factors $\DeltaYukWinoHiggsino$ and $\DeltaYukBinoHiggsino$, which
contain the logarithms of $M_{U3}$ and $M_{Q3}$ multiplied with the
top-Yukawa coupling and large prefactors. These large logarithms are
effective if the one-loop contribution is dominated by $\amuWHnu$ or
$\amuBHmuR$, as in BM1 or BM4. In these cases, the logarithms can
drive the two-loop corrections up to $15\%$ ($30\%$) of the one-loop
contributions for two-loop masses in the 20~TeV (1000~TeV) range.

On the other hand, in the case that $\amuBmuLmuR$ dominates at the one-loop level,
as in BM3, the two-loop corrections are smaller since the leading
logarithms are suppressed by the small gauge coupling $g_1^2$. In this
case, the two-loop corrections remain below $10\%$ of the one-loop
contributions for two-loop masses up to 1000~TeV. The parameter region
where $\amuBmuLmuR$ dominates has also been investigated in
Ref.~\cite{Endo:2013lva}. There, analytical results have been given for the
leading logarithm if not only sfermion masses but also the wino
and higgsino masses $M_2$ and $\mu$ are set to a common, very
large scale.

\subsection{Decomposition of contributions}
\label{sec:decompositioncontributions}

\begin{figure}
  \hspace{-10mm}
  \epsfxsize=0.62\textwidth\epsfbox{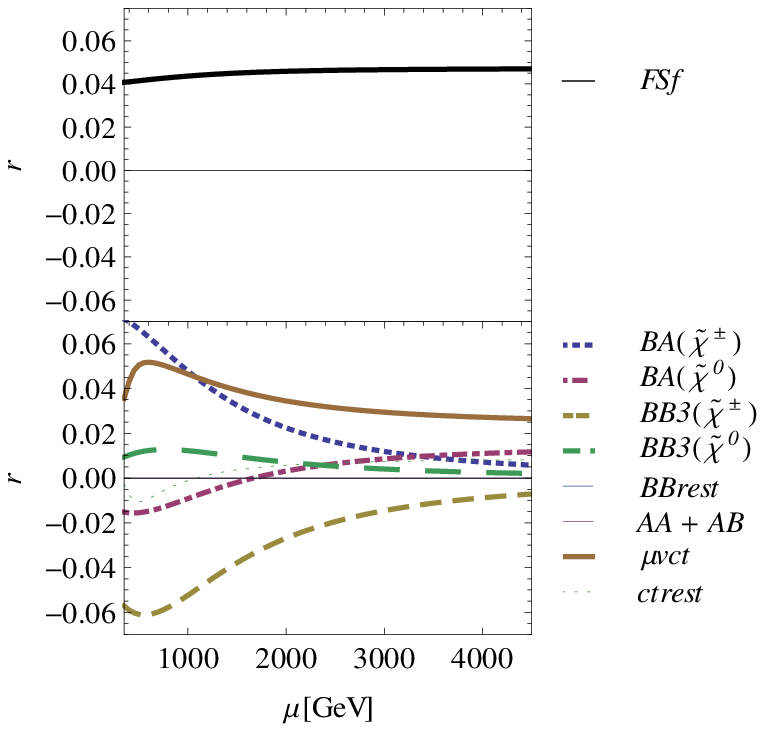}~
  \hspace{-20mm}
  \epsfxsize=0.62\textwidth\epsfbox{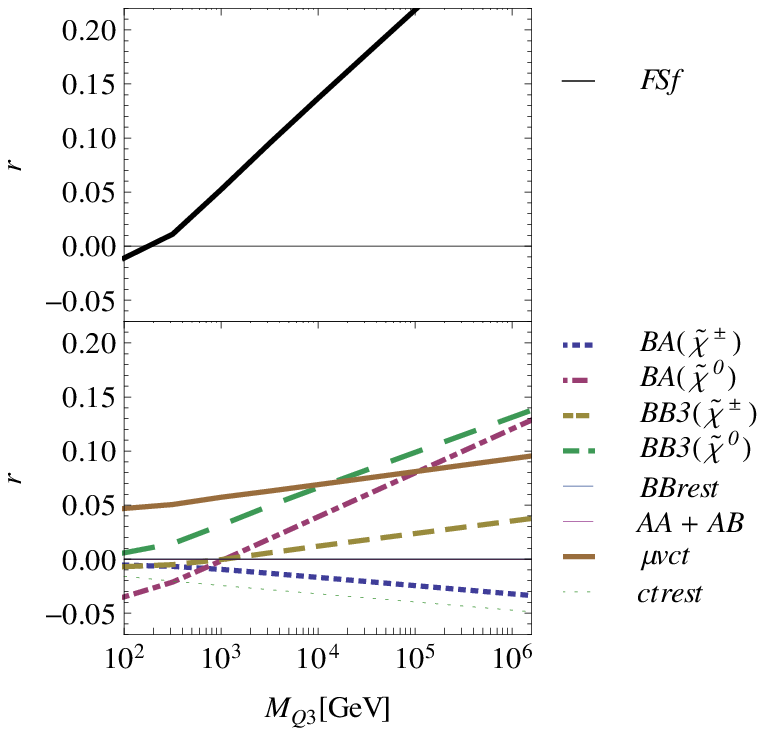}
  \caption{\label{fig:decompositioncontributionsBM}
    Full result $r\equiv\amuFSf/{\amuSUOL}$ and individual contributions as defined in
    Sec.~\ref{sec:decompositioncontributions},
    for two different scenarios. Left: one-loop parameters as in BM1, BM2, BM3
    except that $\mu$ is varied, and $M_{U,D,Q,U3,D3,Q3}=7\mbox{
      TeV}$, $M_{E3,L3}=3\mbox{ TeV}$. Right: one-loop parameters as
    in BM4, two-loop parameters as before, except that
    $M_{U3}=1\mbox{ TeV}$ and $M_{Q3}$ is varied.}
\end{figure}

To deepen the understanding of the fermion/sfermion-loop corrections,
we now show how the full result is decomposed into the individual
two-loop and counterterm contributions. The parameters are still
chosen such that the leading logarithmic approximation $\amuFSfL$ in
Sec.~\ref{sec:leadinglog} is valid. Fig.~\ref{fig:decompositioncontributionsBM}
shows two such parameter scenarios, considered already in Ref.~\cite{fsf2loopA}.
In both panels, the upper half shows the full result for the
fermion/sfermion-loop corrections, and the lower half shows the
following individual contributions:
\begin{itemize}
\item $BA(\tilde{\chi}^\pm), BA(\tilde{\chi}^0)$: all genuine chargino/neutralino two-loop
  contributions with coupling combination ${\cal BA}$, as given in
  Sec.~\ref{sec:twoloop}. These contributions are $\tan\beta$-enhanced due to the
  couplings ${\cal B}$ of the outer loop.
  They do not involve an explicit factor of the inner fermion mass.
\item $BB{\textsl 3}(\tilde{\chi}^\pm),BB{\textsl 3}(\tilde{\chi}^0)$: all genuine chargino/neutralino two-loop
  contributions with couplings ${\cal BB}$, but only from
  third generation fermion/sfermion pairs. These are $\tan\beta$-enhanced
  and proportional to the mass of the inner fermion, due to the coupling combination ${\cal B}$
  of the inner loop.
\item $BBrest$: all remaining contributions of the type ${\cal BB}$. These
  contributions are suppressed by a factor of the first or second generation
  fermion from the inner loop.
\item $AA+AB$: all genuine two-loop contributions involving the coupling
  combinations ${\cal AA}$ or ${\cal AB}$. These are not
  $\tan\beta$-enhanced and expected to be small.
\item $\mu vct$: all counterterm contributions from the external muon
  vertex, i.\,e.~of the classes~(\MuNV) and~(\MuCV). These counterterms
  are individually finite, contain $\Delta\rho$, and are the only
  contributions which involve pure fermion and pure sfermion loops.
\item $ctrest$: all remaining counterterm contributions.
\end{itemize}

The parameters of Fig.~\ref{fig:decompositioncontributionsBM} are chosen as
follows: In the left panel, the one-loop parameters are set as in BM1,
BM2, BM3 except that $\mu$ is varied. The sfermion masses are set to
 $M_{U,D,Q,U3,D3,Q3}=7\mbox{ TeV}$ and $M_{E3,L3}=3\mbox{ TeV}$ (see Tab.~2 of Ref.~\cite{fsf2loopA}).
In the right panel,  the one-loop parameters are
set as in BM4, the two-loop parameters as before, except that
$M_{U3}=1\mbox{ TeV}$ and $M_{Q3}$ is varied (see Fig.~6 of Ref.~\cite{fsf2loopA}). 

The full result in the left panel is always around 4\%, in agreement
with the result of Ref.~\cite{fsf2loopA}, and this result is well
described by the leading logarithmic approximation of Sec.~\ref{sec:leadinglog}.
Among the contributions listed above, only $BA$,
$BB{\textsl 3}$, and the counterterms are sizeable.  This is expected,
as discussed above.  For small values of $\mu$, many individual
contributions are large and there are strong cancellations.  For
larger $\mu$, in the BM2 and BM3 region, the $\amuBmuLmuR$ one-loop
contribution becomes dominant. All two-loop corrections are therefore
governed by the small gauge coupling $g_1^2$, and the individual
two-loop contributions become smaller.  For all $\mu$, the muon-vertex
counterterms $\mu vct$ almost account for the full result.

In the right figure the muon-vertex counterterms do not dominate the
full result. Since in the scenario of BM4 the bino--higgsino exchange
is most important, the neutralino contributions of the type
$BA(\tilde{\chi}^0)$ and $BB{\textsl 3}(\tilde{\chi}^0)$ are most important.
They rise logarithmically, while the chargino contributions are much
smaller and cancel each other to a large extent.

\subsection{Behaviour for small inner sfermion masses}

\begin{figure}
  \centering
  \epsfxsize=0.49\textwidth\epsfbox{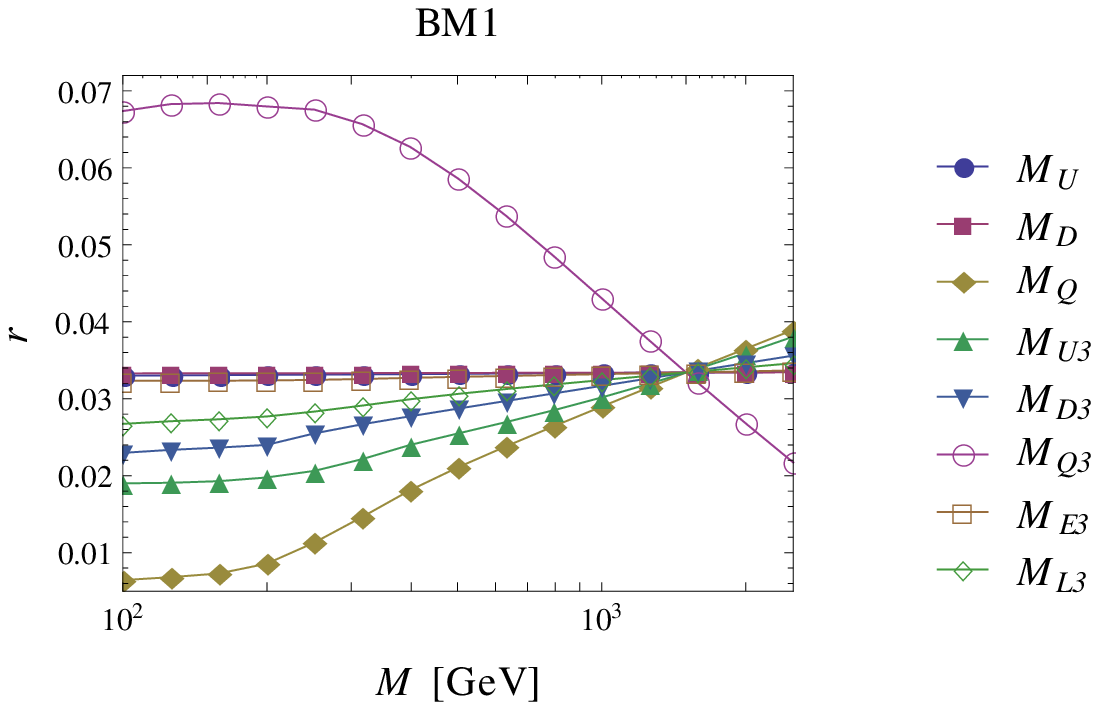}
  \epsfxsize=0.49\textwidth\epsfbox{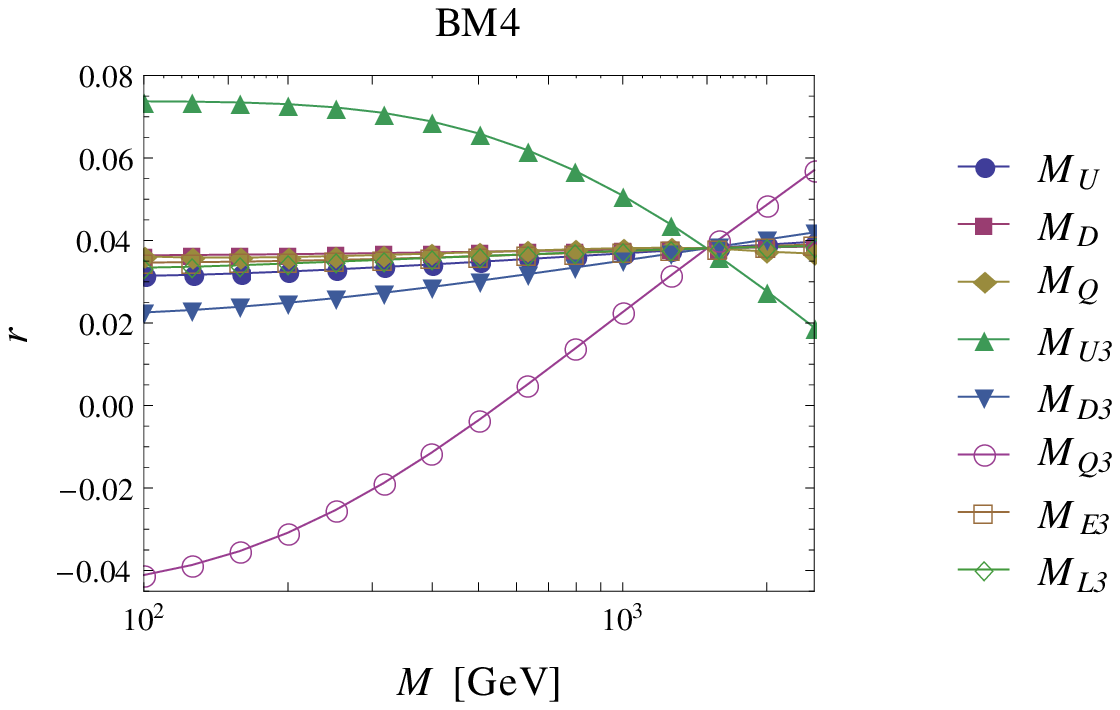}
  \caption{\label{fig:smallsfermionmasses}Relative correction
    $r\equiv\amuFSf/{\amuSUOL}$ from 
  fermion/sfermion loops
for the benchmark points BM1 and BM4 as a function of each sfermion
mass parameter. One sfermion mass is varied at a time; the remaining
sfermion masses are set to 1.5~TeV, and $A_t=0$.
}
\end{figure}

Fig.~\ref{fig:smallsfermionmasses} shows the general behaviour of
$\amuFSf$ if one of the inner sfermion masses is varied at a time
in the range $100\ldots2500$~GeV and can be viewed as an
extension of Fig.~4 of Ref.~\cite{fsf2loopA} to smaller sfermion masses.
Since the approximation of Sec.~\ref{sec:leadinglog} cannot be expected to
be valid over the whole mass range, Fig.~\ref{fig:smallsfermionmasses} quantifies
when and to what extent the exact dependence on the inner sfermion
masses differs from a purely logarithmic one.
In the left (right) panel of Fig.~\ref{fig:smallsfermionmasses} the
one-loop parameters are set to the values of benchmark point BM1 (BM4).
The inner sfermion masses are set to $1.5~\text{TeV}$ as a standard value and
all $A$-parameters are set to zero.

Both plots show the familiar logarithmic dependence on the inner
sfermion masses, as long as the masses remain sufficiently large.
For inner sfermion masses below 500~GeV the 
contributions saturate, and the difference between the leading logarithmic
approximation and the exact result can be quite sizeable. 

The mass scale of 500~GeV can be compared with the typical mass scale
of the one-loop parameters in BM1 and BM4, which is 300~GeV and 200~GeV,
respectively. Generally, therefore, the leading logarithmic
approximation can be expected to work well only as long as the inner
sfermion masses are at least twice as large as the one-loop masses in
the outer loop. For smaller inner sfermion masses, the exact result
can be expected to be smaller than the approximated one.

\subsection{Dependence on stop mixing}

\begin{figure}
  \centering
  \epsfxsize=0.49\textwidth\epsfbox{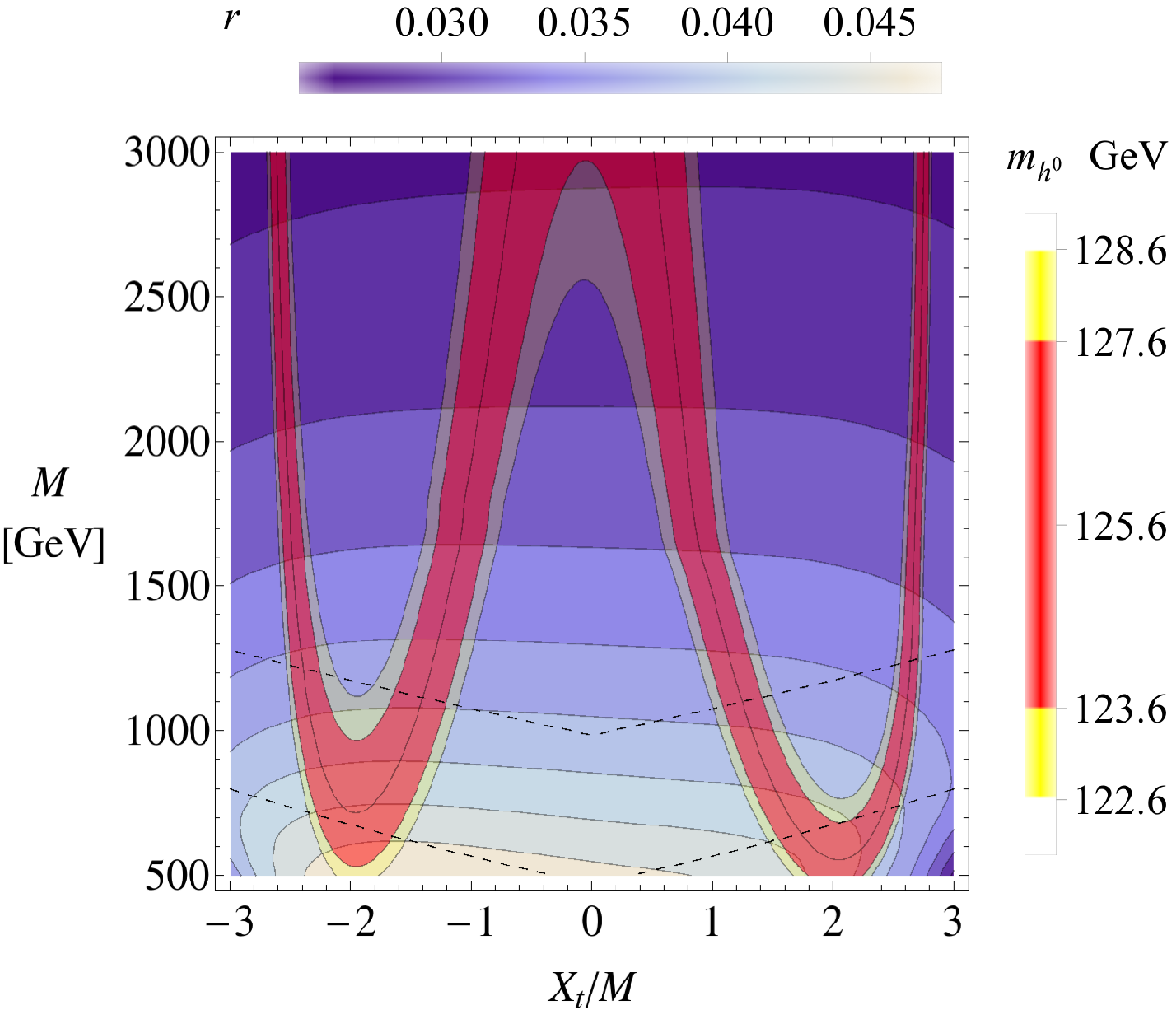}\hfill
  \epsfxsize=0.49\textwidth\epsfbox{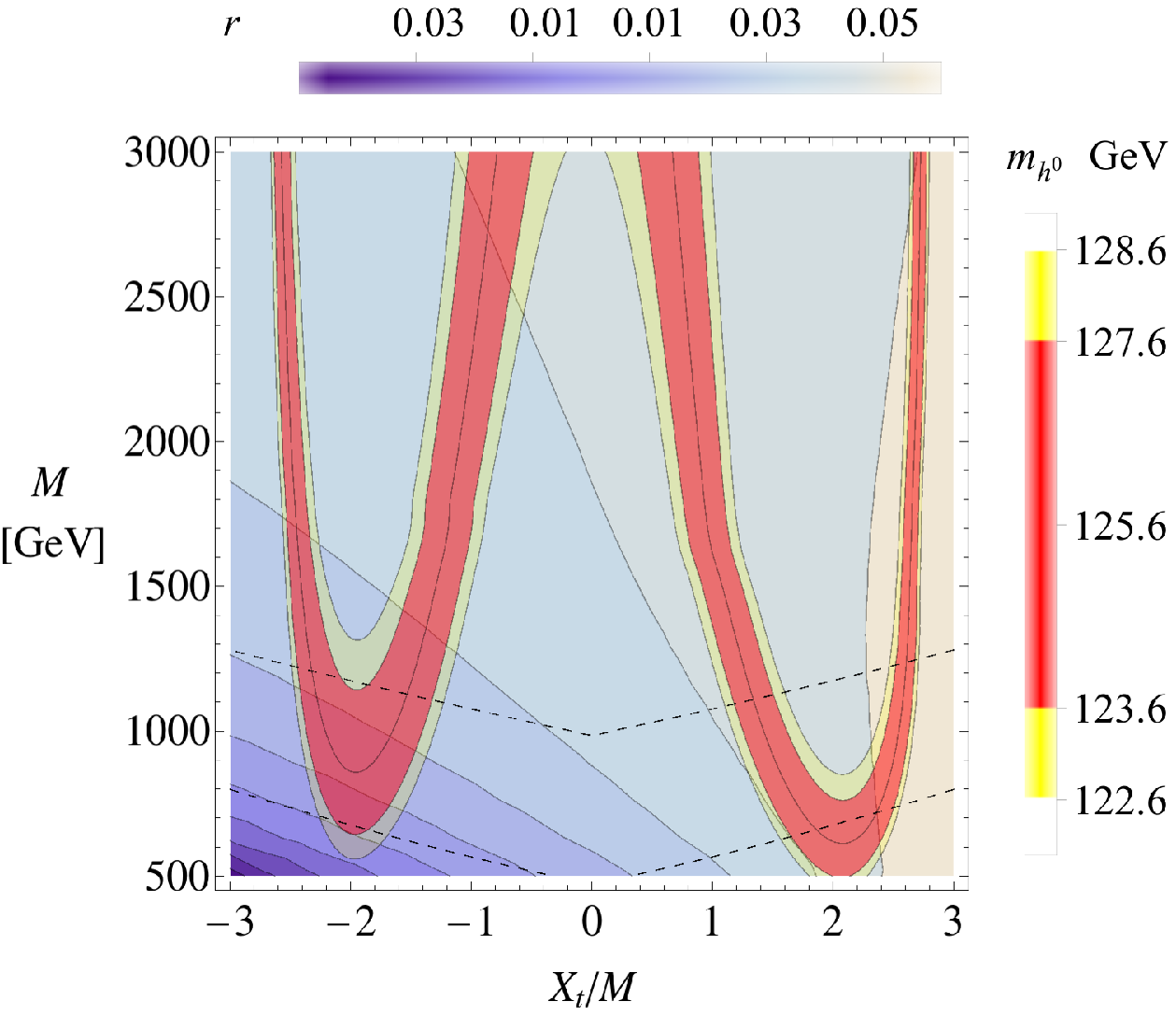}
  \caption{\label{fig:Xt}The dependence of $r\equiv\amuFSf/\amuSUOL$ on $M
    \equiv M_{Q3} = M_{U3} = M_{D3}$ and the ratio $X_{t}/M$. All
    other squark and the third generation slepton masses are set to
    $1.5~\text{TeV}$. The one-loop parameters are set to the values of
    BM1 in the left and of BM4 in the right panel. The red and
    yellow bands indicate the mass of the lightest Higgs boson. The
    dashed lines depict the thresholds for a lighter stop mass of
    $500~\text{GeV}$ or $1~\text{TeV}$.}  
\end{figure}

The discovery of a Higgs-like particle at the LHC constrains the
allowed parameter space of the MSSM. If stop
mixing is not allowed, i.\,e.~$A_{t} = 0$, its rather high mass can be 
accomodated only by very heavy stops with a mass of several~TeV. In
contrast, a non-zero mixing allows for stop masses below
the TeV scale. 

So far, all trilinear $A$-parameters were set to zero in the
discussion of $\amuFSf$. Now, the influence of stop mixing, induced by
$A_t\ne0$, is studied. $A_t$ and the associated stop-mixing parameter
$X_t = (A_t-\mu^*/t_{\beta})$ enter the calculation of $a_{\mu}$ at
the two-loop level through the mixing matrices of the stops in the
inner loop.

Fig.~\ref{fig:Xt} compares the influence of a normalized
$X_t$ on both $\amuFSf$ and the Higgs-boson mass. It shows contour
plots in the plane of the universal SUSY breaking parameter
\mbox{$M\equiv M_{Q3}=M_{U3}=M_{D3}$} and $X_t/M$. Contours are drawn for both
$\amuFSf/\amuSUOL$ and $m_{h^0}$; the latter is computed using
FeynHiggs~\cite{HiggsDRbar,FH2,FH3,FH4}.
The dashed lines at the bottom depict the thresholds where the mass of the lighter
stop falls below $500~\text{GeV}$ and $1~\text{TeV}$, respectively.
The input parameters besides $M$ and $X_t$ in Fig.~\ref{fig:Xt} are
chosen as \mbox{$M_{Q} = M_{U} = M_{D} = M_{E3} = M_{L3} = 1.5~\text{TeV}$};
the one-loop parameters are chosen as in benchmark points BM1 (left) and BM4 (right).

For the Higgs-boson mass we find the well-known dependence on $X_t/M$ as
an approximate fourth order polynomial. In particular the dependence is
approximately symmetric for $X_t\leftrightarrow-X_t$. For $\amuFSf$,
however, the dependence is approximately linear. In the case of BM4,
$X_t/M$ has a pronounced influence. Moving from \mbox{$X_t/M = -2$} to
\mbox{$X_t/M = +2$} changes the $a_\mu$ correction from $1\%$ to $5\%$ at
$M = 750~$GeV, and by an even larger amount for smaller
$M$. In the case of BM1, however, the influence of $X_t$
is tiny.

\begin{figure}
  \hspace{-10mm}
  \epsfxsize=0.62\textwidth\epsfbox{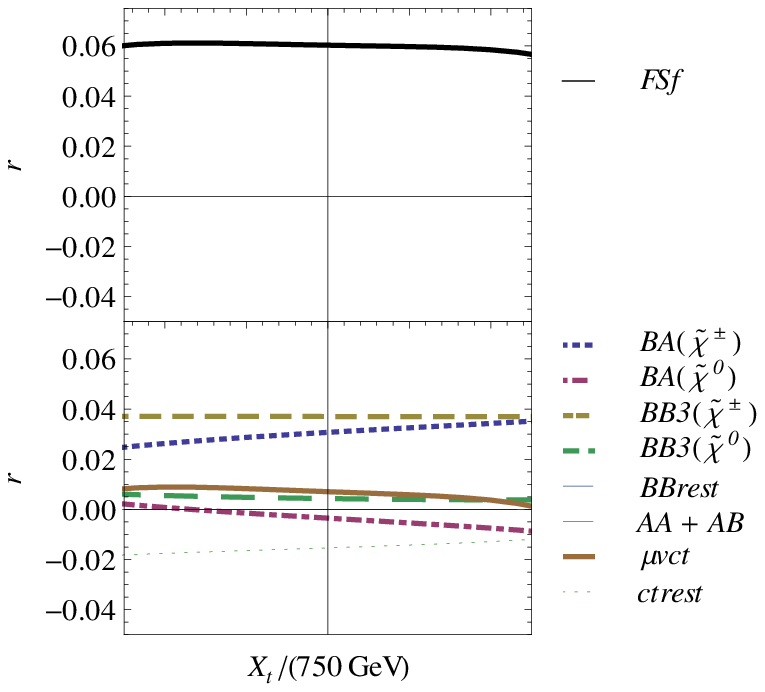}~
  \hspace{-20mm}
  \epsfxsize=0.62\textwidth\epsfbox{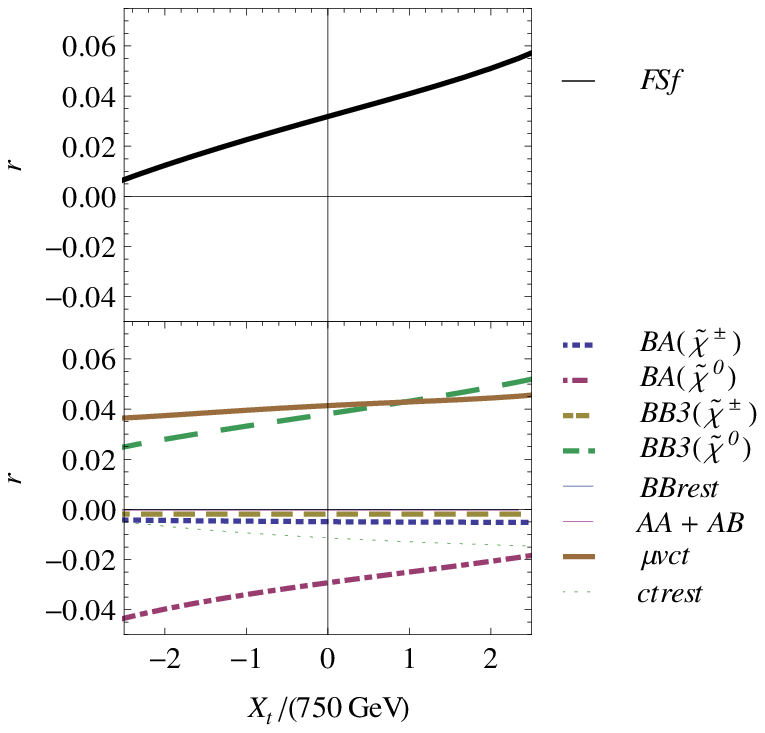}
\caption{\label{fig:decompositioncontributionsXt}
    Full result $r\equiv\amuFSf/{\amuSUOL}$ and individual contributions
    as defined in Sec.~\ref{sec:decompositioncontributions}, for the
    scenarios of Fig.~\ref{fig:Xt}, but with fixed $M=750$~GeV. }
\end{figure}

Again, a deeper understanding of this behaviour can be obtained by
considering the decomposition of contributions introduced in Sec.~\ref{sec:decompositioncontributions}.
Fig.~\ref{fig:decompositioncontributionsXt} shows plots corresponding
to the parameter choices of Fig.~\ref{fig:Xt}, but at fixed
$M=750$~GeV. The style is as in Fig.~\ref{fig:decompositioncontributionsBM}.
In the left plot, based on BM1, wino--higgsino mass-insertion diagrams
dominate. As a result, the $BA$-contributions of charginos and
neutralinos (from stop/bottom and stop/top loops generating a
wino--higgsino transition) have the strongest dependence on the mixing parameter.
Accidentally these contributions cancel out to a large extent,
and the full result is almost insensitive
to the choice of $X_t$. In the right plot, based on BM4 with
bino--higgsino dominance,  only the $BA$ and $BB{\textsl 3}$
contributions of the neutralinos are important. Since they have
positive slopes and add up constructively
the full result is very sensitive to the mixing parameter.

These considerations also show that the behaviour found for BM1 and
BM4 is typical for the behaviour in the larger parameter regions
represented by these benchmark points. 

\subsection{Particular scenarios with extremely small SUSY masses}

In the previous sections the behaviour of $\amuFSf$ has been studied in
a quite generic way. Now, special parameter scenarios are considered
in which particular SUSY masses can be very small, without violating
experimental bounds. We focus on the following three cases:
\begin{itemize}
\item light stop and large stop-mass splitting, in the
  scenario of Ref.~\cite{Delgado}:
  This scenario fixes the stop sector in a particular way, such that
  the lighter stop mass is as small as the one-loop masses; we study
  $\amuFSf$ for several choices of the one-loop parameters. 
\item light stau scenario of Ref.~\cite{HeinemeyerBM}: This scenario essentially
  fixes the two-loop parameters for $\amuFSf$ as well as most
  one-loop parameters, such that both staus can be lighter than the
  one-loop masses.
\item light smuon, chargino and neutralino masses and
  extremely small $\tan\beta$, in the scenario of Ref.~\cite{Batell:2013bka}:
  This scenario fixes all one-loop parameters, such that none of the
  usual hierarchies ($\tan\beta\gg 1$, $M_{\rm SUSY}\gg M_Z$) is
  valid; we study the dependence on the two-loop parameters.
\end{itemize}

\paragraph{Light stop scenario:}
\begin{figure}
  \centering
  \epsfxsize=.62\textwidth\epsfbox{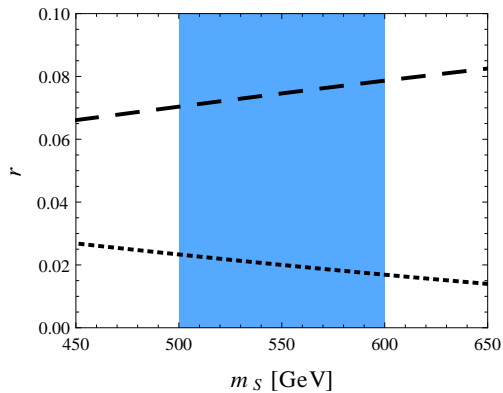}
  \caption{\label{fig:lightstopscenario}
   Full result $r\equiv\amuFSf/{\amuSUOL}$ for the scenario of
   Ref.~\cite{Delgado}, \mbox{$M_{U3}=200$~GeV},
   \mbox{$A^2_t=6 M_{U3} M_{Q3}$},
   as a function of $m_S=(m_{\tilde{t}_1} m_{\tilde{t}_2})^{1/2}$.
   The one-loop parameters are defined for slightly modified benchmark points
   BM1 (dotted) and BM4 (dashed), respectively.}
\end{figure}

In Ref.~\cite{Delgado}, arguments are put forward in favour of a very light
right-handed stop with almost degenerate neutralino, together with
large $A_t$ and a heavier left-handed stop. This parameter choice is
of interest for $\amuFSf$ since a large left/right stop-mass splitting
can lead to particularly large logarithmic corrections. However, the
scenario of Ref.~\cite{Delgado} differs from the scenarios considered
so far in the present paper or in Ref.~\cite{fsf2loopA}, because it
combines large stop-mass splitting with large stop mixing, and because
one stop is so light that the leading logarithmic approximation cannot
be expected to be valid.

Fig.~\ref{fig:lightstopscenario} shows the result for
$r\equiv\amuFSf/{\amuSUOL}$ if the stop and neutralino parameters are chosen
according to this scenario. According to Ref.~\cite{Delgado} the
right-handed stop mass $M_{U3}$ is set to $200$~GeV, and $M_{Q3}$ is
varied such that the quantity
$m_S=(m_{\tilde{t}_1} m_{\tilde{t}_2})^{1/2}$
is in the range $500\dots600$~GeV, depicted by
the blue shaded area in Fig.~\ref{fig:lightstopscenario}. The
trilinear mixing parameter is always set to
$A^2_t=6 M_{U3} M_{Q3} \approx 6 m^2_S$.
All remaining two-loop mass parameters equal $1.5$~TeV.  The one-loop
parameters are not fixed by the scenario of Ref.~\cite{Delgado},
except for the requirement that the lightest neutralino is
$30\dots40$~GeV lighter than the lightest stop. Hence we set the
one-loop parameters to the values of either BM1 or BM4, with the
modifications $M_1=178$~GeV (BM1-like scenario), $M_1=190$~GeV and
$\mu=-220$~GeV (BM4-like scenario).

The results can be understood by comparing with
Fig.~\ref{fig:smallsfermionmasses}, where $A_t=0$.  There, varying
down the 
value of $M_{U3}$ from $1.5$~TeV to $200$~GeV for BM1 leads to a
relative reduction of $r\equiv\amuFSf/{\amuSUOL}$ from $\approx 3\%$
to $\approx 2\%$ for BM1.  In the BM4-like scenario the reduction of
$M_{U3}$ has an even larger impact on the final result and $r$
increases from $\approx 4\%$ up to $\approx 7\%$. The large $A_t$
present in Fig.~\ref{fig:lightstopscenario} and the slightly
different one-loop parameters do not significantly modify the result.

\paragraph{Light stau scenario:}

In Ref.~\cite{HeinemeyerBM}, benchmark scenarios for the MSSM Higgs
sector are proposed which are in agreement with the most recent
experimental data. One of them contains a very light stau. This
scenario is of interest for the evaluation of $\amuFSf$ since it
constitutes an example where the sfermion mass in the inner loop can
be lighter than the one-loop masses in the outer loop, i.\,e.~the
opposite of the situation in which the leading logarithmic
approximation is valid.

According to Ref.~\cite{HeinemeyerBM} we consider the following
parameter choice:
\begin{align}
\begin{alignedat}{2}
  \tan\beta&=25,\quad \mu=500~\text{GeV},& M_2 &= 200~\text{GeV},\quad M_1=95.61~\text{GeV}, \\
  M_{L}&=M_{E}=400~\text{GeV},& M_{L3} &= M_{E3}=245~\text{GeV}, \\
  M_{Q3}&=M_{U3}=M_{D3}=1~\text{TeV},& A_t &= A_b=1620~\text{GeV}.
\end{alignedat}
\end{align}

All other sfermion mass parameters are set to $1.5$~TeV. This leads to the
following results for $\amuSUOL$ and $r\equiv\amuFSf/{\amuSUOL}$:
\begin{align}
\amuSUOL&=26.25\times10^{-10}, \\
r&=0.0289.\label{lightstauresult}
\end{align}
The input parameters are similar to BM1, so the result can be compared
to Fig.\ \ref{fig:smallsfermionmasses}. In that Figure, the BM1 result for
$r$ with very small stau masses is slightly below $0.03$, like in
Eq.\ (\ref{lightstauresult}).

\paragraph{Light one-loop masses and small $\text{tan}\beta$:}

\begin{figure}
  \centering
  \epsfxsize=.75\textwidth\epsfbox{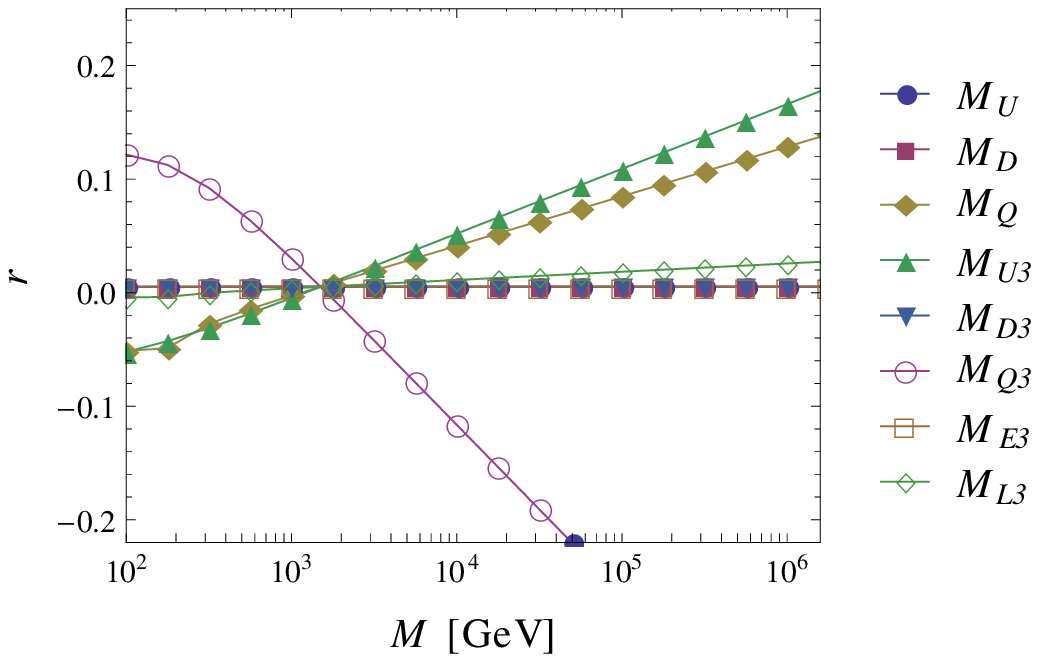}
  \caption{\label{fig:Wagnerscenario}
    Relative correction $r\equiv\amuFSf/{\amuSUOL}$ as in
    Fig.~\ref{fig:smallsfermionmasses},
    except that the one-loop parameters are set to the scenario of
    Ref.~\cite{Batell:2013bka} with light one-loop masses,
    see Eq.~\eqref{BMWagner}.}
\end{figure}

In Ref.~\cite{Batell:2013bka} a scenario is considered where a
chargino, a neutralino, a slepton and a sneutrino are all lighter than
the $Z$ boson, and it is demonstrated that such a scenario cannot be
ruled out by current experimental data. In that scenario $\tan\beta$
takes the very small value of $\tan\beta=1.5$. For $a_\mu$ this scenario
is interesting since all usual approximations are invalid: $\tan\beta$ is
small, and the usually $\tan\beta$-suppressed terms become important;
likewise, the approximation $\text{(1-loop masses)}\gg M_Z$ fails.

We consider the following parameter choice:
\begin{equation}
\begin{aligned}
\label{BMWagner}
\tan\beta&=1.5,&
\mu&=149~\text{GeV},&
M_2&=160~\text{GeV},&
M_1&=1~\text{TeV},\\
&&M_{L}&=76~\text{GeV},&
M_{E}&=1~\text{TeV},
\end{aligned}
\end{equation}
which is similar to the choice made in Ref.~\cite{Batell:2013bka} but
avoids the singularity for $\mu=M_2$ in the chargino-mass
renormalization constants. This choice fixes all one-loop parameters
and thus $\amuSUOL$. The small masses, together with the small
$\tan\beta$, lead to an interesting value in the ballpark of the
deviation~\eqref{eq:deviation},
\begin{align}
\amuSUOL&=16.29\times10^{-10}.
\end{align}

The two-loop corrections can be large, too, as shown in Fig.~\ref{fig:Wagnerscenario}.
Like Fig.~\ref{fig:smallsfermionmasses} it
shows the two-loop corrections if one of the inner sfermion masses is varied at
a time, while all others remain at the standard value of 1.5~TeV.
If all inner sfermion masses are 1.5~TeV, the two-loop corrections
accidentally cancel. But whenever either $M_{Q3}$, $M_{U3}$, or
$M_{Q}$ is varied away from 1.5~TeV, large corrections arise. 

The pattern of the individual slopes in Fig.~\ref{fig:Wagnerscenario}
is similar to the one in Fig.~\ref{fig:smallsfermionmasses} for BM1,
but the slopes are much larger. They cannot be predicted by the
approximation of Sec.~\ref{sec:leadinglog} for the reasons mentioned
above. Positive and negative corrections of around 10\% are possible if
the inner sfermion masses are in the sub-TeV or few-TeV region. For
smaller inner sfermion masses, there is a slight saturation effect,
but less pronounced compared to Fig.~\ref{fig:smallsfermionmasses}. 
\begin{figure}
  \hspace{-10mm}
  \epsfxsize=0.62\textwidth\epsfbox{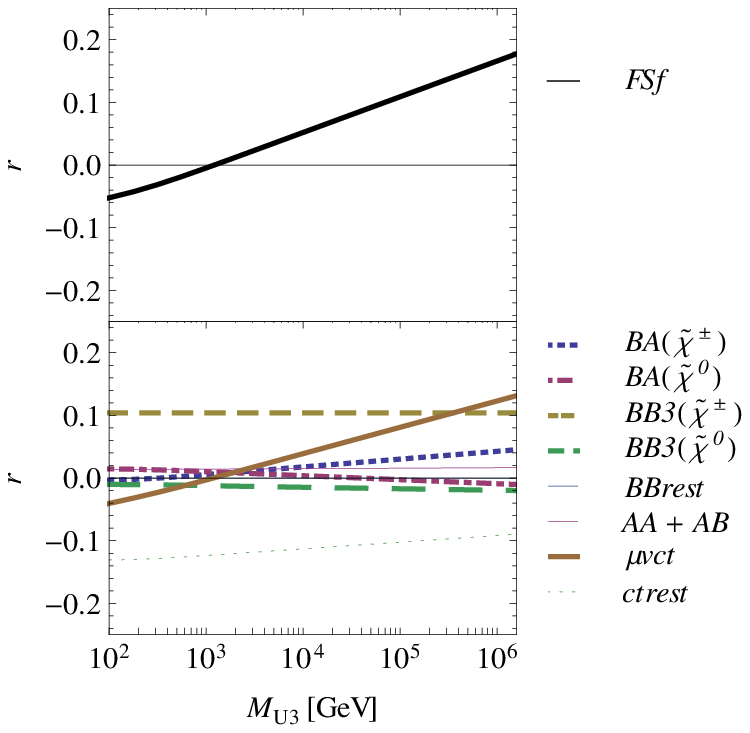}\hfill
  \hspace{-20mm}
  \epsfxsize=0.62\textwidth\epsfbox{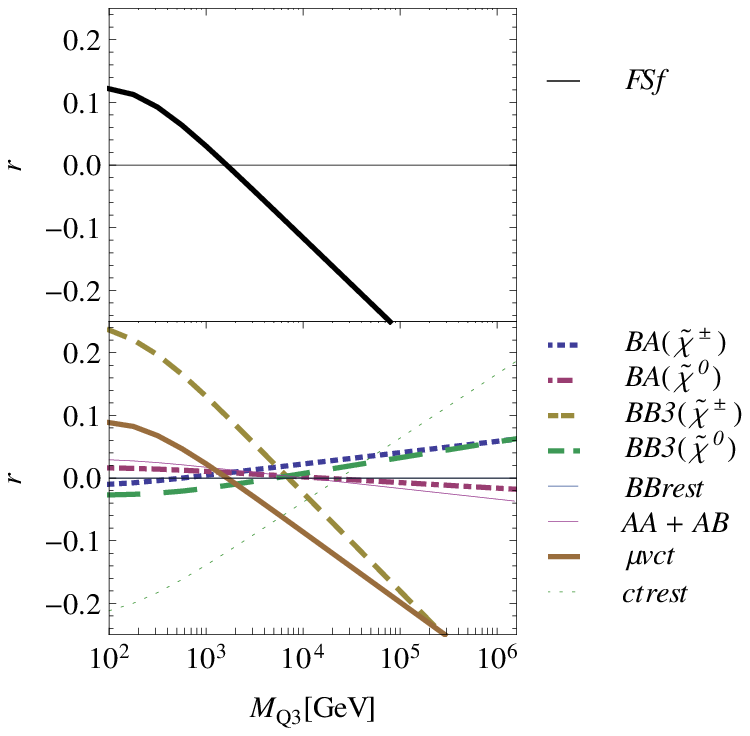}
\caption{\label{fig:decompositioncontributionsWagner}
    Full result $r\equiv\amuFSf/{\amuSUOL}$ and individual
    contributions as defined in Sec.~\ref{sec:decompositioncontributions}, for the
    scenario of Ref.~\cite{Batell:2013bka} and Fig.~\ref{fig:Wagnerscenario}. The one-loop parameters 
    are chosen as in Eq.~\eqref{BMWagner}.
    Left: $M_{U3}$ line of Fig.~\ref{fig:Wagnerscenario},
    Right: $M_{Q3}$ line of Fig.~\ref{fig:Wagnerscenario},}
\end{figure}

Because of the very peculiar nature of the parameter scenario
Eq.~\eqref{BMWagner}, it is instructive to consider again the
decomposition of the contributions, as in
Sec.~\ref{sec:decompositioncontributions}.
Fig.~\ref{fig:decompositioncontributionsWagner} shows two
corresponding plots; the left one corresponds to the $M_{U3}$ line,
the right one to the $M_{Q3}$ line of
Fig.~\ref{fig:Wagnerscenario}. Several features are noteworthy. First,
in contrast to the previous cases, only
$BB{\textsl 3}(\tilde{\chi}^-)$, $\mu vct$ and $ctrest$ are sizeable;
all other contributions are very
small. This observation is particularly interesting as $\tan\beta$ is
very small, and the $\tan\beta$-suppressed \mbox{$AA+AB$} contributions could
have been expected to play a more important role. Furthermore, there
is always a strong cancellation between $ctrest$ and the other
contributions.  In the left plot the $M_{U3}$ dependence is dictated
by the muon-vertex counterterms.
All other contributions are almost insensitive to $M_{U3}$.  In the
right plot the situation is slighly different, and the full $M_{Q3}$
dependence is governed by a combination of the
$BB{\textsl 3}(\tilde{\chi}^-)$, $\mu vct$ and $ctrest$ contributions.
\newpage
\section{Conclusions}
\label{sec:conclusions}

The calculation of the fermion/sfermion-loop contributions extends the
known results of the MSSM two-loop contributions to~$a_{\mu}$ in several
important ways.
\begin{itemize}
\item It is an exact evaluation of the fermion/sfermion-loop
corrections to MSSM one-loop diagrams and all associated
counterterm diagrams. The two-loop diagrams contain the maximum number
of different mass scales possible in the MSSM.
\item It introduces a dependence of~$a_{\mu}$ on the
squarks and sleptons of all generations. As shown already in
Ref.~\cite{fsf2loopA} this sensitivity is strong: The contributions
are logarithmically enhanced 
by heavy sfermions in the inner loop, and they can be the largest SUSY two-loop
contributions. 
\item It eliminates the ambiguity from parametrizing the one-loop
  contributions either in terms of $\alpha$, $\alpha(M_Z)$, or
  $G_{\text{F}}$. This is due to the fact that the counterterms contain in
  particular the leading contributions to the large quantities
  $\Delta\alpha$ and $\Delta\rho$ from light and heavy SM quarks and
  leptons.
\end{itemize}
 

In the present paper, full details of the calculation and analytical
results, as well as a complete survey of the numerical behaviour have
been given.

A very compact approximation formula was provided in Sec.~\ref{sec:leadinglog}.
It can be easily implemented in any code for
numerical evaluation, and it is available as Mathematica code.\footnote{The Mathematica implementation 
  of the approximation formula of Sec.~\ref{sec:leadinglog} can be
  obtained from 
  \url{http://iktp.tu-dresden.de/?id=theory-software}.} This compact
formula is based on the leading logarithms and numerical constants
approximating the non-logarithmic terms. It is a good approximation in
large regions of the parameter space, and it also provides qualitative 
understanding of the parameter dependence of the result.

The largest possible two-loop corrections can be obtained from loops
involving stops or sbottoms due to their large Yukawa couplings; these
Yukawa-enhanced corrections can be positive or negative, depending on
the hierarchy between left- and right-handed stop masses. Large
two-loop corrections can also arise from 1st/2nd generation squarks
due to their large multiplicity and SU(2) gauge coupling. Generally,
for inner sfermion masses in the sub-TeV or few-TeV range, the
two-loop corrections can be around 10\% of the SUSY
one-loop contributions to $a_\mu$. 

Even if certain or all relevant SUSY masses become small, such that
the leading logarithmic approximation fails, the corrections can be
sizeable. We have considered three examples of experimentally allowed
scenarios with extremely  light stop, light stau, or light slepton and
chargino masses. In all cases the total SUSY contribution to $a_\mu$ can be in
the ballpark of the deviation~\eqref{eq:deviation}, and the two-loop
corrections are up to $10\%$.

The computation of the fermion/sfermion-loop correction represents an
important step towards the full two-loop calculation of $a_\mu$ in the
MSSM. On a technical level, the diagrams involve 2 light and up to 5
different heavy mass scales. This is a higher number than for all
previously considered $a_\mu$ two-loop corrections, and it is the
maximum number possible in the MSSM. The standard two-loop techniques
based on integration by parts lead to very cumbersome expressions. Our
second, alternative calculation based on an iterated one-loop
calculation leads to elegant final analytical results in terms of a
one-dimensional Feynman parameter integration.


It is interesting to compare the fermion/sfermion-loop corrections to
other SUSY two-loop contributions to the muon magnetic
moment. Up to now, two classes of corrections to SUSY one-loop diagrams
were known. The first are photonic, or QED corrections~\mbox{\cite{amuPhotonicSUSY,DG98}},
which are dominated by QED logarithms and
amount to around $(-7\ldots-9)\%$ in typical parameter regions. The
second is a universal $(\tan\beta)^2$-enhanced correction arising from
a shift of the muon Yukawa coupling~\cite{Marchetti:2008hw}. In
large regions of the MSSM parameter space, particularly for approximately
degenerate SUSY masses, the $(\tan\beta)^2$-corrections are positive
(for positive $\amuSUOL$) and can partially or fully
compensate the photonic corrections for large $\tan\beta$. 
Further, two-loop corrections to SM one-loop diagrams from SUSY
particle loops have been fully evaluated in Ref.~\cite{HSW03,HSW04},
and they amount to around $2\%$ of $\amuSUOL$ for
degenerate masses. However, these corrections decouple for heavy SUSY
particles.

Hence, the fermion/sfermion-loop corrections can be as large as any of
the previously known corrections. For all these corrections either the
exact result or a useful approximation formula can be easily
implemented. Numerical comparisons between all these known two-loop
results can be found in Ref.~\cite{fsf2loopA}.

The remaining MSSM two-loop corrections to $a_\mu$ comprise SUSY
one-loop diagrams with a second loop with gauge or Higgs boson or
neutralino/chargino exchange. These remaining corrections depend on a
subset of parameters of the fermion/sfermion-loop corrections, hence
their parameter dependence will be more straightforward. Nevertheless,
their evaluation will be important to reduce the theory error of the
SUSY prediction of $a_\mu$ below the experimental uncertainty of the
future $a_\mu$ experiments.

\section*{Acknowledgments} 
We acknowledge financial support by the German Research Foundation DFG through
Grant No. STO876/1-1, by DAAD and by CNPq.
HF thanks TU Dresden and IKTP for their hospitality.

\begin{appendix}
\section{Loop functions for one-loop diagrams\label{app:loopf}}
 The one-loop functions with a single mass ratio read
\begin{align} 
\mathcal{F}^{C}_{i}(x) & = F^{C}_{i}(x)[1 - \epsilon \,
  L(m_{\tilde{\nu}_{\mu}}^2)]  + 
\epsilon \, F^{C}_{i \epsilon}(x) ,\\ 
\mathcal{F}^{N}_{i}(x) & = F^{N}_{i}(x)[1 - \epsilon \,
  L(m_{\tilde{\mu}}^2)]  + 
\epsilon \, F^{N}_{i \epsilon}(x) ,
\end{align}
where we have used the abbreviation
\begin{align}
L(m^2) &= \log\frac{m^2}{\muDR^2}
\label{LDef}
\end{align}
with the dimensional-regularization scale $\muDR$,
and the well-known functions
\begin{align}
F_1^C(x) &= \frac{ 2 }{(1-x)^4 }\Big[
2+3x-6x^2+x^3+6x\log x
\Big],\\
F_2^C(x) &= \frac{ 3 }{2(1-x)^3 }\Big[
-3+4x-x^2-2\log x
\Big],\\
F_1^N(x) &= \frac{ 2 }{(1-x)^4 }\Big[
1-6x+3x^2+2x^3-6x^2\log x
\Big],\\
F_2^N(x) &= \frac{ 3 }{(1-x)^3 }\Big[
1-x^2+2x\log x
\Big],
\end{align}
normalized such that $F_i^j(1)=1$. The functions for the
$\epsilon$-dependent parts are defined as 
\begin{align}
F^{C}_{1\epsilon}(x) & = F^{C}_{1}(x) \left(
\frac{-x^3 + 6 x^2 + 15 x + 2 - 6 x \log x}{12 x} \right) +
\frac{x^2 - 8 x - 4}{6 x}
 \label{loopeqfceps1}
,\\
F^{C}_{2\epsilon}(x) & = F^{C}_{2}(x)  \left(
\frac{-2 x^2 + 8 x + 6 - 4 \log x}{8} \right) + \frac{3 x -
  15}{8} \label{loopeqfceps2}
,\\
F^{N}_{1\epsilon}(x) & = F^{N}_{1}(x) \left( \frac{2
  x^3 +  15 x^2 + 6 x - 1 - 6 x ^2 \log x}{ 12 x^2} \right) + \frac{1
  - 8 x - 4 x^2}{6 x^2}, \label{loopeqfneps1}\\ 
F^{N}_{2\epsilon}(x) & = F^{N}_{2}(x) \left(
\frac{x^2 + 4 x + 1 - 2 x \log x}{4 x} \right) - \frac{3 x + 3}{4 x},
 \label{loopeqfneps2}
\end{align}
and are normalized to $F^j_{i\epsilon}(1)=0$.

The one-loop functions with two mass ratios can be related
to the loop functions of single mass ratios. For $k=1,2$, we have
\begin{align}
{\cal F}_k^C(x_i,x_j)
&= \frac{{\cal G}_k^C(x_i) - {\cal G}_k^C(x_j)}{x_i-x_j}
&(k=1,2),
\end{align}
where 
\begin{align}
{\cal G}_k^C(x)&=\int{\cal F}_k^C(x)
&(k=1,2).
\end{align}
In this way the ${\cal G}_{1,2}^C$ are defined up to irrelevant
constants. The third chargino one-loop function can be expressed
in terms of new one-variable functions as
\begin{align}
{\cal F}_3^C(x_i,x_j)&=
\frac{{\cal G}_{3a}^C(x_i)+{\cal G}_{3a}^C(x_j)}{x_i-x_j}
+\frac{{\cal G}_{3b}^C(x_i)-{\cal G}_{3b}^C(x_j)}{(x_i-x_j)^2}
\end{align}
with
\begin{align}
\begin{split}
{\cal G}_{3a}^C(x)&=
-\frac{x 
[1 - \epsilon \,
  L(m_{\tilde{\nu}_{\mu}}^2)]
}{8 (x-1)^2}
\Big[
-2 (-1+x+(x-2) \log x)
\\&
+\epsilon  
\left(3-3 x
-(x-4) \log x
+(x-2) \log ^2x\right)\Big]
\end{split}
,\\
{\cal G}_{3b}^C(x) &=
\frac{x^2 \log x 
[1 - \epsilon \,
  L(m_{\tilde{\nu}_{\mu}}^2)]}{4 (x-1)}
\Big[ -2-3 \epsilon+\epsilon  \log x\Big]
.
\end{align}
Similarly to the case with only one mass ratio, the one-loop functions
can be decomposed into terms of ${\cal O}(\epsilon^0,\epsilon^1)$, as
\begin{align} 
\mathcal{F}^{C}_{k}(x_{i},x_{j}) & = F^{C}_{k}(x_{i},x_{j})[1 - \epsilon \,
  L(m_{\tilde{\nu}_{\mu}}^2)]  + 
\epsilon \, F^{C}_{k \epsilon}(x_{i},x_{j}) ,\\ 
 \mathcal{G}^{C
}_{k}(x) & = G^{C}_{k}(x)[1 - \epsilon \, L(m_{\tilde{\nu}_{\mu}}^2)]  + \epsilon \, G^{C}_{k \epsilon}(x),\end{align}
with
\begin{align}
  F^{C}_{k}(x_{i},x_{i}) &= F^{C}_{k}(x_{i}), & F^{C}_{k \epsilon}(x_{i},x_{i}) &= F^{C}_{k \epsilon}(x_{i}), & (k \in \{1,2,3\})\\
  F^{C}_{3}(x_{i}) &= 0, & F^{C}_{3 \epsilon}(x_{i}) &= 0.
\end{align}
For reference we list the explicit expressions for the $G^{C}_{k}$ and $G^{C}_{k \epsilon}$ are as follows:
\begin{align}
  G^{C}_{1}(x) &= \frac{2 x (-2 + (2 - 3\log x) x + x^{2}\log x)}{(-1 + x)^{3}},\\
  G^{C}_{1 \epsilon}(x) &= \frac{x (-22 + (22 - 27 \log x +9 \log^{2} x) x + (5 \log x - 3 \log^{2} x) x^{2})}{3(-1 + x)^{3}},\\
  G^{C}_{2}(x) &= \frac{3 (-1 + x - 2 x \log x + x^{2} \log x)}{2 (-1 + x)^{2}},\\
  G^{C}_{2 \epsilon}(x) &= \frac{3 (-3 + (3 - 4 \log x + 2 \log^{2} x) x + (\log x - \log^{2} x) x^{2})}{4 (-1 + x)^{2}},\\
  G^{C}_{3a}(x) &= \frac{x (- 1 - 2 \log x + (1 + \log x) x)}{4 (-1 + x)^{2}},\\
  G^{C}_{3a \epsilon}(x) &= \frac{x (-3 - 4 \log x + 2 \log^{2} x + (3 + \log x - \log^{2} x) x)}{8 (-1 + x)^{2}},\\
  G^{C}_{3b}(x) &= -\frac{x^{2} \log x}{2 (-1 + x)},\\
  G^{C}_{3b \epsilon}(x) &= \frac{x^{2}(-3 \log x + \log^{2} x)}{4 (-1 + x)}.
\end{align}

\end{appendix}


\begin{thebibliography}{AA}
  

\bibitem{Bennett:2006}G.W. Bennett, et al.,
  (Muon $(g-2)$ Collaboration), Phys. Rev. D {\bf 73}, 072003 (2006).
  
\bibitem{Kinoshita2012}
  T.~Aoyama, M.~Hayakawa, T.~Kinoshita and M.~Nio,
  Phys.\ Rev.\ Lett.\  {\bf 109} (2012) 111808
  [arXiv:1205.5370 [hep-ph]].
  
\bibitem{Davier}
  M.~Davier, A.~Hoecker, B.~Malaescu and Z.~Zhang,
  Eur.\ Phys.\ J.\ C {\bf 71} (2011) 1515
  [Erratum-ibid.\ C {\bf 72} (2012) 1874]
  [arXiv:1010.4180 [hep-ph]].
  
\bibitem{HMNT}
  K.~Hagiwara, R.~Liao, A.~D.~Martin, D.~Nomura and T.~Teubner,
  J.\ Phys.\ G G {\bf 38} (2011) 085003
  [arXiv:1105.3149 [hep-ph]].
  
\bibitem{Benayoun:2012wc}
  M.~Benayoun, P.~David, L.~DelBuono and F.~Jegerlehner,
  Eur.\ Phys.\ J.\ C {\bf 73} (2013) 2453
  [arXiv:1210.7184 [hep-ph]].
  
\bibitem{JegerlehnerSzafron}
  F.~Jegerlehner and R.~Szafron,
  Eur.\ Phys.\ J.\ C {\bf 71} (2011) 1632
  [arXiv:1101.2872 [hep-ph]].
  
\bibitem{JegerlehnerNyffeler}
  F.~Jegerlehner and A.~Nyffeler,
  Phys.\ Rept.\  {\bf 477} (2009) 1
  [arXiv:0902.3360 [hep-ph]].
  
\bibitem{dRPV}
  J.~Prades, E.~de Rafael and A.~Vainshtein,
  (Advanced series on directions in high energy physics. 20)
  [arXiv:0901.0306 [hep-ph]].
  
\bibitem{Goecke:2010if}
  T.~Goecke, C.~S.~Fischer and R.~Williams,
  Phys.\ Rev.\ D {\bf 83} (2011) 094006
  [Erratum-ibid.\ D {\bf 86} (2012) 099901]
  [arXiv:1012.3886 [hep-ph]].
  
\bibitem{Bijnens:2012an}
  J.~Bijnens and M.~Z.~Abyaneh,
  EPJ Web Conf.\  {\bf 37} (2012) 01007
  [arXiv:1208.3548 [hep-ph]].
  
\bibitem{Masjuan:2012qn}
  P.~Masjuan and M.~Vanderhaeghen,
  arXiv:1212.0357 [hep-ph].
  
\bibitem{Blum:2013qu}
  T.~Blum, M.~Hayakawa and T.~Izubuchi,
  PoS LATTICE {\bf 2012} (2012) 022
  [arXiv:1301.2607 [hep-lat]].
  
\bibitem{ATLAS:2013mma}
  [ATLAS Collaboration],
  ATLAS-CONF-2013-014.
  
\bibitem{CMS:yva}
  [CMS Collaboration],
  CMS-PAS-HIG-13-005.
  
\bibitem{Gnendiger:2013pva}
  C.~Gnendiger, D.~St{\"o}ckinger and H.~St{\"o}ckinger-Kim,
  Phys.\ Rev.\ D {\bf 88} (2013) 053005
  [arXiv:1306.5546 [hep-ph]].


\bibitem{CKM1} 
  A.~Czarnecki, B.~Krause and W.~J.~Marciano,
  Phys.\ Rev.\ D {\bf 52} (1995) 2619
  [hep-ph/9506256].
  
\bibitem{CKM2}
  A.~Czarnecki, B.~Krause and W.~J.~Marciano,
  Phys.\ Rev.\ Lett.\  {\bf 76} (1996) 3267
  [hep-ph/9512369].

\bibitem{CzMV}
  A.~Czarnecki, W.~J.~Marciano and A.~Vainshtein,
  Phys.\ Rev.\ D {\bf 67} (2003) 073006
  [Erratum-ibid.\ D {\bf 73} (2006) 119901]
  [hep-ph/0212229].
  
\bibitem{HSW04}
  S.~Heinemeyer, D.~St\"ockinger and G.~Weiglein,
  Nucl.\ Phys.\ B {\bf 699} (2004) 103
  [hep-ph/0405255].
  
  
\bibitem{Gribouk}
  T.~Gribouk and A.~Czarnecki,
  Phys.\ Rev.\ D {\bf 72} (2005) 053016
  [hep-ph/0509205].
  
\bibitem{MdRRS}
  J.~P.~Miller, E.~d.~Rafael, B.~L.~Roberts and D.~St\"ockinger,
  Ann.\ Rev.\ Nucl.\ Part.\ Sci.\  {\bf 62} (2012) 237.
  
\bibitem{Carey:2009zzb}
  R.~M.~Carey, K.~R.~Lynch, J.~P.~Miller, B.~L.~Roberts, W.~M.~Morse, Y.~K.~Semertzides, V.~P.~Druzhinin and B.~I.~Khazin {\it et al.},
  FERMILAB-PROPOSAL-0989.
  
  
\bibitem{Roberts:2010cj}
  B.~L.~Roberts,
  Chin.\ Phys.\ C {\bf 34} (2010) 741
  [arXiv:1001.2898 [hep-ex]].
  
\bibitem{Iinuma:2011zz}
  H.~Iinuma [J-PARC New g-2/EDM experiment Collaboration],
  J.\ Phys.\ Conf.\ Ser.\  {\bf 295} (2011) 012032.
  
\bibitem{fsf2loopA}
  H.~Fargnoli, C.~Gnendiger, S.~Pa{\ss}ehr, D.~St{\"o}ckinger and H.~St{\"o}ckinger-Kim,
  arXiv:1309.0980 [hep-ph]; Phys.\ Lett.\  B, in press.
  

\bibitem{CzM}
  A.~Czarnecki and W.~J.~Marciano,
  Phys.\ Rev.\ D {\bf 64} (2001) 013014
  [hep-ph/0102122].
  
\bibitem{review}
  D.~St\"ockinger,
  J.\ Phys.\ G {\bf 34} (2007) R45
  [hep-ph/0609168].
  
\bibitem{Benbrik:2012rm} 
  R.~Benbrik, M.~Gomez Bock, S.~Heinemeyer, O.~Stal, G.~Weiglein and L.~Zeune,
  Eur.\ Phys.\ J.\ C {\bf 72}, 2171 (2012)
  [arXiv:1207.1096 [hep-ph]].
  
\bibitem{Arbey:2012dq} 
  A.~Arbey, M.~Battaglia, A.~Djouadi and F.~Mahmoudi,
  JHEP {\bf 1209}, 107 (2012)
  [arXiv:1207.1348 [hep-ph]].
  
\bibitem{Endo:2013bba}
  M.~Endo, K.~Hamaguchi, S.~Iwamoto and T.~Yoshinaga,
  arXiv:1303.4256 [hep-ph].
  
\bibitem{Ibe:2012qu}
  M.~Ibe, S.~Matsumoto, T.~T.~Yanagida and N.~Yokozaki,
  JHEP {\bf 1303} (2013) 078
  [arXiv:1210.3122 [hep-ph]].
  
\bibitem{1304.2508}
  G.~Bhattacharyya, B.~Bhattacherjee, T.~T.~Yanagida and N.~Yokozaki,
  arXiv:1304.2508 [hep-ph].
  
\bibitem{Cheng:2013hna}
  T.~Cheng and T.~Li,
  Phys.\ Rev.\ D {\bf 88} (2013) 015031
  [arXiv:1305.3214 [hep-ph]].
  
\bibitem{Ibe:2013oha}
  M.~Ibe, T.~T.~Yanagida and N.~Yokozaki,
  JHEP {\bf 1308} (2013) 067
  [arXiv:1303.6995 [hep-ph]].
  
\bibitem{Mohanty:2013soa}
  S.~Mohanty, S.~Rao and D.~P.~Roy,
  arXiv:1303.5830 [hep-ph].
  
\bibitem{Akula:2013ioa}
  S.~Akula and P.~Nath,
  Phys.\  Rev.\  D 87, {\bf 115022} (2013)
  [arXiv:1304.5526 [hep-ph]].
  
\bibitem{Evans:2012hg}
  J.~L.~Evans, M.~Ibe, S.~Shirai and T.~T.~Yanagida,
  Phys.\ Rev.\ D {\bf 85} (2012) 095004
  [arXiv:1201.2611 [hep-ph]].
  
\bibitem{Endo:2013lva}
  M.~Endo, K.~Hamaguchi, T.~Kitahara and T.~Yoshinaga,
  arXiv:1309.3065 [hep-ph].
  
\bibitem{Bechtle:2012zk} 
  P.~Bechtle, T.~Bringmann, K.~Desch, H.~Dreiner, M.~Hamer, C.~Hensel, M.~Kramer and N.~Nguyen {\it et al.},
  JHEP {\bf 1206}, 098 (2012)
  [arXiv:1204.4199 [hep-ph]].
  
\bibitem{Balazs:2012qc}
  C.~Balazs, A.~Buckley, D.~Carter, B.~Farmer and M.~White,
  arXiv:1205.1568 [hep-ph].
  
\bibitem{Buchmueller:2012hv}
  O.~Buchmueller, R.~Cavanaugh, M.~Citron, A.~De Roeck, M.~J.~Dolan, J.~R.~Ellis, H.~Flacher and S.~Heinemeyer {\it et al.},
  Eur.\ Phys.\ J.\ C {\bf 72} (2012) 2243
  [arXiv:1207.7315 [hep-ph]].

\bibitem{Endo:2013xka}
  M.~Endo, K.~Hamaguchi, S.~Iwamoto, T.~Kitahara and T.~Moroi,
  arXiv:1310.4496 [hep-ph].

\bibitem{WhitePaper}
  D.~W.~Hertzog, J.~P.~Miller, E.~de Rafael, B.~Lee Roberts and D.~St\"ockinger,
  arXiv:0705.4617 [hep-ph].
  
\bibitem{Adam:2010uz}
  C.~Adam, J.~-L.~Kneur, R.~Lafaye, T.~Plehn, M.~Rauch and D.~Zerwas,
  Eur.\ Phys.\ J.\ C {\bf 71} (2011) 1520
  [arXiv:1007.2190 [hep-ph]].
  
  
\bibitem{moroi}
  T.~Moroi,
  Phys.\ Rev.\ D {\bf 53} (1996) 6565
  [Erratum-ibid.\ D {\bf 56} (1997) 4424]
  [hep-ph/9512396].

\bibitem{MartinWells}
  S.~P.~Martin and J.~D.~Wells,
  Phys.\ Rev.\ D {\bf 64} (2001) 035003
  [hep-ph/0103067].
  

\bibitem{Cho:2011rk}
  G.~-C.~Cho, K.~Hagiwara, Y.~Matsumoto and D.~Nomura,
  JHEP {\bf 1111} (2011) 068
  [arXiv:1104.1769 [hep-ph]].

\bibitem{HSW03}
  S.~Heinemeyer, D.~St{\"o}ckinger and G.~Weiglein,
  Nucl.\ Phys.\ B {\bf 690} (2004) 62
  [arXiv:hep-ph/0312264].
  
\bibitem{ArhribBaek}
  A.~Arhrib and S.~Baek,
  Phys.\ Rev.\ D {\bf 65} (2002) 075002
  [hep-ph/0104225].
  
\bibitem{ChenGeng}
  C.~H.~Chen and C.~Q.~Geng,
  Phys.\ Lett.\ B {\bf 511} (2001) 77
  [arXiv:hep-ph/0104151].
  
\bibitem{amuPhotonicSUSY}
  P.~von Weitershausen, M.~Sch\"afer, H.~St\"ockinger-Kim and D.~St\"ockinger,
  Phys.\ Rev.\ D {\bf 81} (2010) 093004
  [arXiv:1003.5820 [hep-ph]].
  
\bibitem{DG98}
  G.~Degrassi and G.~F.~Giudice,
  Phys.\ Rev.\ D {\bf 58} (1998) 053007
  [arXiv:hep-ph/9803384].
  
\bibitem{Marchetti:2008hw}
  S.~Marchetti, S.~Mertens, U.~Nierste and D.~St{\"o}ckinger,
  Phys.\ Rev.\  D {\bf 79}, 013010 (2009)
  [arXiv:0808.1530 [hep-ph]].
  
\bibitem{Feng1}
  T.~F.~Feng, L.~Sun and X.~Y.~Yang,
  Phys.\ Rev.\  D {\bf 77} (2008) 116008
  [arXiv:0805.0653 [hep-ph]].
  
\bibitem{Feng2}
  T.~F.~Feng, L.~Sun and X.~Y.~Yang,
  Nucl.\ Phys.\  B {\bf 800} (2008) 221
  [arXiv:0805.1122 [hep-ph]].
  
\bibitem{Feng3}
  T.~F.~Feng and X.~Y.~Yang,
  Nucl.\ Phys.\  B {\bf 814} (2009) 101
  [arXiv:0901.1686 [hep-ph]].
  
\bibitem{FengLM06}
  T.~-F.~Feng, X.~-Q.~Li, L.~Lin, J.~Maalampi and H.~-S.~Song,
  Phys.\ Rev.\ D {\bf 73} (2006) 116001
  [hep-ph/0604171].
  
\bibitem{BarrZee}
  S.~M.~Barr and A.~Zee,
  Phys.\ Rev.\ Lett.\  {\bf 65} (1990) 21
  [Erratum-ibid.\  {\bf 65} (1990) 2920].

\bibitem{Yamanaka:2012qn}
  N.~Yamanaka,
  arXiv:1212.5800 [hep-ph].

\bibitem{Yamanaka:2012ia}
  N.~Yamanaka,
  Phys.\ Rev.\ D {\bf 87} (2013) 011701
  [arXiv:1211.1808 [hep-ph]].


  
\bibitem{flippingrules}
  A.~Denner, H.~Eck, O.~Hahn and J.~Kublbeck,
  Nucl.\ Phys.\ B {\bf 387} (1992) 467.

\bibitem{Takagi}
  T. Takagi,
  Japanese J. Math. 1 (1927) 83.


\bibitem{Fritzsche:2011nr} 
  T.~Fritzsche, S.~Heinemeyer, H.~Rzehak and C.~Schappacher,
  Phys.\ Rev.\ D {\bf 86}, 035014 (2012)
  [arXiv:1111.7289 [hep-ph]].
  
\bibitem{tf}
  T.~Fritzsche and W.~Hollik,
  Eur.\ Phys.\ J.\  C {\bf 24}, 619 (2002)
  [arXiv:hep-ph/0203159].
  
\bibitem{Heidi}
  W.~Hollik and H.~Rzehak,
  Eur.\ Phys.\ J.\  C {\bf 32} (2003) 127
  [arXiv:hep-ph/0305328].
  
\bibitem{HeinemeyerNew}
  T.~Fritzsche, T.~Hahn, S.~Heinemeyer, H.~Rzehak and C.~Schappacher,
  [arXiv:1309.1692 [hep-ph]].
  
\bibitem{Heinemeyer:2010mm}
  S.~Heinemeyer, H.~Rzehak and C.~Schappacher,
  Phys.\ Rev.\ D {\bf 82} (2010) 075010
  [arXiv:1007.0689 [hep-ph]].
  
\bibitem{Denner93}
  A.~Denner,
  Fortsch.\ Phys.\  {\bf 41} (1993) 307
  [arXiv:0709.1075 [hep-ph]].
  
\bibitem{HKRRSS}
  W.~Hollik, E.~Kraus, M.~Roth, C.~Rupp, K.~Sibold and D.~St\"ockinger,
  Nucl.\ Phys.\  B {\bf 639} (2002) 3
  [arXiv:hep-ph/0204350].
  
\bibitem{FeynArts}
  T. Hahn,
  Comput.~Phys.~Commun. {\bf 140} (2001) 418--431
  [arXiv:0012260 [hep-ph]].
  
\bibitem{FormCalc}
  T. Hahn,
  J.Phys.Conf.Ser. {\bf 368} (2012) 012054
  [arXiv:1112.0124 [hep-ph]].
  
\bibitem{Freitas:2002um}
  A.~Freitas and D.~St\"ockinger,
  Phys.\ Rev.\ D {\bf 66} (2002) 095014
  [hep-ph/0205281].
  
\bibitem{Baro:2008bg}
  N.~Baro, F.~Boudjema and A.~Semenov,
  Phys.\ Rev.\ D {\bf 78} (2008) 115003
  [arXiv:0807.4668 [hep-ph]].
  
\bibitem{HiggsDRbar}
  M.~Frank, T.~Hahn, S.~Heinemeyer, W.~Hollik, H.~Rzehak and G.~Weiglein,
  JHEP {\bf 0702} (2007) 047
  [hep-ph/0611326].
  
\bibitem{Sperling:2013eva}
  M.~Sperling, D.~St{\"o}ckinger and A.~Voigt,
  JHEP {\bf 1307} (2013) 132
  [arXiv:1305.1548 [hep-ph]].

\bibitem{Sperling:2013xqa} 
  M.~Sperling, D.~St{\"o}ckinger and A.~Voigt,
  arXiv:1310.7629 [hep-ph].

\bibitem{Heinemeyer:2011gk}
  S.~Heinemeyer, F.~von der Pahlen and C.~Schappacher,
  Eur.\ Phys.\ J.\ C {\bf 72} (2012) 1892
  [arXiv:1112.0760 [hep-ph]].

\bibitem{Bharucha:2012re}
  A.~Bharucha, S.~Heinemeyer, F.~von der Pahlen and C.~Schappacher,
  Phys.\ Rev.\ D {\bf 86} (2012) 075023
  [arXiv:1208.4106 [hep-ph]].

\bibitem{Chatterjee:2011wc}
  A.~Chatterjee, M.~Drees, S.~Kulkarni and Q.~Xu,
  Phys.\ Rev.\ D {\bf 85} (2012) 075013
  [arXiv:1107.5218 [hep-ph]].

\bibitem{Baro:2009gn}
  N.~Baro and F.~Boudjema,
  Phys.\ Rev.\ D {\bf 80} (2009) 076010
  [arXiv:0906.1665 [hep-ph]].

\bibitem{OneCalc}
  G.~Weiglein, R.~Scharf and M.~B{\"o}hm,
  Nucl.\ Phys.\ B {\bf 416} (1994) 606
  [hep-ph/9310358].
  
\bibitem{HIRS1}
  W.~Hollik, J.~I.~Illana, S.~Rigolin and D.~St{\"o}ckinger,
  Phys.\ Lett.\ B {\bf 416} (1998) 345
  [arXiv:hep-ph/9707437].

  
\bibitem{ChangKP98}
  D.~Chang, W.~Y.~Keung and A.~Pilaftsis,
  Phys.\ Rev.\ Lett.\  {\bf 82} (1999) 900
  [Erratum-ibid.\  {\bf 83} (1999) 3972]
  [arXiv:hep-ph/9811202].
  
\bibitem{Pilaftsis98}
  A.~Pilaftsis,
  Phys.\ Lett.\ B {\bf 471} (1999) 174
  [arXiv:hep-ph/9909485].
  
\bibitem{ChangCCK00}
  D.~Chang, W.~F.~Chang, C.~H.~Chou and W.~Y.~Keung,
  Phys.\ Rev.\ D {\bf 63} (2001) 091301
  [arXiv:hep-ph/0009292].
  
\bibitem{CheungCK01}
  K.~m.~Cheung, C.~H.~Chou and O.~C.~W.~Kong,
  Phys.\ Rev.\ D {\bf 64} (2001) 111301
  [arXiv:hep-ph/0103183].


\bibitem{PDG2012}
  J.~Beringer et al. (Particle Data Group)
  Phys.\ Rev.\ D {\bf 86} (2012) 010001. 
  
\bibitem{SPSDef}
  B.~C.~Allanach {\it et al.},
in {\it Proc. of the APS/DPF/DPB Summer Study on the Future of Particle Physics (Snowmass 2001) } ed. N.~Graf,
  Eur.\ Phys.\ J.\ C {\bf 25} (2002) 113.
  [eConf {\bf C010630} (2001) P125]
  [arXiv:hep-ph/0202233].

\bibitem{Grothaus:2012js}
  P.~Grothaus, M.~Lindner and Y.~Takanishi,
  JHEP {\bf 1307} (2013) 094
  [arXiv:1207.4434 [hep-ph]].


\bibitem{FH2}
  G.~Degrassi, S.~Heinemeyer, W.~Hollik, P.~Slavich and G.~Weiglein,
  Eur.\ Phys.\ J.\ C {\bf 28} (2003) 133
  [hep-ph/0212020].

\bibitem{FH3}
  S.~Heinemeyer, W.~Hollik and G.~Weiglein,
  Eur.\ Phys.\ J.\ C {\bf 9} (1999) 343
  [hep-ph/9812472].

\bibitem{FH4}
  S.~Heinemeyer, W.~Hollik and G.~Weiglein,
  Comput.\ Phys.\ Commun.\  {\bf 124} (2000) 76
  [hep-ph/9812320].

\bibitem{Delgado}
  A.~Delgado, G.~F.~Giudice, G.~Isidori, M.~Pierini and A.~Strumia,
  Eur.\ Phys.\ J.\ C {\bf 73} (2013) 2370
  [arXiv:1212.6847 [hep-ph]].

\bibitem{HeinemeyerBM}
  M.~Carena, S.~Heinemeyer, O.~Stal, C.~E.~M.~Wagner and G.~Weiglein,
  Eur.\  Phys.\  J.\ C {\bf 73} (2013) 2552
  [arXiv:1302.7033 [hep-ph]].

\bibitem{Batell:2013bka}
  B.~Batell, S.~Jung and C.~E.~M.~Wagner,
  arXiv:1309.2297 [hep-ph].

\end{thebibliography}
\end{document}